Thèse

# Plasmonic-Photonic Hybrid Nanodevice

Présentée devant
Ecole Centrale de Lyon

Pour obtenir
Le grade de docteur

École doctorale : Electronique, Electrotechnique et Automatisme de Lyon

Par
Taiping Zhang

Soutenue le 22 novembre 2012

**Jury**

| | | |
|---|---|---|
| Rapporteur | Thierry GROSJEAN | Chargé de Recherche CNRS |
| Rapporteur | David PEYRADE | Chercheur |
| Examinateur | Natalia DEL FATTI | Professeur des Universités |
| Directeur de Thèse | Xavier LETARTRE | Directeur de Recherche CNRS |
| Co-directeur de Thèse | Ali BELAROUCI | Chargé de recherche CNRS |
| Co-directeur de Thèse | Ségolène CALLARD | Professeur des Universités |

Institut des Nanotechnologies de Lyon (INL) – CNRS UMR 5270

# Acknowledgment

I would first to give my deepest gratitude to my Directeurs de thèse: Xavier Letartre, Ali Belarouci, Ségolène Callard for giving me the opportunity of doing doctoral research under their supervision.

I would like to thank all the colleagues who give me kind helps in technologies and theory during my thesis research: Cécile jamois, Céline Chevalier, Radoslaw Mazurczyk, Pedro Rojo-Romeo, Pierre Cremillieu, Jean-Louis Leclercq, Di Feng, Thanh-Phong Vo, Corrado Sciancalepore, Koku Kusiaku, Philippe Regreny, Brice Devif, Aziz Benamrouche, Regis Orobtchouk, Nicolas Chauvin, Pierre Viktorovitch, Michel Garrigues.

I would like to thank Prof. Natalia Del Fatti, Mr. Thierry Grosjean and Mr. David Peyrade as accepted to be my committee and for insightful comments and discussions.

Thank you all the colleagues who work at 4th floor of F7, Especially Thérèse Matin and Yves Robach for their kind helps. And same acknowledge to Nanophotonics group's memberss.

Last but not the least, I would like to thank my family: my parents for spiritually supporting me through my life.

# Contents

# General Introduction

Many of the breakthroughs in human technology have resulted from a deeper understanding of the properties of materials. In the last few decades one goal in this case is to control the optical properties of materials [1]. The development of the nanotechnologies opens the way to artificial materials that allows for a tight control of the spatial-temporal trajectory of photons. This approach, the so-called light harnessing, leads to numerous functionalities: strong confinement or slow-down of light, super prism phenomena… One of the most interesting possibility is to confine the light within a tiny space (as compared to the wavelength) and for long time (as compared to the period), while achieving efficient light addressing (to), or collection (from) the wave-length scale photonic structures where they are meant to be confined. The general approach to achieve strong confinement of photons consists in high index contrast structuring of space at the wavelength scale. Materials commonly in use for that purpose are structured metals or high optical index dielectrics, immersed in low index dielectric material such as e.g. silica or air [2].

High index dielectric materials, such as semiconductors, have been widely used for photon confinement, along the so called refraction and Photonic Crystal (PC) based diffraction schemes: for the former, use is made of the total or partial internal reflection of photons at the semiconductor-low index material interface; for the latter, diffractive phenomena occurring in periodically structured materials are exploited to control the spatial-temporal trajectory of photons [2, 3].

On the other hand, metals have attracted considerable attention in the context of plasmonics: very strong spatial confinement of photons can be attained in plasmonic nano-structures, such as in metallic nano-particles or nanoantennas (NAs) [4, 5], at the expense however of a limited resonance strength (or confinement time) as a result of optical losses induced by metal absorption and of radiation to free space continuum. In recent years, we have witnessed a flurry of activity in the fundamental research and development of surface plasmon based structures and devices [6]. Their unique properties enable a wide range of practical applications, including optical devices [7], optical energy transport [8], near field scanning optical microscopy [9, 10], surface-enhanced spectroscopies [11], and chemical and biological sensors [12].

Although it may seem that metallic nanoparticules (or optical antennas) are almost an off-the-self product, some issues need special attention and specific research. A key challenge is to address and collect light from those nano-scale systems. The tiny active area of the optical antenna is both an advantage for its miniaturization, and a real limit for the level of the collected signal. Therefore, one needs to reconsider how to drive efficiently such NAs. NAs are usually illuminated from the far field in a large, at best diffraction limited, focus. However, this procedure is not very efficient to address metallic micro-nano-structures or NAs. The reason lies in the difficulty to achieve a sufficient coupling rate between the incoming optical beam and the NA in order to compensate for the rather large optical losses (radiation to free space and metallic absorption). A new approach is proposed in the present work for the optimum addressing of NA with a free space optical beam via the use of an intermediate coupling resonator structure, which is aimed at providing the appropriate modal conversion of the incoming beam, in the time domain [2, 13].

In this thesis, we propose to tackle this important issue by designing and realizing a novel nano-optical device based on the use of a photonic crystal (PC) structure to generate an efficient coupling between the external source and a NA. On one hand, the PC structure can take the place of intermediate coupling resonator structure to address the NA better. On the other hand, the NA can localize the optical field of the PC cavity in a very tiny space to create a subwavelength light source. Details of the PC-NA coupling analysis and its optimization were published elsewhere [13].

Some researches with similar kinds of plasmonic-photonic hybrid nanodevices were approached by other research groups [14, 15, 16, 17, 18, 19]. They had studied far-field optical characterizations of the hybrid nanodevices and had proposed a wide range of applications. In our research, we expected to extend the study at the near-field level and to demonstrate a plasmonic-photonic crystal laser. This device may apply as a nano-laser source [20, 21, 22, 23].

In this dissertation, the content is arranged into three charpters. Chapter 1 introduces the theoritical background of this research including surface plasmon and photonic crystal concepts. This chapter also shows the design of the hybrid devices and demonstrates the numerical simulation of their optical properties. Chapter 2 mainly describes the process and the fabricated samples. The nanodevices are fabricated on an InP membrane substrate. The critical technology for the fabrication is complex electron beam lithography. With this technology the alignment of the positions of PC structure and NA is well controlled. Chapter 3 demonstrates the optical characterizations of the hybrid nanodevices including far-field characterizations and near-field characterizations. The far-field measurement is performed by micro-photoluminescence spectroscopy at room temperature. The results show that for the defect PC cavities, the presence of the NA influences the optical properties of the laser, such as lasing threshold and laser wavelength. The near-field measurement is performed by near-field scanning microscopy, at room temperature also. The investigation shows that the NA modifies the optical field distribution of the laser mode. The modification depends on the position and direction of the NA and it is sensitive to the polarization of the optical field.

**Bibliography**

[1] J.D. Joannopoulos, S.G. Johnson, J.N. Winn and R.D. Meade. *Photonic Crystals: Molding the Flow of Light,* 2nd Edn, (Princeton University Press).

[2] T. Zhang, A. Belarouci, S. Callard, P.R. Romeo, X. Letartre and P. Viktorovitch. *Plasmonic-photonic hybrid nanodevice,* International Journal of Nanoscience, 11(4) : 1240019 (2012).

[3] P. Viktorovitch, E. Drouard, M. Garrigues, J.L. Leclercq, X. Letartre, P.R. Romeo and C. Seassal. *Photonic Crystals: basic concepts and devices,* C. R. Physique, 8 : 253-266 (2007).

[4] S.A. Maier. *Effective mode volume of nanoscale plasomn cavities,* Optical and Quantum Electronics, 38 : 257-267 (2006).

[5] P. Mühlschlegel, H.-J. Eisler, O.J.F. Martin, B. Hecht and D.W. Pohl. *Resonant Optical Antennas,* Science, 308(5728) : 1607-1609 (2005)

[6] E. Ozbay. *Plasmonics : Merging Photonics and Electronics at Nanoscale Dimensions,* Science, 311(5758) : 189-193 (2006).

[7] Y. Dirix, C. Bastiaansen, W. Caseri and P. Smith. *Oriented pearl-necklace arrays of metallic nanoparticals in polymers: A new route toward polarization-dependent color filters,* Adv. Mater., 11 : 223-227 (1999).

[8] J.B. Pendry. *Playing tricks with light,* Science, 285 : 1687-1688 (2002).

[9] B. Knoll and F. Keilmann. *Near-field probing of vibrational absorption for chemical microscopy,* Nature, 399 : 134-137 (1999).

[10] B. Razavi. *A 5.2-GHz CMOS receiver with 62-dB image rejection,* IEEE J. Solid-State Circuits, 36(5) : 810-815 (2001).

[11] S. Nie and S.R. Emory. *Probing Single Molecules and Single Nanoparticals by Surface-Enhanced Raman Scattering,* Science, 277 : 1102-1106 (1997).

[12] R. Elghanian, J.J Storhoff, R.C. Mucic, R.L Letsinger and C.A. Mirkin. *Selective Colorimetric Detection of Polynucleotides Based on the Distance-Dependent Optical Properties of Gold Nanoparticles,* Science, 277(5329) : 1078-1081 (1997).

[13] A. Belarouci, T. Benyattou, X. Letartre and P. Viktorovitch. *3D light harnessing based on couping engineering between 1D-2D Photonic Crystal memberanes and metallic nano-antenna,* Optics Express, 18(S3) : A381-A394 (2010).

[14] F. De Angelis, M. Patrini, G. Das, I. Maksymov, M. Galli, L. Businaro, L.C. Andreani and E. Di Fabrizio. *A Hybrid Plasmonic-Photonic Nanodevice for Label-Free Detection of A Few Molecules,* Nano Lett., 8 : 2321-2327 (2008).

[15] M. Barth, J. Stingl, J. Kouba, N. Nüsse, B. Löchel and O. Benson. *A hybrid approach towards nanophotonic devices with enhanced functionality,* Phys. Status Solidi B, 246(2) : 298-231 (2009).

[16] M. Barth, S. Schietinger, S. Fischer, J. Becker, N. Nüsse, T. Aichele, B. Löchel, C. Sönnichsen and O. Benson. *Nanoassembled Plasmonic-Photonic Hybrid Cavity for Tailored Light-Matter Coupling,* Nano Lett., 10 : 891-895 (2010).

[17] F. De Angelis, G. Das, P. Candeloro, M. Patrini, M. Galli, A. Bek, M. Lazzarino, I. Maksymov, C. Liberale, L.C. Andreani and E. Di Fabrizio. *Nanoscale chemical mapping using three-dimensional adiabatic compression of surface plasmon polaritons,* Nature Nanotechnology, 5 : 67-72 (2010).

[18] F.J. Gonzalez and J. Alda. *Optical Nanoantennas Coupled to Photonic Crystal Cavities and Waveguides for Near-Field Sensing,* IEEE Journal of Selected Topics in Quantum Electronics, 16(2) : 446-449 (2010).

[19] I.S. Maksymov. *Optical switching and logic gates with hybrid plasmonic-photonic crystal nanobeam cavities,* Physics Letters A, 375 : 918-921 (2011).

[20] R.F. Oulton, V.J. Sorger, T. Zentgraf, R-M. Ma, C. Gladden, L. Dai, G. Bartal and X. Zhang. *Plasmon lasers at deep subwavelength scale,* Nature, 461 : 629-632 (2009).

[21] S-H. Kwon, J-H. Kang, C, Seassal, S-K. Kim, P. Regreny, Y-H. Lee, C. M. Lieber and H-G. Park. *Subwavelength Plasmonic Lasing from a Semiconductor Nanodick with Sliver Nanopan Cavity,* Nano Lett., 10 : 3679-3683 (2010).

[22] V.J. Sorger and X. Zhang. *Spotlight on Plasmon Lasers,* Science, 333 : 709-710 (2011).

[23] R-M. Ma, X. Yin, R.F. Oulton, V.J. Sorger and X. Zhang. *Multiplexed and Electrically Modulated Plasmon Laser Circuit,* Nano. Lett., 12(10) : 5396-5402 (2012).

# Chapter 1: Research Background and Device Design

## Introduction

Photonics at the nanoscale, or *nanophotonics* might be defined as: the science and engineering of light-matter interactions that take place on wavelength and subwavelength scales where the physical, chemical, or structural nature of natural or artificial nanostructure matter controls the interactions. Broadly speaking, over the next ten years nanophotonic structures and devices promise dramatic reductions in energies of device operation, densely integrated information systems with lower power dissipation, enhanced spatial resolution for imaging and patterning, and new sensors of increased sensitivity and specificity. Both photonics and plasmonics concern investigations into building, manipulating, and characterizing optically active nanostructures with a view to creating new capabilities in instrumentation for the nanoscale, chemical and biomedical sensing, information and communications technologies, enhanced solar cells and lighting, disease treatment, environmental remediation, and many other applications. Photonics and plasmonics share the characteristic that at least some of their basic concepts have been known for 40–50 years, but they have come into their own only in the last ten years, based on recent discoveries in nanoscience. Photonic materials and devices have played a pervasive role in communications, energy conversion, and sensing since the 1960s and 1970s.

Many of the critical concepts for today's rapidly growing and diverse field of nanophotonics were established in past decades, but recent progress in the ability to control materials at the nanoscale in multiple dimensions has allowed the validation of those concepts and the anticipation of yet more intriguing and powerful photonic behaviors. For example, an early paper by Yablonovitch (1987) discussed dielectric materials with spatial variations in index of refraction on the order of a wavelength of light. He anticipated that these *photonic crystals* would have a dramatic effect on the spontaneous emission of light within these structures. Negative index materials, a component of a class of *metamaterials,* were anticipated as early as the 1960s by Veselago (1968). *Surface plasmons* have played an important role in surface-enhanced

Raman spectroscopy (SERS), an active area of research since the late 1970s. Today's research in plasmonics encompasses an even broader range of structures and applications, as is the case for photonics. Plasmonics aims to exploit the unique optical properties of metallic nanostructures to enable routing and active manipulation of light at the nanoscale. It has only been in the past 10 years that the young field of plasmonics has rapidly gained momentum, enabling exciting new fundamental science as well as groundbreaking real-life applications in the coming 10 years in terms of targeted medical therapy, ultrahigh-resolution imaging and patterning, and control of optical processes with extraordinary spatial and frequency precision. In addition, because plasmonics offers a natural integration compatibility with electronics and the speed of photonics, circuits and systems formed of plasmonic and electronic devices hold promise for next-generation systems that will incorporate the best qualities of both photonics and electronics for computation and communication at high speed, broad bandwidth, and low power dissipation.

## 1.1 Advances in the last 10 years and current status

### 1.1.1 Photonics

Over the last 5–10 years we have witnessed an enormous increase in the number and diversity of photonics applications. This has resulted from the significant advances in computational design tools and their accessibility, the emergence of new nanofabrication techniques, and the realization of new optical and structural characterization methods. At the forefront of these advances are the developments in the area of micro- and nano-photonic devices that have dimensions on the order of or below the wavelength of light.

Advances in electromagnetic and electronic device simulation tools have been tremendous over the last decade. Good commercial and freeware codes (e.g., finite difference time domain (FDTD), discrete dipole approximation (DDA), boundary element methods (BEM), or finite-difference frequency domain (FDFD) codes are inexpensive and available to virtually everyone. Improvements in nanofabrication techniques, greater accessibility of high-resolution patterning (e.g., electron beam lithography), and pattern transfer processes (e.g., low-damage ion-assisted etching) have produced photonic crystal, microdisk, and ring resonator devices with exceptional performance.

Discovery in nanophotonics has been enabled by the accessibility of optical nanoscale characterization tools such as scanning near-field optical microscopy (SNOM), atomic force microscopy (AFM), nano-Auger, nano-secondary ion mass spectrometry (nano-SIMS), scanning electron microscopes (SEMs), and transmission electron microscopes (TEM). These instruments have had a major impact on the ability to correlate the size, atomic structure, and spatial arrangements of nanostructure to the observed optical properties.

The progress in both simulation and fabrication of nanophotonic structures has resulted in the formation of ultrahigh-quality (Q, meaning low-optical-loss) optical structures [1]. These structures, in turn, have allowed researchers to engineer distinctive optical states, localize and slow the velocity of light, and create efficient light-emitting sources and strongly coupled light–matter interactions, resulting in new quantum mechanical states:

• *Slowing of light* has been observed in photonic crystal waveguides [2, 3] and in coupled ring resonators [4]. This is an achievement not only scientifically, but also for the implication in controlled delay and storage of light in compact, on-chip information processing.

• *Strong coupling between dielectric nanocavities and quantum dots* has by now been observed by a number of research groups [5, 6, 7]. The exciton-photon (polariton) states that result can form the basis of quantum information schemes, or produce ultra-low threshold lasers.

## 1.1.2 Plasmonics

Although surface-enhanced Raman spectroscopy (SERS), one of the first "killer applications" of metallic nanostructures, was discovered in the 1970s [8, 9, 10], the young field of plasmonics only started to rapidly spread into new directions in the late 1990s and early 2000s. At that time it was demonstrated in rapid succession that metallic nanowires can guide light well below the seemingly unsurpassable diffraction limit [11], that metal films with nanoscale holes show extraordinarily high optical transmission [12], and that a simple thin film of metal can serve as an optical lens [13]. Plasmonic elements further gained importance as popular components of *metamaterials*, i.e., artificial optical materials with rationally designed geometries and arrangements of nanoscale building blocks. The burgeoning field of transformation optics elegantly demonstrates how such materials can facilitate unprecedented control over light [14].

As these novel phenomena captured the imagination of a broad audience, researchers in the lab started to also gain respect for some of the severe limitations of metals. The most important challenge that still persists today is that metals exhibit substantial resistive heating losses when they interact with light. For this reason, it will be valuable to explore ways to get around that issue. In some cases, local heat generation may be used to advantage, and in some cases heating losses can be neglected. A very diverse set of plasmonics applications has emerged in the last ten years. Early applications included the development of high-performance near-field optical microscopy (NSOM) and biosensing methods. More recently, many new technologies have emerged in which the use of plasmonics seems promising, including thermally assisted magnetic recording [15], thermal cancer treatment [16], catalysis and nanostructure growth [17], solar cells [18, 19], and computer chips [20, 21]. It now has been established that modulators and detectors can be achieved that meet the stringent power, speed, and materials requirements necessary to incorporate plasmonics with CMOS technology. Plasmonic sources capable of efficiently coupling quantum emitters to a single, well-defined optical mode may first find applications in the field of quantum plasmonics and later in power-efficient chip-scale optical sources [22, 23]. In this respect, the recent prediction [24] and realization [25, 26, 27] of coherent nanometallic light sources constitutes an extremely important development.

## 1.1.3 Exploit Synergies between Plasmonics, Photonics, and Electronics

Over the last decade, it has gradually become clear what role plasmonics can play in future device technologies and how it can complement electronics and conventional photonics. Each of these device technologies can perform unique functions that play to the strength of the key materials. The electrical properties of semiconductors enable the realization of truly nanoscale elements for computation and information storage; the high transparency of dielectrics (e.g., glass) facilitates information transport over long distances and at very high data rates. Unfortunately, semiconductor electronics is limited in speed by interconnect (RC) delay-time issues, and photonics is limited in size by the fundamental laws of diffraction. Plasmonics offers precisely what electronics and photonics do not have: the size of electronics and the speed of photonics. Plasmonic devices might therefore naturally interface with similar-speed photonic devices and with similar-size electronic components, increasing the synergy between these technologies. The assembly of hybrid nanophotonic devices from different fundamental photonic entities—such as single molecules, nanocrystals, semiconductor quantum dots, nanowires and metal nanoparticles—can yield functionalities that exceed those of the individual subunits. The next generation of challenges in the design and applications of nanophotonic circuits, fueled by the continuing development of powerful and flexible top-down and bottom-up nanofabrication techniques, lies in understanding and manipulating collective phenomena — phenomena that emerge from the interactions of the individual photonic, plasmonic, electronic and mechanical components. The scattering, radiative and mechanical properties of structures and materials dominated by collective phenomena can differ significantly from those of individual components. Additional degrees of freedom offered by complex heterogeneous nanostructures can be used to obtain new device functionality through coupling-induced tailored control of fundamental physical processes.

Although merging these different systems on a single hybrid platform is at present challenging, it promises improved performance and novel devices. Particularly rapid progress is seen in the combination of plasmonic–dielectric constituents with quantum emitters that can be assembled on demand into fundamental model systems for future optical elements. Photonic and plasmonic nanostructures open up fascinating opportunities for controlling and harvesting light-matter interactions on the nanoscale. In particular, a capability of optical microcavities and plasmonic nanostructures to manipulate the local density of electromagnetic states could open up a variety of practical applications in optical communications and biomedical research. Properly-designed photonic and plasmonic nanostructures could also provide a useful testbed for the exploration of novel physical regimes in atomic physics and quantum optics.

In the scope of this thesis, we approach this technological relevant challenge by considering the design, fabrication and characterization of hybrid photonic–plasmonic devices that exploit the coupling between localized surface plasmon resonance (LSPR) of a bowtie antenna and a photonic mode provided by photonic crystal structures. In the next section the basic building blocks of the optoplasmonic structure will be highlighted.

## 1.2 Basic building blocks of the hybrid device:

### 1.2.1 Metallic nanostructures

#### 1.2.1.1 The world of Plasmon

Many of the fundamental electronic properties of the solid state can be described by the concept of single electrons moving between an ion lattice. The motion of the electron gas in metals can be described by a classical equation of dynamics where the electron cloud follows a harmonic oscillatory behaviour. If we consider the metal as a three-dimensional crystal of positive ions and a free electron gas moving on the periodic potential created by the ions, when an electric or electromagnetic excitation is applied to the metal, the electron cloud will react to the corresponding electric field displacing from its equilibrium position. On the other hand, the positive ionic background induces a restoring force attracting the electron cloud back to its original position. The action of those two forces will cause an oscillation of the electron gas at a frequency $\omega_p$ around the equilibrium position:

$$\omega_p = \sqrt{\frac{4\pi n e^2}{m_e}} \qquad (1.1)$$

Where $\omega_p$ refer to the volume or bulk plasma frequency, n is the electron density, and $m_e$ is the effective mass of the electron. The specific value of $\omega_p$ for most metals lies in the ultraviolet region, which is the reason why they are shiny and glittering in the visible spectrum. A light wave with $\omega < \omega_p$ is reflected, because the electrons in the metal screen the light. On the other hand if $\omega > \omega_p$ the light wave gets transmitted (the metal becomes transparent), since the electrons in the metal are too slow and cannot respond fast enough to screen the field.

### 1.2.1.2 Surface Plasmon Polaritons

Surface plasmon polaritons are similar to bulk plasmons however they are typically bound to and propagate along metal-dielectric interfaces. In this case, a combined excitation of free electrons i.e. a plasmon and a photon form a propagating mode bound to the metal-dielectric interface. A necessary condition for surface plasmon polaritons is that materials at the interface must have a Re($\varepsilon$) of opposite sign. Metals, which are strongly absorbing, inherently have a Re($\varepsilon$) < 0 and when interfaced to a dielectric this condition is often met. Upon excitation of the surface plasmon polaritons, transverse and longitudinal electromagnetic fields propagate where a maximum intensity is present at the interface. Beyond and away from the interface, the field exhibits near-field or evanescent behavior with exponential decay from the interface (Figure 1.1). The frequency of surface plasma polaritons, $\omega_{sp}$ is then:

$$\omega_{sp} = \frac{\omega_p}{\sqrt{2}} \tag{1.2}$$

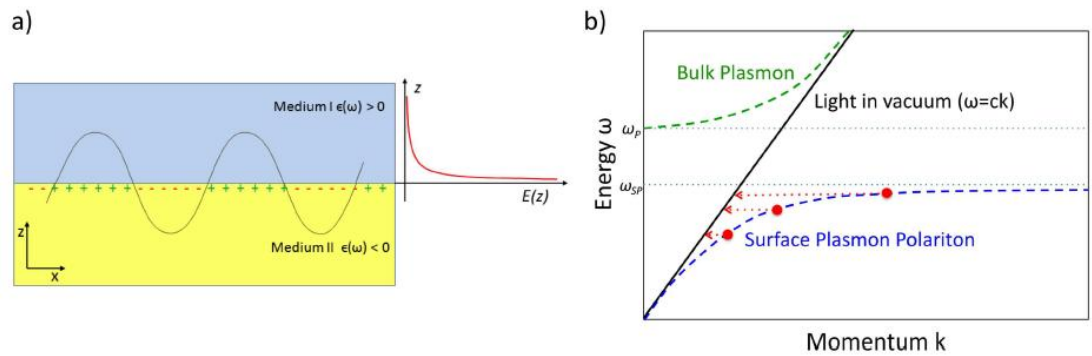

*Figure 1.1: a) Schematics of SPP at an interface separating two semi infinite regions with two dielectric functions of different sign, propagating along x direction and confined in z direction. b) Energy (ω) versus momentum (k) dispersion line of the propagating surface plasmon polariton (blue-dashed line). The bulk plasma frequency $\omega_p$ and the surface plasmon energy $\omega_{sp} = \omega_p/\sqrt{2}$ are marked as dotted horizontal lines.*

### 1.2.1.3 Localized surface plasmons (LSP)

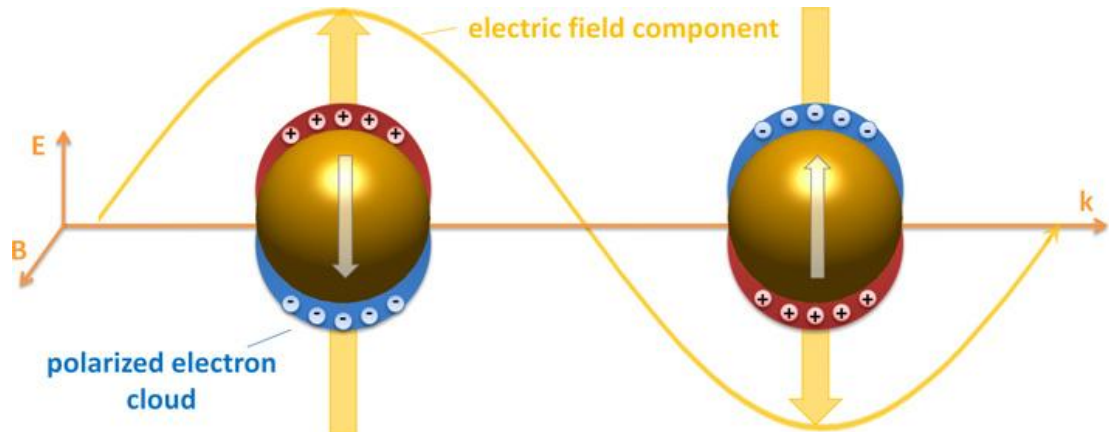

*Figure 1.2: Excitation of particles plasmons through the polarization of metallic nanoparticles. At the resonance frequency the plasmons are oscillating with a 90° phase difference (180° above resonance).*

In general, we have seen that plasmons arise from an interplay of electron density oscillations and the exciting electromagnetic fields. In this sense, we should talk about surface plasmon polaritons and also distinguish the propagating (evanescent) modes at the interface of a metal and a dielectric from their localized counterpart at the surface of metallic particles (so called *particle plasmon polaritons*). If an electromagnetic wave impinges on a metallic nanoparticle (whose spatial dimension is assumed to be much smaller than the wavelength of light), the electron gas gets polarized (polarization charges at the surface) and the arising restoring force again forms a plasmonic oscillation (Figure 1.2)

The metallic particle thus acts like an oscillator and the corresponding resonance behaviour determines the optical properties. Since we solely discuss particle plasmon polaritons in this work, we will furthermore always refer to them.

LSPs exhibit some distinct and special properties which have generated promising prospects in a variety of fields of science and technology. We summarize some of these properties:

• *Localization of the electromagnetic fields:* The fields induced by surface plasmons decay exponentially as we move away from the interface where they are excited. Thus, all the electromagnetic effects induced by those fields are reduced to a small region close to the surface (typically some nanometers). Complementarily, plasmons have shorter wavelength compared to the incoming source that excites them. As a consequence of these two effects, it is possible to concentrate light on regions smaller than the diffraction limit. Thanks to this property, for example greater resolution can be achieved if plasmons are used for imaging purposes. On the other hand, this feature is also essential for spectroscopic applications where a small amount of molecules located near a surface can be sufficient for the detection purpose.

• *Field enhancement:* The plasmonic resonances induce charge pile up at the surface of the metals and those charges induce electromagnetic fields on the surroundings of the surface. Therefore, the fields are enhanced compared to electro-magnetic fields in free space. This field enhancement is a key property for many photonic applications, such as spectroscopy and sensing.

• *Tunability:* Plasmons have also the capability of tuning the energy of their resonances through the modification of both the geometry and/or the coupling between different systems [28, 29, 30]. This opens up the door to engineer the optical spectral response in optically active systems to optimize the properties of plasmons for the requirements of a particular application. In addition, plasmons are extremely sensitive to the environment that surrounds them. Thus, a small change on the dielectric function of the surrounding medium can produce large spectral shifts on the far-field radiation of the system.

Such metallic nanostructures with specific length can act as efficient resonant antennas with strong potential for field enhancement schemes. Among diverse nanoantenna designs, closely placed particle pairs recently have attracted extensive research interest, and they are usually referred to as the nanoparticle dimmers. Early studies of the two interacting metal nanoparticles included a theoretical study in [31], an experimental observation of gold spherical particles in [32] and systematic studies of separation distance of gold nanodisks in [33] and elliptical pairs in [34]. These studies of nanoparticle pairs were important because their coupling effect will lead to stronger field enhancement as well as much more effective confinement compared with a single particle [35, 36, 37]. In particular, obvious concentrations of luminance were observed near the sharp tips or inside the gap between the two particles of these dimmer nanoantennas [38]. The resonance can be further controllable by changing certain dimensional parameters and thus has potential engineering applications [39, 40]. The resonance's tuning characteristics depending on the gap are investigated in [41, 42]. Most intensively studied coupled particles structures include the bow-tie antenna, nanorods antenna and dipole antenna [38, 43, 44].

### 1.2.2 Photonic element

The general approach to achieve strong confinement of photons consists in structuring space with high index contrast at the wavelength scale, i.e. in the sub-micron range for the optical domain. Materials commonly used for this purpose are structured metals, including metal nano-particules, or high optical index dielectrics, immersed in low index dielectric material such as e.g. silica or air. We will concentrate on the later; the former being relevant to plasmonic has been described in the previous section.

The most commonly used high index dielectrics are semiconductor materials, including III-V compound semiconductors and silicon. For those materials, two confinement strategies are usually applied or combined:

- The refractive confinement scheme exploits the total or partial internal reflection of photons at the semiconductor-low index material interface. This strategy, widely used in traditional Optoelectronics, is principally devoted to devices operating solely in the waveguided regime. Strong confinement of optical modes can be achieved with this approach, owing to the high index contrast between the confining semiconductor micro-structures and the surrounding cladding material.

- The diffractive confinement strategy exploits diffractive phenomena in periodically structured materials to control the spatial-temporal trajectory of photons. This strategy is at the heart of quite a few of the recent developments in the field of Micro-Nanophonics, along the line of the Photonic Crystal (PC) approach. A photonic crystal, in which a periodic modulation of the index of refraction of a dielectric can guide light tightly and control its dispersion They are considered to day as fundamental building block for the ultimate control of light in spatial and spectral domains (trapping of light in the photonic band gap regime, slowing down of photons in the slow Bloch mode regime). In the following, the basic concepts of photonic crystal will be reviewed

### 1.2.2.1 Photonic crystal: a brief overview of basic concepts

A Photonic Crystal is a medium which the optical index shows a periodical modulation with a lattice constant on the order of the operation wavelength. The specificity of Photonic Crystals, inside the wider family of periodic photonic structures, lies in the high contrast of the periodic modulation: this specific feature is central for the control of the spatial-temporal trajectory of photons at the scale of their wavelength and of their periodic oscillation duration. Figure 1.3 shows schematic views of a variety of photonic crystals with dimensions ranging from 1 to 3.

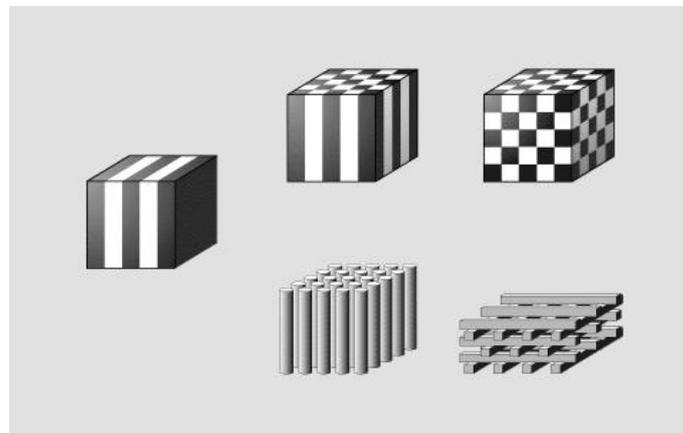

*Figure 1.3: schematic view of Photonic Crystals with different dimensionalities*

One dimensional photonic crystals (1D PC) have been around for quite a long time in the very fertile field of thin film optics; however, the periodic structuration being limited to only one direction of space, devices based on 1D PC suffer from a limited angular resolution and their 'lateral' size cannot be made compact. The concept of 3D PC was demonstrated experimentally for the first time in the microwave regime. The initial motivation was the full control of the spontaneous emission of an active emitting material. 3D PCs are potentially the best candidates for this purpose and for many other applications, where the best control of photons in space and time is requested. However fabrication technology of 3D PC in the optical domain is extremely complex.

In between, 2D PCs are far more accessible than 3D PC, since they may be fabricated using planar technological schemes which are familiar to the world of integrated optics and micro-electronics. A real 2DPC consists in considering a 2D structuring of a planar dielectric waveguide where photons are "index guided", that is to stay vertically confined by the vertical profile of the optical index. Figure 1.4 shows schematic and SEM views of a real typical 2DPC, consisting in a triangular lattice of holes formed in a semiconductor slab.

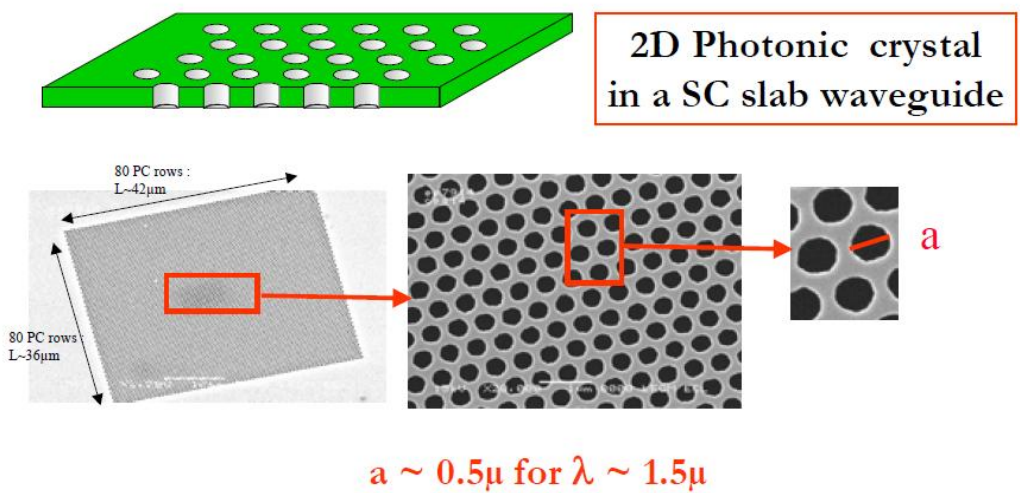

*Figure 1.4: schematic and SEM views of a real 2DPC, consisting in a triangular lattice of holes formed in a semiconductor slab*

In the rest of this chapter we will concentrate on the so called membrane approach, where the vertical confinement is achieved in a high index semiconductor membrane surrounded with low index cladding or barrier layers (for example an insulator like silica or simply air).

### 1.2.2.2 Optical mode confinement schemes in 2D PC membranes

The principal characteristics of the photonic crystal manifest themselves in the so called dispersion characteristics of the periodically structured medium, which relate the pulsation ω (eigen value) to the propagation constants $k$ (eigen vector) of optical modes, and which are the eigen solutions of Maxwell equations, corresponding to a stationary spatial distribution of the electromagnetic field. It is appropriate here to speak in terms of dispersion surfaces ω (k)= ω (k$_x$, k$_y$), real space being two-dimensional. Strong diffraction coupling between optical modes occurs; these diffraction processes affect significantly the surface dispersion characteristics, or the so called band structure, according to the solid state physics terminology. The essential manifestations of these disturbances consist in (see Figure 1.5) the opening of multidirectional and large photonic band gaps (PBG):

- The opening of multidirectional and large photonic band gaps (PBG)

- The presence of flat photonic band edge extremes (PBE), where the group velocity vanishes, with low curvature (second derivative) $\alpha \approx \dfrac{1}{PBG}$

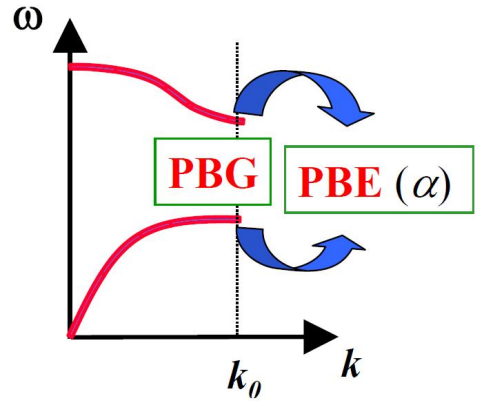

*Figure 1.5: schematic representation of a photonic band gap (PBG) and of related photonic band edges (PBE) in the dispersion characteristics of a photonic crystal.*

*PBG confinement scheme using localized "defect" or cavity modes*

In the PBG scheme, the propagation of photons is forbidden at least in certain directions. This is in particular true when they are trapped in a so called localized defect or microcavity and the related optical modes are *localized*: in this case the propagation of photons is fully prohibited. Opening of large PBG (in the spectral range) provided by the PC, allows for a very efficient trapping of photons, which can be made strongly localized in free space. The basic building blocks of photonic components for in plane operation along the PBG scheme are micro-cavities [45] and waveguides [46]. These building blocks were among the earliest structures based on 2DPC reported in the field: they have been the matter of a large number of publications since the late nineteen nineties and have resulted in the production of a variety of devices including very high Q factor nanoresonators for Quantum Electrodynamics [47], very low loss wave-guides [48], micro-lasers [49], channel drop filters [50], etc…

*PBE confinement scheme using delocalized slow Bloch modes*

In the PBE scheme, the PC operates around an extreme of the dispersion characteristics where the group velocity of photons vanishes. It is more appropriate to speak in terms of *slowing down* of optical modes (so called Bloch modes for a periodical structure): it can be shown that the lateral extension of the area *S* of the slowing down Bloch mode during its lifetime τ is proportional to ατ [51]. As mentioned above, one essential virtue of PC is to achieve a very low curvature α at the band edge extremes, thus resulting in strong slowing down and a very efficient PBE confinement of photons. Although the PBE scheme provides weaker confinement efficiency than with the PBG approach, it results in an improved control over the directionality or spatial/angular resolution of the light. Active as well as passive devices have been demonstrated. For the former, Bloch mode micro-lasers designed for in-plane emission have been reported (see, for example [52, 53]). Passive structures as channel drop filters making use of the PBE scheme have also been proposed: the principal advantage over their counterpart based on the PBG scheme, lies in their 'natural' propensity to provide a directional dropping [54, 55].

*The issue of vertical confinement in 2DPC: below and above light-line operation*

It has been explained previously that the vertical confinement of photons is based on refraction phenomena. However, full confinement of photons in the membrane wave-guiding slab is achieved only for those optical modes which operate below the light-line (see figure 1.6a). This mode of operation is restricted to devices which are meant to work in the sole wave-guided regime, where wave-guided modes are not allowed to interact or couple with radiative modes. For waveguided modes with dispersion characteristics lying above the light line, coupling with the radiated modes is made possible, the waveguided 'state' of the related photons is transitory, and the photonic structure can operate in both waveguided and free space regimes (see figure 1.6b).

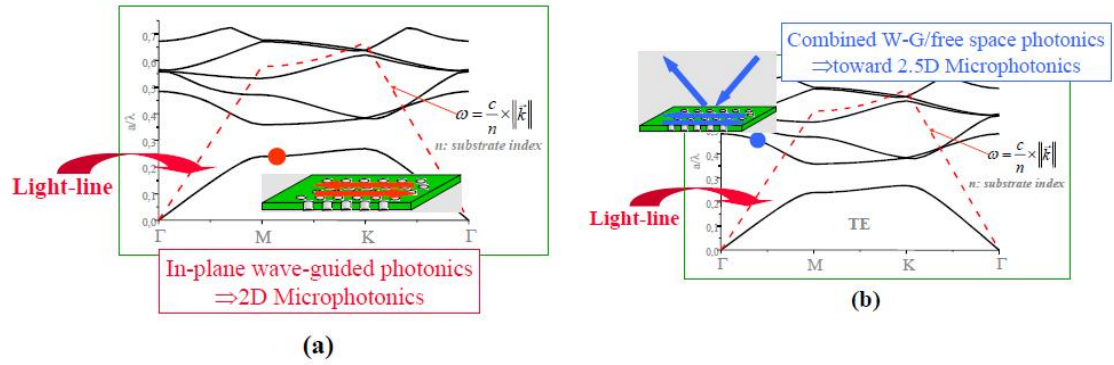

*Figure 1.6: below (a) and above (b) light-line operation of photonic structures based on 2D photonic crystals*

### 1.2.3 Hybrid Plasmonic-Photonic Crystal structure

Hybrid resonant metallo-dielectric structures allow combining the best of two worlds: multiple-frequency and high-Q properties of photonic components with the nanoscale dimensions of plasmons, leading to a wealth of novel effects. In particular, interaction of high-Q photonic modes with the localized SP resonances on noble-metal nanostructures results in giant cascaded field enhancements within subwavelength volumes in hybrid optoplasmonic devices. Enhanced information and energy transfer in optoplasmonic materials due to greatly reduced dissipative losses when compared to plasmonic waveguides is only one example where hybrid plasmonic-photonic crystal creates new functionalities beyond the capabilities of the individual building blocks. We anticipate that optoplasmonic components will also facilitate active nanoplasmonic circuit elements for field modulation and frequency switching, as photon recycling in microcavities; translates into narrow linewidths and thus greatly enhances mode sensitivities to external stimuli and environmental changes.

In the following I will place a particular emphasis on a model system combining a metallic nanoantenna and a dielectric photonic crystal to efficiently collect radiation over an extended area and funnel it into a single well-defined nanoscale focal spot.

### 1.2.3.1 The question of optimum coupling to the nanoworld

Ultimate spatial confinement of photons can be attained in plasmonic nano-structures, such as in metallic nano-particules or nano-antennas (NA), at the expanse however of a limited resonance strength (or confinement time) as a result of optical losses induced by metal absorption and of radiation to free space continuum. However, excitation of individual plasmonic NA using free-space optics is not efficient, and the signal-to-noise-ratio is very small. The reason lies in the difficulty to achieve a sufficient coupling rate between the incoming optical beam and the NA in order to compensate for the rather large optical losses (radiation to free space and metallic absorption) [56]. It has been shown that a darkfield microscope can be used to interrogate individual plasmonic nanoresonators by using scattering spectroscopy [57]. This still requires a very sensitive detector and a bulky microscope system with a precise alignment control. The inefficient coupling of the lightwave to each individual plasmonic nanoresonator reduces the signal-to-noise-ratio in sensing and spectroscopy applications; and it also limits the level of possible field enhancement, which is necessary for efficient light-matter interaction. To harness the advantages of plasmonic antenna for practical applications, it is essential to improve the coupling of the lightwave to the localized surface plasmon resonance modes.

A new approach is proposed in the present work for the optimum addressing of NA with a free space optical beam via the use of an intermediate coupling resonator structure, which is aimed at providing the appropriate modal conversion of the incoming beam, in the time domain. The basic idea consists in relaxing the stringency of the coupling conditions by increasing the optical energy density achieved in the vicinity of the NA, owing to the photon storage capabilities of the intermediate resonator. More specifically, the intermediate resonator is a PC membrane resonant structure based on surface addressable slow Bloch modes. Those structures are currently recognized as very versatile photonic platform in a sense that there are not only restricted to operate in the in plane wave-guiding regime, but can be opened to the third space dimension by controlling the coupling between wave-guided and radiation modes. They have the capability to provide the optimum conversion of the incoming free space optical beam into a wave-guided slow Bloch mode with spatial and temporal confinement characteristics accurately designed to achieve the finest coupling conditions to metallic NA.

We consider here a 1D PC membrane which consists in a periodic array of low index material slits (silica) formed in a high index silicon membrane. The structure, which is designed in order to accommodate a slow Bloch mode Fano resonance, with a large quality factor, addressable in the vertical direction, is shown in Figure 1.7, where its topological parameters (membrane thickness, period and silica filling factor of the slit array) are given. The transmission spectrum obtained by 2D FDTD simulation (here we assume that the size of the structure along the slit direction is infinite with respect to the wavelength) is also reported and reveals a resonance near 1.506 μm. The lateral size of the incoming excitation beam is 12μm and its polarization is TM (electric field perpendicular to the slits: blue arrow in Figure 1.7).

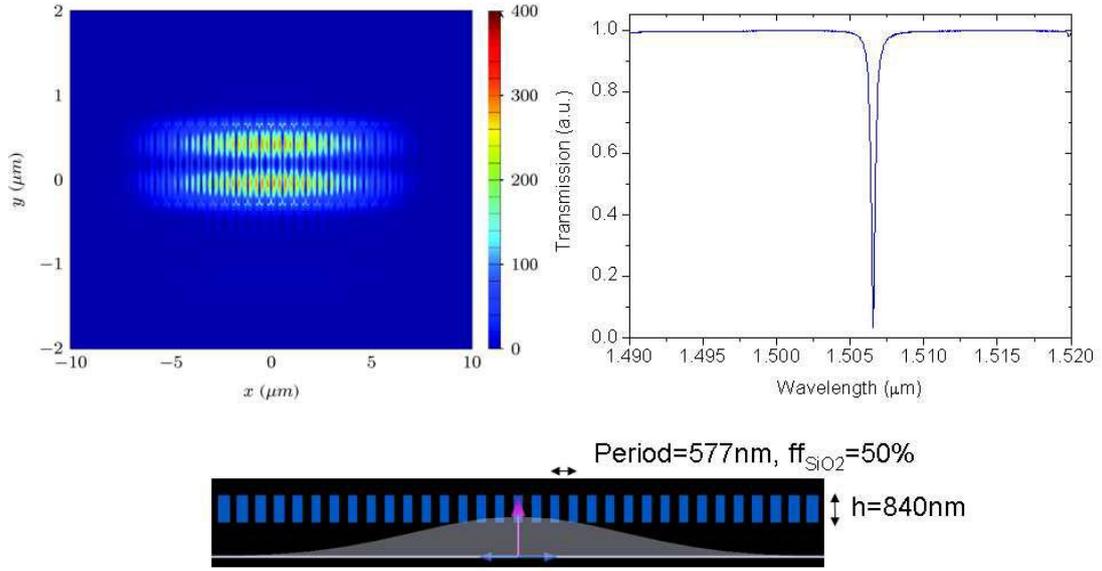

*Figure 1.7: Schematic view of the 1D PC membrane reflector used as an intermediate coupling resonator. (The topological parameters of the structure are given in the figure). The field intensity distribution at resonance and the transmission spectrum are also shown.*

The strength of the resonance is given by the spectrum bandwidth δλ which is related to its quality factor, $Q_{PC} = \dfrac{\lambda_0}{\delta\lambda} = \omega_0\tau_c$ , where λ₀ is the Fano resonance wavelength. $Q_{PC}$ is around 4800, which corresponds to a rather strong resonance. The reflectivity of the structure is close to 100% at the resonance wavelength: this indicates that the above discussed optimum coupling, or impedance matching, conditions between the incoming optical beam and the wave-guided slow Bloch mode are achieved (in-plane *k*-vector and lateral size matching). The field intensity distribution of the Fano resonance is also reported.

Let's consider now the coupling scheme to the NA based on the use of the intermediate PC membrane resonator, as shown in Figure 1.8. The NA is a gold dipole which has been designed to match the wavelength of the PC resonator with the following geometrical parameters: 50nm thickness, 200nm arm length and 40nm feed gap width. The gold dielectric function was obtained by fitting on the optical bulk experimental data reported in [58], for the frequency of interest. Other metals (eg silver) could have been chosen as well to illustrate the proposed coupling scheme. The PC membrane is addressed with a Gaussian optical beam (diameter 12μm) from the bottom and the NA is located above the membrane at a variable distance, which results in a variable coupling time constant τ between the NA and the PC membrane resonator: the topological parameters of the structure are given in the figure. An example of transmission spectrum of the system is also shown (NA put directly on top of the PC membrane): the narrow Fano resonance response shows up within the much wider background of the NA spectral response, which cannot be made visible in the narrow wavelength range of the spectrum. This is a clear indication that indeed, the PC membrane resonance is much stronger than the one of the NA

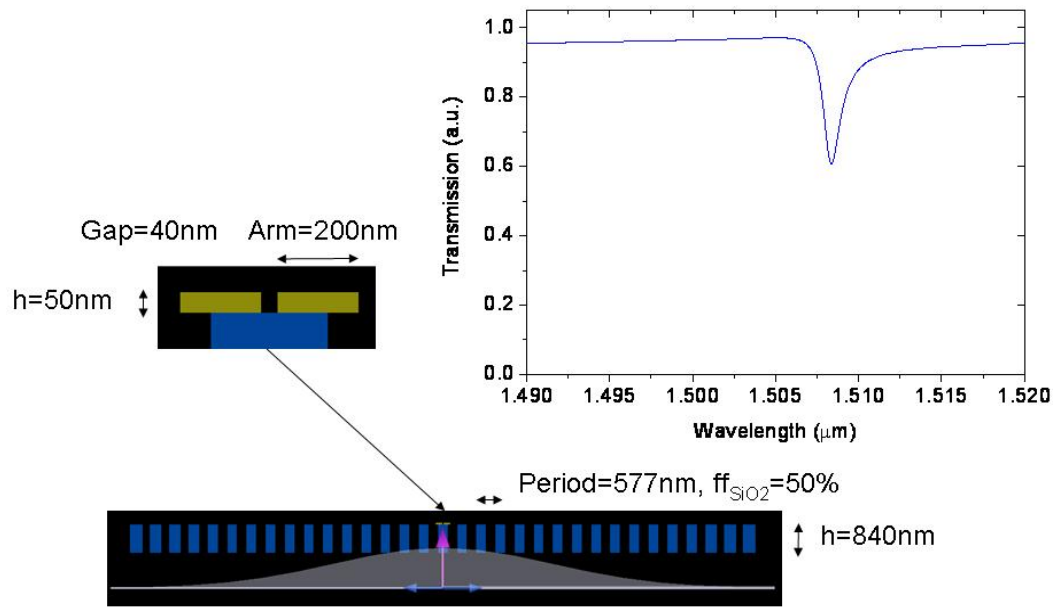

*Figure 1.8: Schematic view of the 1D PC membrane reflector used as an intermediate coupling resonator. (Topological parameters of the structure are given in the figure). The transmission spectrum is also shown.*

A plot of the maximum electric field intensity in the NA as a function of the distance between the latter and the PC membrane is shown in Figure 1.9, at the resonance wavelength (1.51μm). As expected from coupled mode theory, the electric field reaches a maximum for an optimum coupling constant [56], achieved for an optimum distance between the NA and the PC membrane resonator. The field intensity distributions of the structure are also shown in Figure 1.9. The gloss of the NA reaches its maximum under the optimum coupling conditions. For decreasing coupling rates, that is for increasing coupling distances, the NA darkens to the PC membrane advantage, which brightness increases gradually, before levelling off at a saturation value.

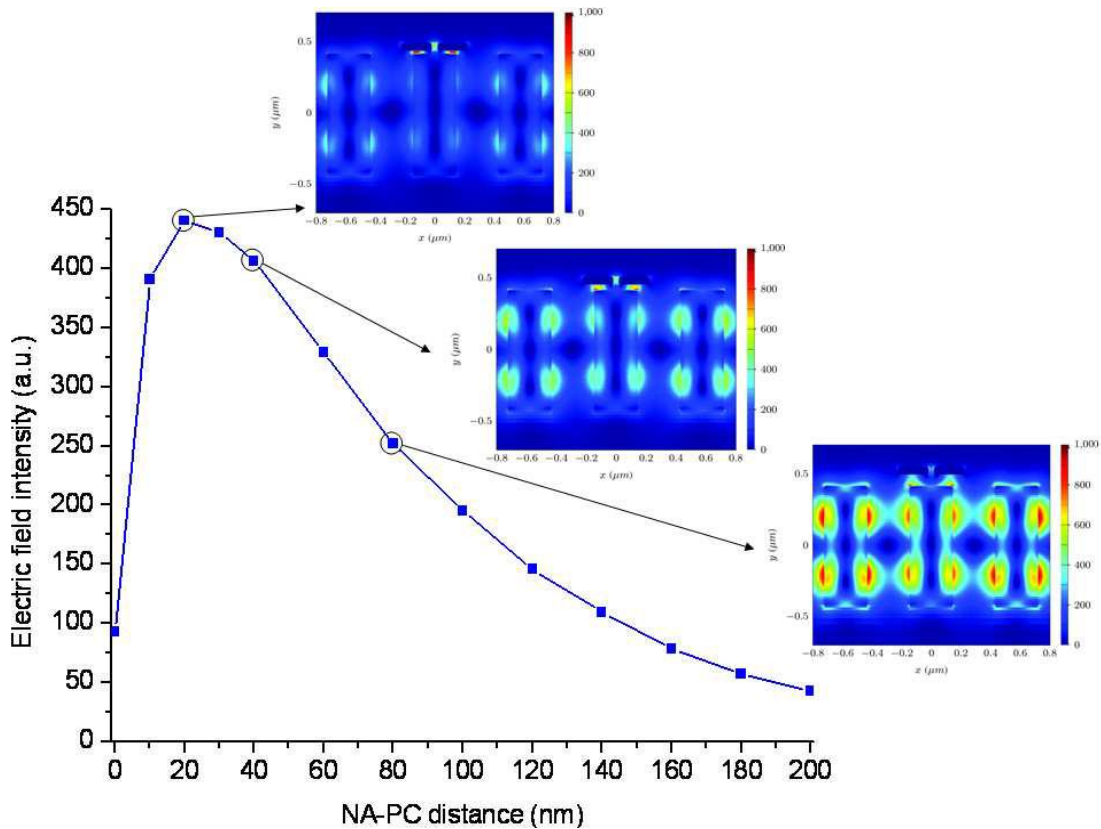

*Figure 1.9: Plot of the maximum electric field in the NA as a function of the distance between the latter and the PC membrane. The field intensity distributions in the structure at resonance are also shown.*

For sake of comparison, we have carried out FDTD simulations of the direct addressing of the NA, positioned at the waist of a focused optical beam under the diffraction limit conditions (waist size down to 1.5μm): the optimum coupling conditions could not be attained; the maximum field intensity (normalized to the total input optical power) achievable in the NA was a factor 3 below the one achieved with the help of the PC membrane, which does not require light focusing. This comparison is illustrated in Figure 1.10, where the electric field intensity distributions are shown for direct addressing and PC membrane mediated addressing of the NA, respectively: for the case of direct addressing, the field colour scale intensity has been artificially multiplied by a factor of 3, in order to make clearly visible (i) the focused incoming free space beam and (ii) the factor 3 enhancement of the maximum achievable electric field intensity in the NA, with the help of the PC membrane.

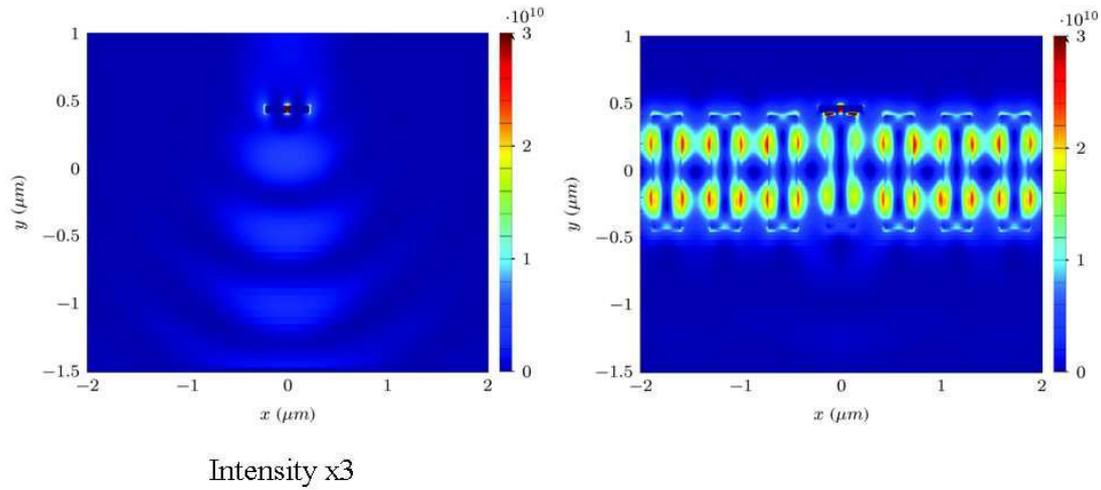

Intensity x3

*Figure 1.10: Field intensity distributions at resonance for a NA directly addressed by a focused Gaussian beam (left) and PC membrane mediated addressing of the NA (right).*

In the previous section, a generic approach for 3D light harnessing based on the coupling engineering between Photonic Crystal membranes and metallic optical antenna has been discussed. No only does this approach allow for very efficient light insertion in tiny micro-nano-resonator structures, but also it opens the way to the production of a wide range of functionality. In the present contribution, this approach is illustrated with the demonstration of efficient light addressing of low quality factor and very high Purcell factor metallic NA; it can be fruitfully extended to the addressing of high quality factor micro-nano-resonators, thus widening the range of accessible coupling regimes, including the strong coupling regime. Attractive outcomes of the proposed scheme can be contemplated for such applications where strong light concentration and/or accurate beam shaping is requested with sub-wavelength spatial resolution: they include low optical power consuming non-linear optics, sensing and bio-photonics, highly sensitive resonant near-field spectroscopy and imaging, light trapping with sub-wavelength spatial resolution, etc…

In the next section we will highlight the design of hybrid plasmonic-photonic structures combining a metallic bowtie antenna with:

-   A slow Bloch mode PC. As discussed formerly they can be used to address one or several NA from free space, improving the signal-to-noise ratio and the field enhancement

-   A defect state based PC. The ability to confine light into ultrasmall volumes is essential for enhancing the interaction of light and matter in the emerging field of nanophotonics. Surface plasmons provide a route to such strong optical confinement in the subwavelength regime and are of high interest for (cavity) quantum electrodynamics applications. Because of their ability to concentrate electromagnetic energy in volumes much smaller than the corresponding wavelength, they provide a very strong interaction between (quantum) emitters and photon fields. However, plasmonic cavity Q factors have so far been limited to values less than 100 both for visible and near-infrared wavelengths [59, 60, 61]. An optoplasmonic structure can thus exploit this field compression exhibiting both high quality factors and pronounced hot spot of the electromagnetic field, potentially enhancing the interaction of the cavity mode with emitters or other types of active materials.

### 1.2.3.2 The antenna design

Optical and infrared antennas based on metal nanostructures allow for efficient conversion of propagating light into nanoscale confined and strongly enhanced optical fields, and vice versa. This special capability enables a variety of cuttingedge applications. In order to maximize the benefits of optical antennas it is necessary to tailor them in order to satisfy the specific needs of each application. In that sense, the control over the localization of the areas where the near-field is maximum (hot-spots) and the tuning of the resonant frequencies of the antennas are central to reach the best optical coupling condition with the PC structure. Bowtie nanoantennas have attracted much attention in the scope of nano-optics [62, 63, 64]. It consists of two opposing tip-to-tip metal triangles, separated by a small gap (see figure 1.11). The sharp corner of each arm leads to strong charge concentration at the edge of each triangle inside the gap. Moreover, localization of fields inside the gap is more likely to increase while two sharp edges with opposite charges are closely spaced [65]. Similar to a dipole antenna, the ratio between the physical length and the incident wavelength determines their impedance. The impedance of a bowtie antenna with large $\alpha$ can be denoted as broad-band impedance, which make them useful for a larger frequency range [66]. Experimental observations on bowtie antennas show that the resonance peak is dramatically influenced by the gap size [64].

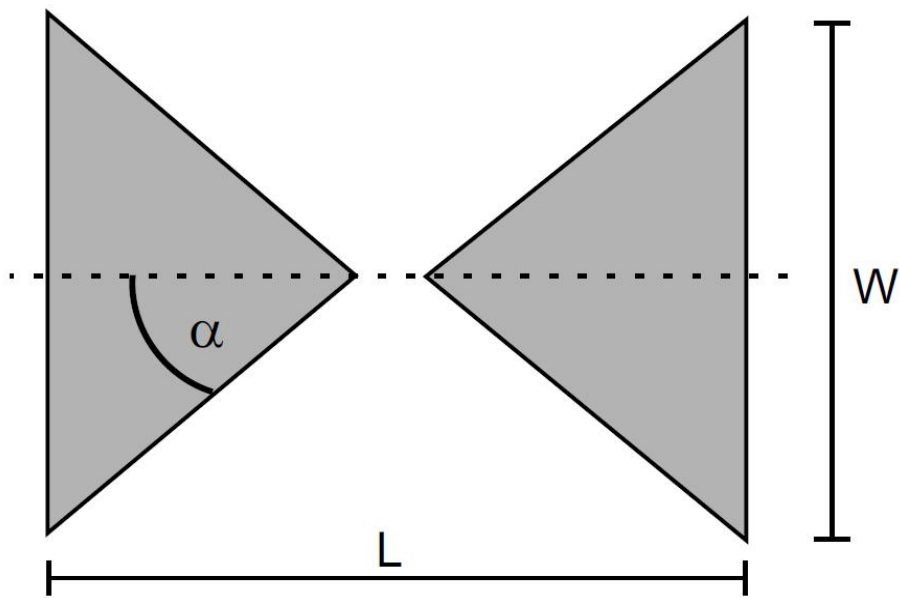

*Figure 1.11.: Sketch of a bowtie antenna*

It is difficult to predict how the antenna characteristics (e.g. resonance length) are influenced, when an antenna is scaled to the optical regime. At optical frequencies metals are no longer perfect conductors and the assumption made by classical antenna theory, that e.m. fields are restricted to the outside of an antenna is no longer valid. Depending on the permittivity of the antenna material the fields penetrate the surface in an extent given by the skin depth. At optical frequencies the skin depth for many metals is quite large and is, even for thick antennas, comparable with the antenna diameter. Seeing the antenna as e.m. boundary problem, the large skin depth at optical frequencies increases the complexity of the problem.    The current distribution and hence the antenna input impedance of optical antennas will differ to some degree from the predictions made by classical antenna theory. This will affect the resonance length as well as the achievable field enhancement in the feed gap of an OA. The presence of a substrate also modifies the position of the resonances. To predict in more detail the function of an OA one has to perform computer simulations that take into account the antenna shape and the finite and frequency dependant permittivity of the antenna material. To perform the design and to support the interpretation of the experimental data a commercial software (LUMERICAL) based on the finite-time-domain (FDTD) method has been used. The method requires that at least the structure itself is discretized in a computational grid. The grid size (dx; dy; dz $\leq\lambda/(50n)$) must be small compared to the wavelength and smaller than the smallest feature in the computation volume. Structures with very fine details require a larger computational domain, resulting in much longer solution times.

We performed FDTD simulations for a gold bowtie geometry on a flat InP substrate. The bowtie geometry ($W$=140nm, $L$ =270nm, height= 40 nm, feed gap size=20 nm) chosen for the simulation has been optimized to hold a plasmonic resonance exactly matching the spectral position of the photonic mode. A sketch of the geometry used for the simulation is shown in Figure 1.12. The dielectric function for gold, titanium adhesion layer, and InP was given by the simulation software. The discretization cells were chosen to be 3x3x3nm$^3$ cubes. Exemplarily two monitor planes ($x = 0$ nm and $y = 0$ nm) are indicated. The bowtie is addressed with a plane wave coming from the substrate and polarized in the y or x direction.

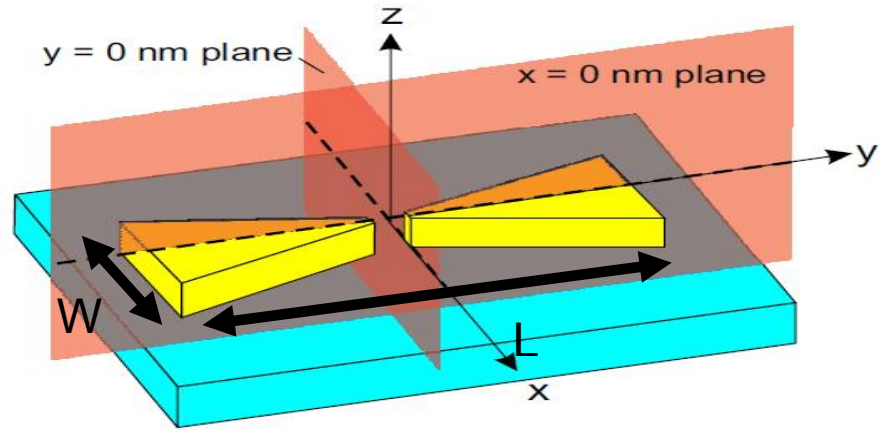

*Figure 1.12.: Sketch of bowtie geometry. Bowtie dimensions used for FDTD simulations, W=140nm, L=270nm, gap=20nm, height =40nm.*

The near-field intensity enhancement factor ($= \left| E / E_0 \right|^2$) 10 nm above the antenna as a function of wavelength is reported on Figure 1.13. The plane wave is polarized in the y-direction (see Figure 1.12) and $E_0$ refers to the incident evanescent field in the absence of the antenna.

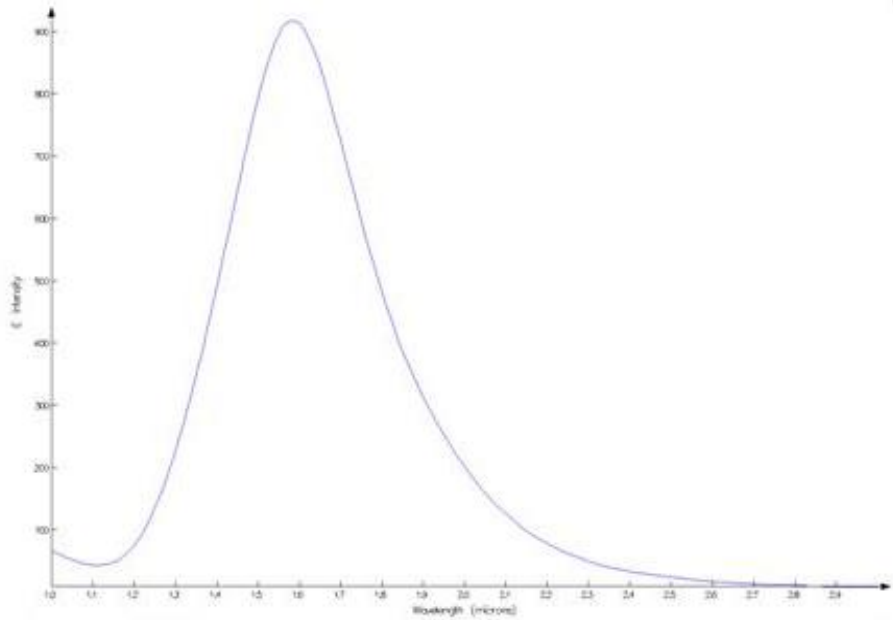

*Figure 1.13.: Near-field intensity enhancement factor versus wavelength*

There is a very large field enhancement (over 400) around 1.5μm corresponding to the x-polarization. The spectrum is broad and the quality factor of the resonance is estimated to be close to unity.

To understand the field confinement and enhancement in bowties a theoretical simulation was done to model the electromagnetic field enhancements in the different polarization directions as shown in Figure 1.14. The y-direction polarization enhances the field in the center of the bowtie gap while y-direction exhibits strong enhancement along the outside tips of the bowtie. The enhancement is about 1000 when the bowtie is launch with a polarization along y and 100 in the perpendicular direction.

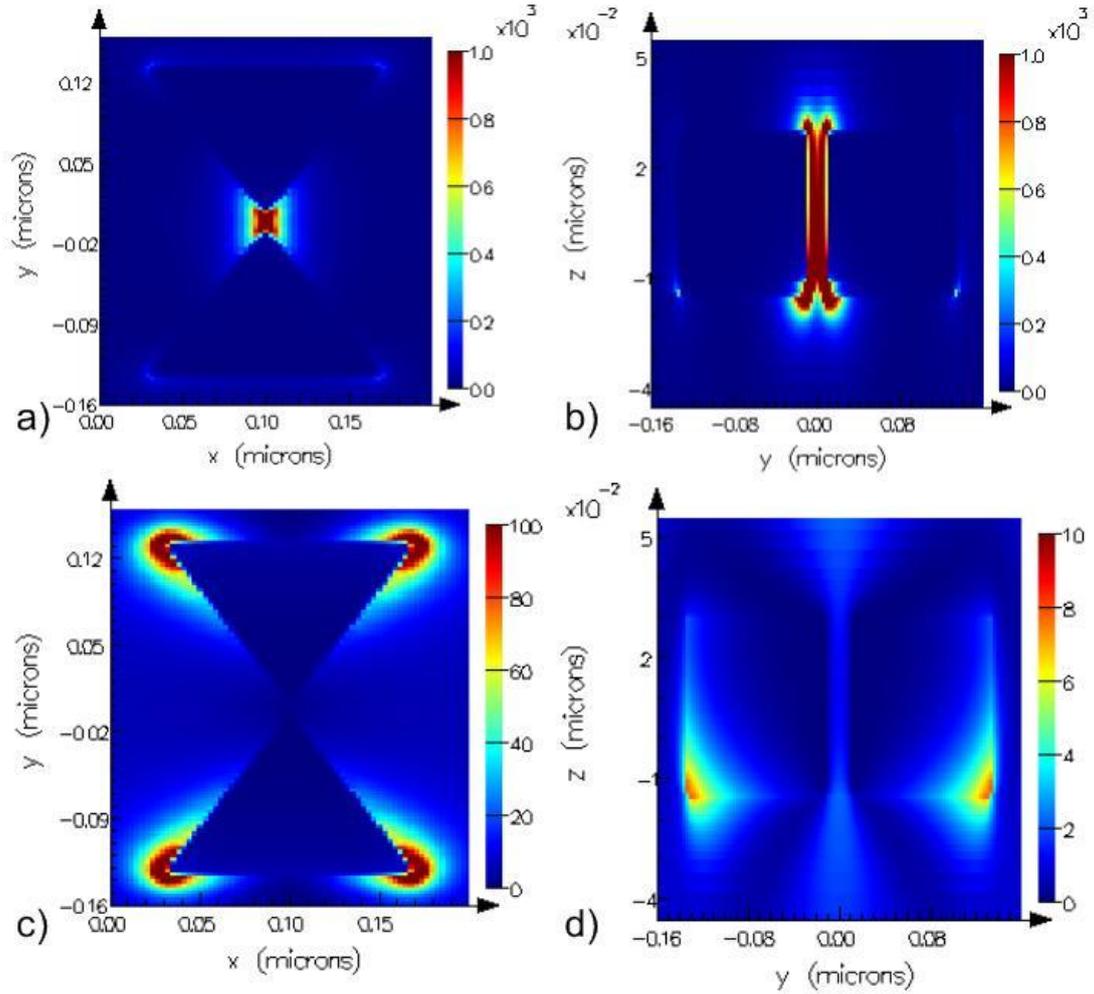

*Figure 1.14 Near-field intensity mapping ( $\left| E / E_0 \right|^2$ ) of the bowtie antenna with incident polarization in the y-direction (a and b) et the x-direction (c and d)*

### 1.2.3.3 Slow Bloch mode based PC design

In this part, we theoretically study the electromagnetic properties of a PC microcavity supporting a slow Bloch Mode (SBM) located at the Γ-point in the first Brillouin zone. The choice of this configuration is motivated by two main reasons:

- the k-space localization of the mode is exploited to get directive vertical emitting devices.

- Its small group velocity is used to get a better temporal confinement for a given

cavity size

As a basic structure to build the slow light resonator, we use a InP-based 2D PC slab that consists of a regular array of air holes, drilled in a membrane, and which exhibits a slow light mode around λ=1.5μm. In the remainder of this thesis, we will consider a honeycomb (graphite) 2DPC structure for several reasons. (i) Compared to a triangular or a square lattice, the surface filling factor of the semiconductor is higher for a honeycomb structure which makes it easier to position the NA in the nanofabrication process. In an active structure a majority of emitters (qdot or qwell) will be located far from the hole interfaces, reducing the effect of surface recombination (ii) The honeycomb structure exhibits several flat bands at the Γ-point. Photons in such modes exhibit a low group velocity and the lateral light confinement is then improved. Active 2D-PCs were designed to support band-edge modes at the Γ-point of the first Brillouin zone. The structure consisted of a 240nm-thick monomode slad InP (n=3.17 at λ=1.55μm) bonded on SiO2 transparent substrate. At the center of the InP slab, four quantum wells of InAsP separated by InP barriers (see Figure 1.15) were embedded to provide internal photoluminescence from 1.3μm to 1.65μm. An array of honeycomb-like cylindrical air holes (e=1) was drilled on InP slab (Figure 1.15-a/b) to provide the periodical dielectric structure with a high contrast of refractive index. The two basic parameters of the honeycomb PC are the lattice parameter a (center-to-center distance of two nearest hexagons) and the hole radius (r). The surface air filling factor was calculated from the following formula:

$$ff = \frac{4\pi}{\sqrt{3}}\left(\frac{r}{a}\right)^2 \hspace{3cm} 1.3$$

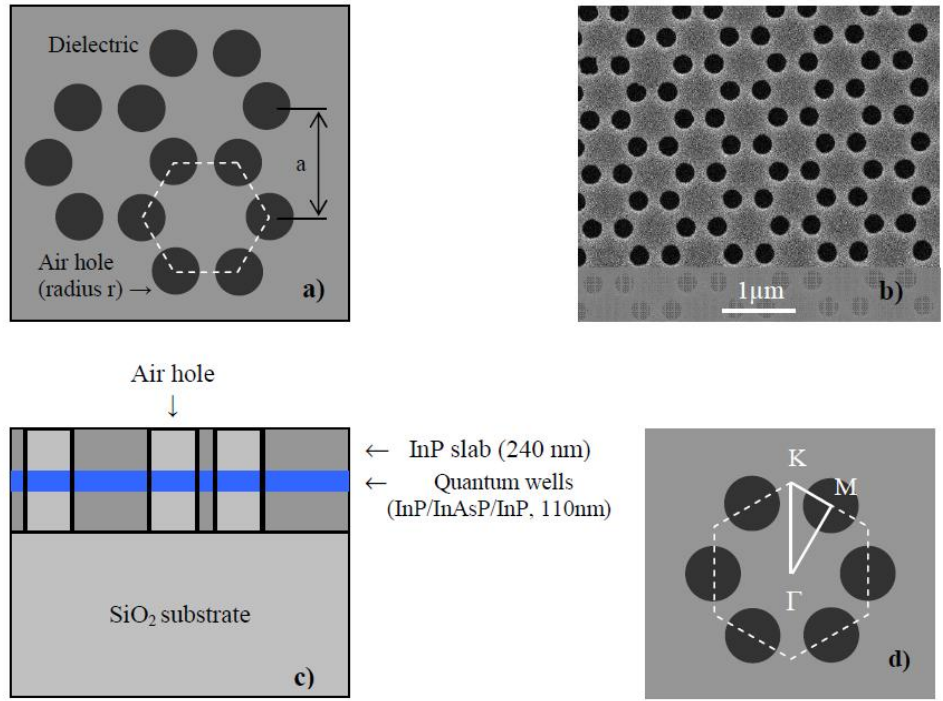

*Figure 1.15: Model of defect–free honeycomb 2D PC. a) (Top view). An array of cylinder air holes (white circles, e=1) centred at the vertex of periodically arranged hexagons on InP slab (grey area, e=10 at λ=1.55μm). b) (Top view) SEM image of a typical 2D-PC. Real lattice parameters measured a=730nm, 2r=245nm, then corresponding ff_air=0.2. c) (Side view). Sample includes 240 nm thick InP slab bonded on top of a SiO2 transparent substrate with four layers of InAsP quantum wells (separated by InP barrier) are embedded at the center of the InP slab. d) The first Brillouin zone with high-symmetry points (Γ-K-M) in the reciprocal space with respect to the real space unit cell.*

The Plane Wave Expansion method (PWE) developed by the MIT [67] has been used to determine the band diagram of the PC structure. It is a well known method in frequency-domain that allows extraction of the Bloch wave functions and frequencies by direct diagonalization.

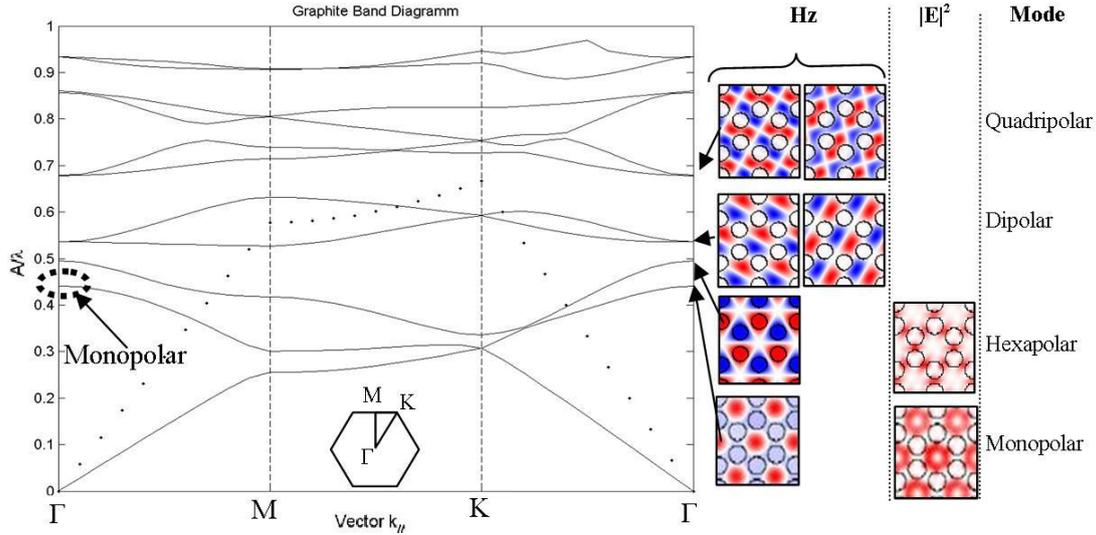

*Figure 1.16: Band diagram (TE polarization) of the graphite 2D PC (dashed line: light line), z-component of the magnetic field and intensity distributions of several Γ-point modes.*

Figure 1.16 shows typical map of the z-component of the magnetic field $H_z$ and electric field intensity map of the different Γ-point band-edge modes. We consider even modes (TE-like modes), with respect to a mirror plane located in the centre of the slab. To get a large $β$- factor, we must ensure that the emitters couple to only one resonant mode. Accordingly, the Γ-point Bloch mode must be non-degenerate and sufficiently spectrally isolated from the other slow Bloch modes. We see in figure 1.16 that two modes satisfy this requirement: the monopolar and hexapolar modes. The spatial distribution intensity of the monopolar mode is mainly located in the material as compared to the intensity of the hexapolar mode: 82% of the intensity is in the InP material. This promotes the interaction of the modes with the emitters and also relaxes the constraint of positioning the NA on the backbone of the PC lattice.

A PC lattice parameter $A$ = 730nm is determined so as to operate around $λ{\sim}1.5μm$. This corresponds to a distance between the closest holes of 240nm. An air filling factor of $ff$ = 20% selected and results from a balance between 2 criteria: (i) to ensure a good spectral isolation of the monopolar mode with respect to other Γ- point slow Bloch modes; (ii) to keep reasonable technological constraints.

3D-FDTD with perfectly matched layers boundary conditions has been used for rigorous prediction of Q-factor, mode intensity and polarization patterns. The numerical simulation is performed on a finite 2D PC (20x20 μm$^2$) with a lattice parameter (A) of 730nm and 245nm hole- diameter (20% of air filling factor). The membrane consists of a 240nm-thick InP slab on a SiO2 substrate (Figure 1.17a). Computational meshes are 30 nm for x, y, and z. The monopolar mode is excited by a dipole placed near the center of the structure. The simulation in Figure 1.17-b presents the total electric intensity of mode M on the InP slab surface. It also shows that the mode surface reaches the edges of the PC-structure. The mode intensity is maximum in the centre area of the PC and its global envelop is Gaussian. Moreover, in Figure 1.17-c, a closer view into the unit cell indicates that the mode intensity is distributed in each honeycomb lattice unit cell as a "doughnut" with an inner and outer radius of about 90nm and 260 nm, respectively. The electric polarization pattern in Figure 1.17-d shows that the field is azimuthally polarized in each unit cell: the zero of the field intensity in the centre region of each hexagon is due to a phase singularity [68]. Figure 1.17-e/f present the mode intensity mapping for polarized electric field along x ($E^2_x$) or y ($E^2_y$).

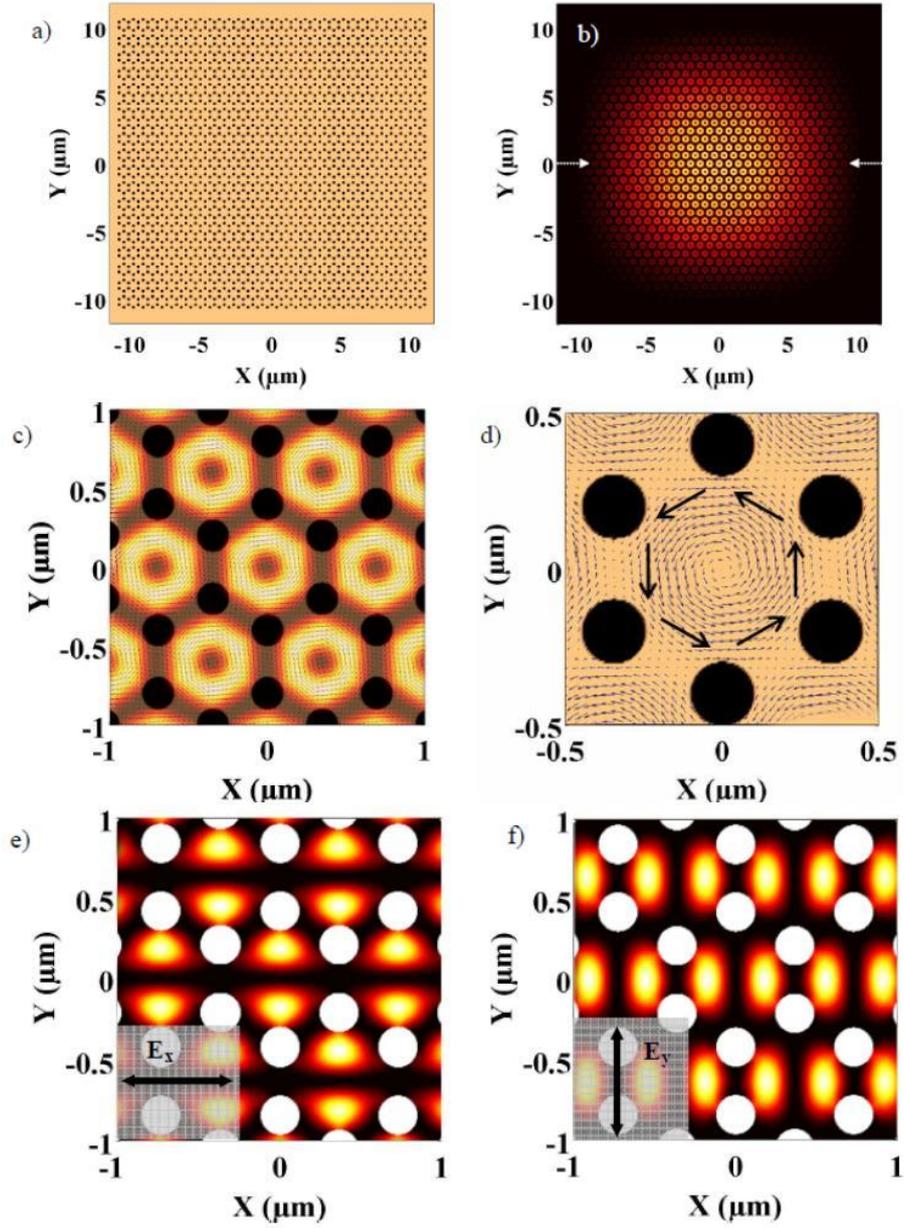

*Figure 1.17: Intensity patterns of mode M on InP slab surface by 3D-FDTD. (a) Finite honeycomb 2D PC structure (20x20μm²) with lattice parameter of 730nm and radius of 245nm (20% air filling factor). (b) Normalized field intensity ($E_x^2 + E_y^2$) at the surface of the PC at 1602 nm. Mode M is excited by a dipole placed near the center of PC structure. The intensity pattern presents the Gaussian-shape expansion limited at the boundary of PC structure. (c) Zoom in the central part of mode pattern in unit cells with air holes in black clearly reveals the mode intensity in doughnut-like shape. (d) Azimuthal polarization map of electric field in unit cell. (e) Normalized field intensity polarized along x ($E_x^2$) and (f) y direction ($E_y^2$).*

### 1.2.3.4 Defect mode based PC design

The CLx cavity, which basically consists of x missing holes in a row, is one of the most extensively studied types of PC cavities. This is mainly due to the fact that it supports a spectrally well separated fundamental mode with small mode volume whose quality factor can easily be optimized to relatively high values. We consider a linear defect nanocavity whose design is shown in figure 1.18. The photonic crystal consists of a triangular array of cylindrical holes (period 420 nm and hole radius 100 nm) patterned on a thin InP slab (thickness 250 nm) positioned on top of a SiO2 substrate The cavity is formed by introducing a linear defect (omitting 7 holes) into the 2D-PC. This defect is referred to as the LC7 microcavity. The cavity structure is designed so as to obtain several spectrally and spatially distinct modes, around 1.5µm. To achieve reasonably high quality factors, the radiation losses were minimized by engineering the holes positions at both edges of the cavity [69]. In our case, the two holes on either side of the cavity were shifted by 80nm outward. The result is an increase by up to 60 percent of the quality factor compared to unmodified microcavity.

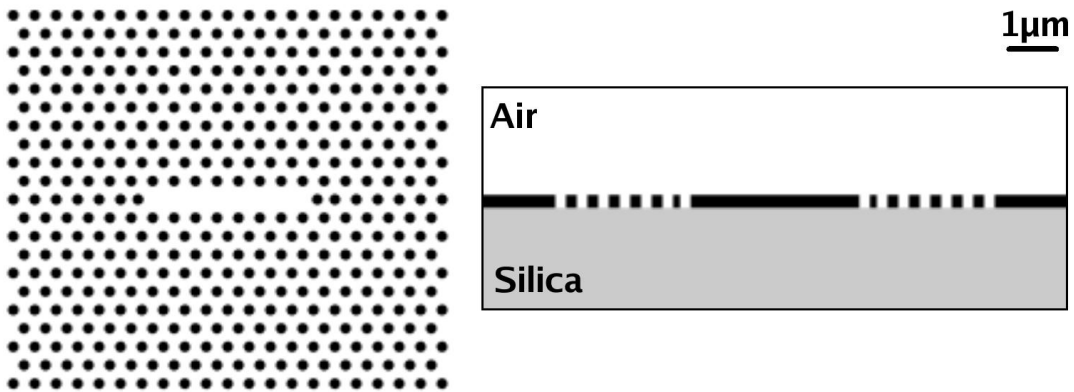

*Figure 1.18: Views of the calculation domain (a) in the plane of the InP membrane (b) Perpendicular to the InP membrane*

The structure was simulated using 3D finite-difference time-domain (FDTD) method with the perfectly matched layers boundary conditions. The full structure lateral size is 11µm x 11µm and consists of the crystal slab on a SiO2 substrate. Computational meshes were 42 nm for x, y, and z, corresponding to 10 cells for a crystal period. They show that the microcavity supports several modes, four of them having a quality factor above 1000, the largest found being 5800. Figure 1.19a, 1.19b, 1.19c and 1.19d present the mode cartographies computed at the surface of the InP slab, for the four modes of lowest energy, labeled A ($\lambda$ = 1444 nm Q = 1615), B ($\lambda$ = 1483 nm Q = 1000), C ($\lambda$ = 1517nm Q = 1550) and D ($\lambda$ = 1530 nm Q = 5800). The fundamental mode is mode D. The highest quality factor was obtained for this mode which is highly confined into the cavity with the intensity of the mode concentrated principally at the center of the cavity, as it is shown by the vertical cross-section cartography along the longitudinal axis of the cavity (Figure 1.19e). The profile of the central lobe is shown Fig. 1.19f. The other modes exhibit a weaker confinement (quality factor around 1000) and different spatial distributions.

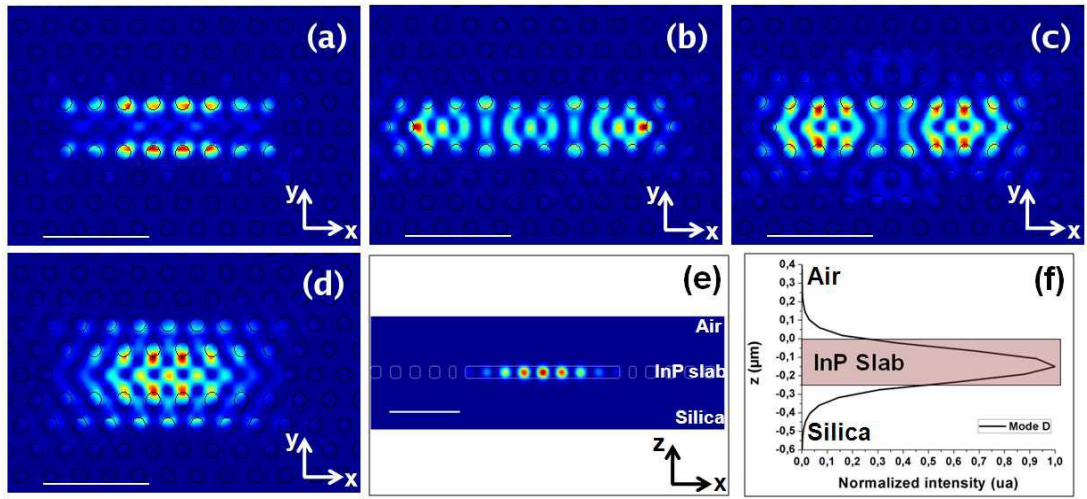

*Figure 1.19: Horizontal cartography at the surface (z = 0) (a) Mode A: $\lambda$ = 1444 nm Q = 1615 (b) Mode B: $\lambda$ = 1483nm Q = 1000 (c) Mode C: $\lambda$ = 1517nm Q = 1550 (d) Mode D: $\lambda$=1530nm Q = 5800 (e) Vertical cross-section cartography of the mode D along the longitudinal axis of the cavity (f) Profile of the central lobe. White bar is 1.5 µm.*

## 1.3 FDTD simulation for the hybrid devices

### 1.3.1 FDTD simulation for the CL5 PC cavity without NA

The optical field distribution of CL5 PC cavity is simulated via 3D-FDTD by Dr. Ali Beraouci. The simulation result is shown in Figure 1.20. The optical fields $Ey^2$ are in the centre of the cavity along the long axis of the cavity and optical fields $Ex^2$ distribute along the side of the cavity. The laser emission is at 1679nm and the peak is very narrow. Theoretically, the result of CL7 PC cavity should show the same regular.

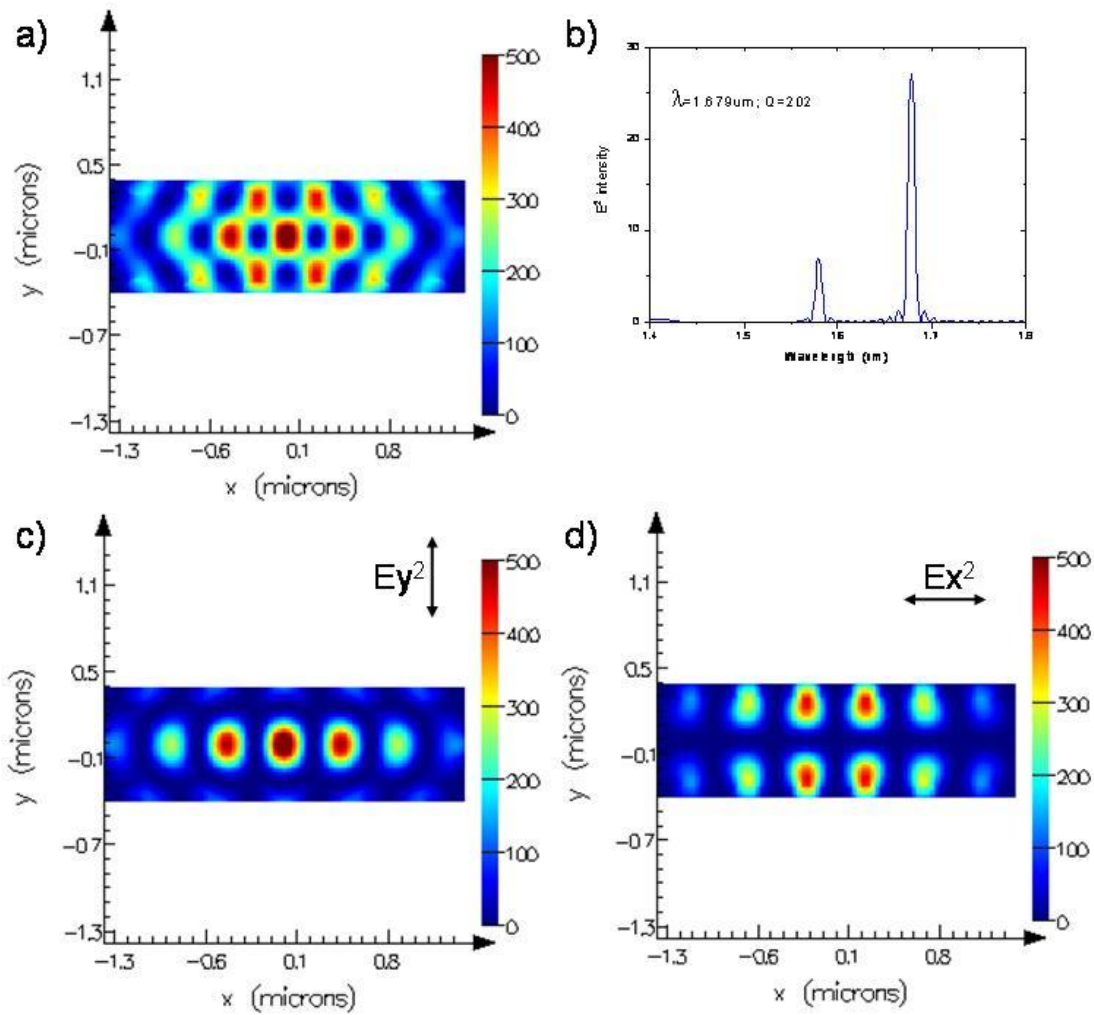

*Figure 1.20 Numerical simulation of optical field distribution of CL5 PC cavity: a) general distribution of optical field; b) spectrum of the laseremission; c) optical field domain of $Ey^2$; d) optical field domain of $Ex^2$.*

### 1.3.2 FDTD simulation for the bowtie NA

We modeled optical bowtie antennas on InP/SiO$_2$ substrate using 3D-FDTD calculation. The plasmonic structure consists of two coupled gold triangles separated by a 20nm gap. The bottom edge of the triangle is 140nm and it is 125nm high. Geometrical parameters are optimized to tune the optical response to 1.5µm (Figure 1.21). The resonant peak is very wide, means that the NA can couple to the PC cavity in a wide wavelength range. The optical field distribution results show that the resonant of NA is sensitive to the polarization. When the polarization is horizontal to the NA, the NA can concentrate the optical field in the gap centre. And when the polarization is vertical to the NA, the NA can not concentrate the optical field. But the corners of the NA are bright. The electric field enhancement in the gap normalized to the incident intensity can reach few hundreds.

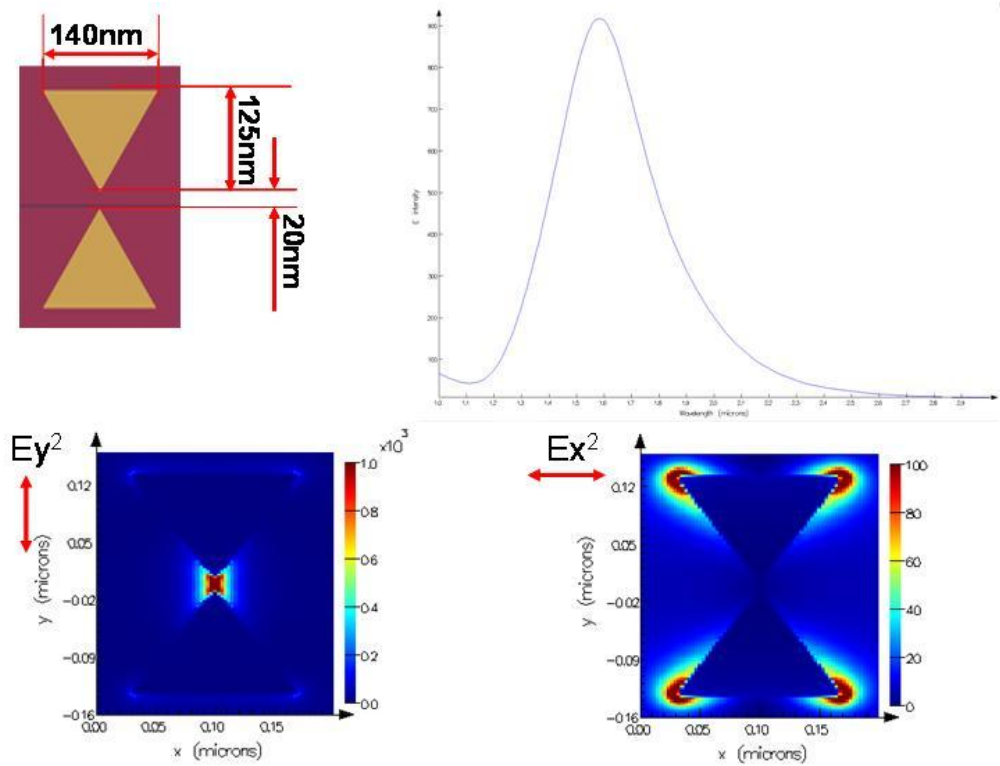

*Figure 1.21 Numerical simulation of the NA optical response.*

### 1.3.3 FDTD simulation for the CL5 hybrid structures

For the CL5 and CL7 PC cavities, the NAs are put at 3 positions on the cavity. First position (P1) is in the centre of the cavity, the direction of the NA is vertical to the cavity to get coupling with the optical field maintained by $Ey^2$ polarization. Second position (P2) is near the above edge or below edge of the cavity, and near the hole to get coupling with the optical field maintained by $Ex^2$ polarization. The last position (P3) is in the centre of the cavity again, but the direction was horizontal to the cavity. This design is to see whether the NA can be coupled with the optical field maintained by $Ex^2$ polarization at the corner of the triangle. The direction of the NA is horizontal to the cavity. The simulation results of the hybrid structures are demonstrated here. The results of CL5 structures and CL7 structures show the same regular.

### 1.3.3.1 FDTD simulation for the CL5 hybrid structures for NA at P1

When the NA is at P1, the optical field distribution is shown in Figure 1.22. 3D-FDTD simulation for the hybrid devices was done. The NA couples to the optical fields especially the optical field in the centre of the cavity. The optical field is localized in the gap of the NA. And the intensity of the optical field in the gap is much stronger than other fields (c)). The emission wavelength is 1676nm. Compare with the PC cavity without NA ($\lambda$ = 1679nm), in this case, the emission wavelength has a blue-shift (b)) of 3nm.

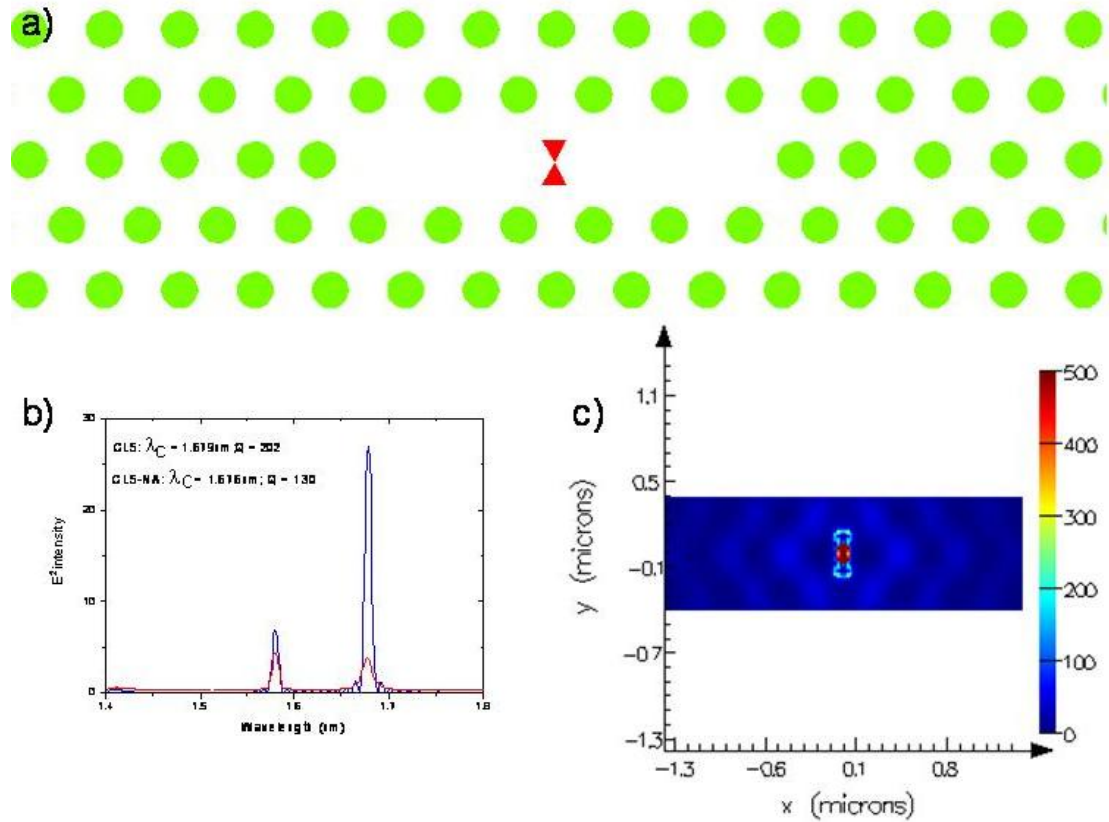

*Figure 1.22 Numerical simulation of optical field distribution of CL5 PC cavity with NA in the centre and vertical to the cavity.*

### 1.3.3.2 FDTD simulation for the CL5 hybrid structures for NA at P2

When the NA is at P2, the optical field distribution is shown in Figure 1.23. 2D-FDTD simulation for these hybrid devices was demonstrated. Hence one can not see the wavelength shift in this case. The NA couples to the optical fields especially the optical field maintained by $Ex^2$ polarization under it. The optical field is localized in the gap of the NA. And the intensity of the optical field in the gap is much stronger than other fields.

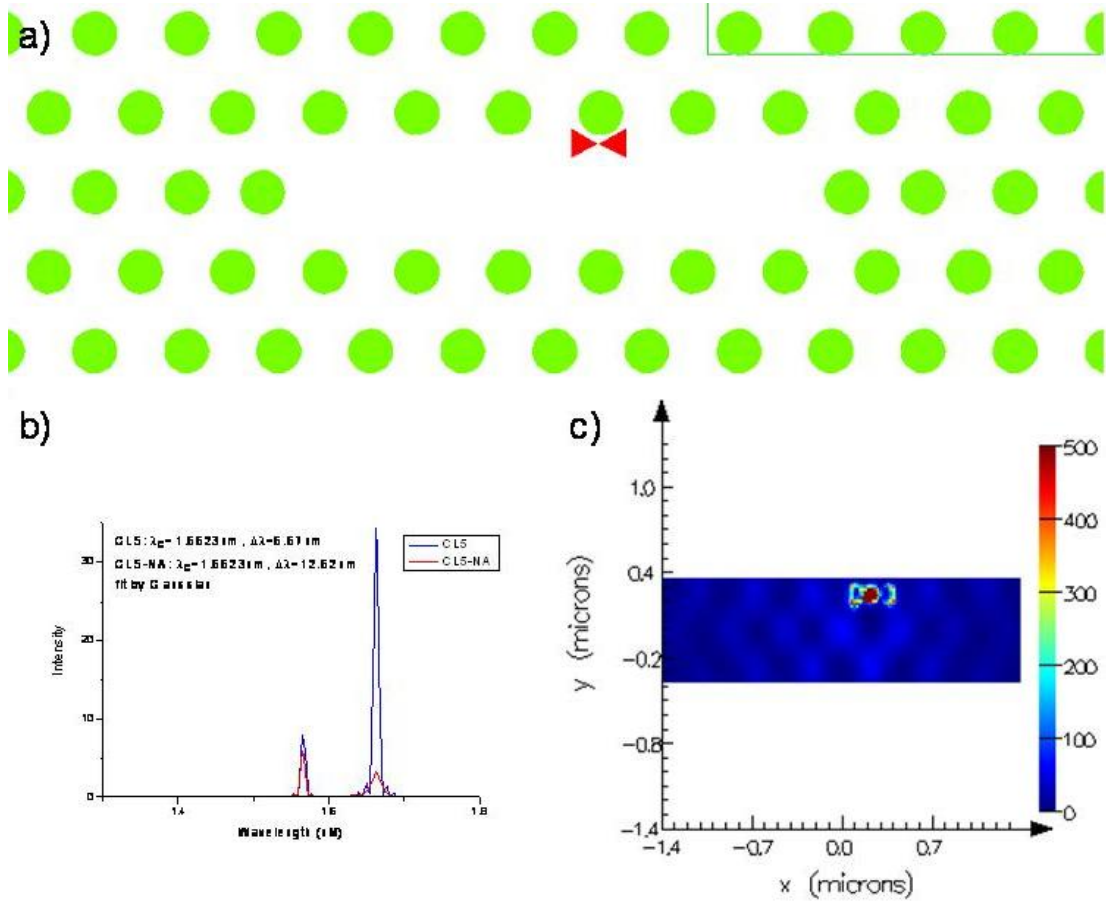

*Figure 1.23 Numerical simulation of optical field distribution of CL5 PC cavity with NA in the side and horizontal to the cavity.*

### 1.3.4 FDTD simulation for the CL7 hybrid structures

The simulation for CL7 PC cavity without NA is not done, but it should similar to the result of CL5 PC cavity. For the hybrid CL7 structures, only the case of NA at P2 is simulated. This work is done by Prof. Di Feng.

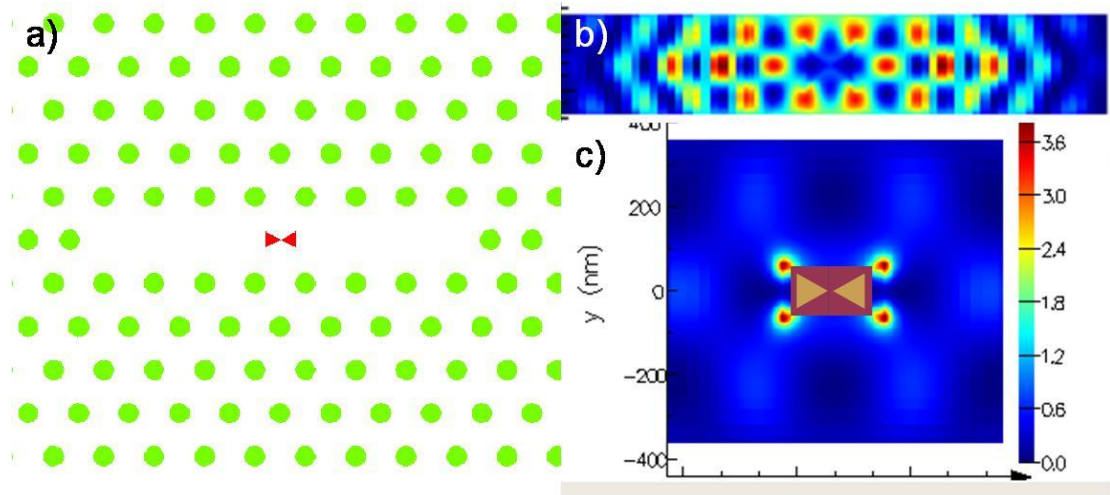

*Figure 1.24 Numerical simulation of optical field distribution of CL7 PC cavity with NA in the centre and horizontal to the cavity.*

When the NA is at P3, the optical field distribution of CL7 hybrid structure is shown in Figure 1.24. The NA couples to the optical filed of the cavity. However, the coupling situation is complex. The gap of the NA is dark. Since the direction of the NA is not along the polarization of the optical field under it. So the optical field is not confined in the gap. The optical field is induced to the four corners outside. The four corners also couple to the optical field near them. Therefore, the four corners of the NA are bright but the gap is dark. This result indicates the coupling between the PC cavity and the NA is sensitive to the polarization.

**Conclusion**

In this chapter, we discussed the theoretical design of a novel photonic-plasmonic hybrid resonator in which photonic crystal modes are efficiently coupled to plasmonic NA modes. To achieve a completely resonant situation, the length and width of the NA have been tailored to display a resonance exactly matching the spectral position of the photonic mode. This structure is not alignment sensitive, and the excitation and detection does not require bulky free-space optics. The hybrid structure can be used as an efficient functional lab-on-chip platform for biochemical sensing, energy harvesting, on-chip signal processing, and communications. The proposed ultra-compact device can be implemented on a chip in an integrated platform. This is the purpose of chapter 2

More specifically, 2 hybrid platforms have been studied in this chapter, both demonstrating an efficient coupling between a PC resonator and a localized plasmonic mode. In the first configuration, we take advantage of the efficient addressing of a PC Bloch mode by a free space beam in order to funnel the light to the NA. In the second one, which is investigated experimentally in this thesis, the NA is inserted in a PC based microcavity. The choice of this coupling scheme is motivated by 2 main purposes:

-   From a physical point of view, it can be considered as a canonical example of the coupling between dielectric and metallic resonators. Moreover, it opens the way to optical characterisations under a lasing regime. Indeed, gain material, like III-V semiconducting quantum structures, can be easily inserted in the PC cavity, leading to an efficient optical feedback and to stimulated emission. In that case, the goal is to study how the localized plasmonic mode will capture photons from the lasing mode to create a hot light spot at the NA position.

-   From a device point of view, this configuration leads to an efficient addressing of the NA from a laser cavity. Therefore, a peculiar attention must be paid to the design of the coupling strength. Indeed, the photons stored into the plasmonic mode will experience a large loss rate due to the absorption and diffusion properties of the NA. To summarize, the trade-off is to get a sufficient coupling between the NA and the PC modes in order to store light into the former, but avoiding a too strong decrease of the Q factor of the hybrid mode which can result in the killing of the stimulated emission.

**Bibliography**

[1] Song, B.-S. S. Noda, T. Asano, and Y. Akahane. 2005. Ultra-high-Q photonic double-heterostructure nanocavity. *Nat. Materials* 4:207–210.

[2] Notomi, M.Y., K. Yamada, A. Shinya, J. Takahashi, C. Takahashi, and I. Yokohama. 2001. Extremely large group-velocity dispersion of line-defect waveguides in photonic crystal slabs. *Phys. Rev. Lett.* 87:253902/1-4.

[3] Gersen, H.K., T.J. Karle, R.J.P. Engelen, W. Bogaerts, J.P. Korterik, N.F. van Hulst, T.F. Krauss, and L. Kuipers. 2005. Real-space observation of ultraslow light in photonic crystal waveguides. *Phys. Rev. Lett.* 94(7):073903/1-4.

[4] Vlasov, Y.A., M. O'Boyle, H.F. Hamann, and S.J. McNab. 2005. Active control of slow light on a chip with photonic crystal waveguides. *Nature* 438:65–69.

[5] Yoshie, T., A. Scherer, J. Hendrickson, G. Khitrova, H.M. Gibbs, G. Rupper, C. Ell, O.B. Shchekin, and D.G. Deppe. 2004. Vacuum Rabi splitting with a single quantum dot in a photonic crystal cavity. *Nature* 432:200–203.

[6] Reithmaier, J.S. 2004. Strong coupling in a single quantum dot-semiconductor microcavity system. *Nature* 432:197–200.

[7] Hennessy, K.B., A. Badolato, M. Winger, D. Gerace, M. Atatüre, S. Gulde, S. Fält, E.L. Hu, and A. Imamoglu. 2007. Quantum nature of a strongly coupled single quantum dot-cavity system. *Nature* 445:896–899.

[8] Fleischmann, M.H., P.J. Hendra, and A.J. McQuillan 1974. Raman spectra of pyridine adsorbed at a silver electrode. *Chem. Phys. Lett.* 26:163–166.

[9] Jeanmaire, D.L., and R.P. Van Duyne. 1977. Surface Raman spectroelectrochemistry, 1. Heterocyclic, aromatic, and aliphatic amines adsorbed on the anodized silver electrode. *J. Electroanal. Chem.* 82(1):1–20.

[10] Moskovits, M. 1978. Surface-roughness and enhanced intensity of Raman-scattering by molecules adsorbed on metals. *J. Chem. Phys.* 69:4159–4162.

[11] Takahara, J., S. Yamagishi, H. Taki, A. Morimoto, and T. Kobayashi. 1997. Guiding of a one-dimensional optical beam with nanometer diameter. *Optics Lett.* 22(7):475–477.

[12] Ebbesen, T.L., H.J. Lezec, H.F. Ghaemi, T. Thio, and P.A. Wolff. 1998. Extraordinary optical transmission through sub-wavelength hole arrays. *Nature* 391:667–669.

[13] Pendry, J. 2000. Negative refraction makes a perfect lens. *Phys. Rev. Lett.* 85:3966–3969.

[14] Shalaev, V.M. 2008. Transforming light. *Science* 322:384–386.

[15] Challener, W.P., C. Peng, A.V. Itagi, D. Karns, W. Peng, Y. Peng, X.M. Yang, X. Zhu, N.J. Gokemeijer, Y.-T. Hsia, G. Ju, R.E. Rottmayer, M.A. Seigler, and E.C. Gage. 2009. Heat-assisted magnetic recording by a near-field transducer with efficient optical energy transfer. *Nat. Photonics* 3:220–224.

[16] Hirsch, L.S., R.J. Stafford, J.A. Bankson, S.R. Sershen, B. Rivera, R.E. Price, J.D. Hazle, N.J. Halas, and J.L. West. 2003. Nanoshell-mediated near-infrared thermal therapy of tumors under magnetic resonance guidance. *Proc. Natl. Acad. Sci. U. S. A.*100(23):13549–13554.

[17] Cao, L.B., D.N. Barsic, and A.R. Guichard. 2007. Plasmon-assisted local temperature control to pattern individual semiconductor nanowires and carbon nanotubes. *Nano Lett.* 7(11):3523–3527.

[18] Pala, R.W., J. White, E. Barnard, J. Liu, and M.L. Brongersma. 2009. Design of plasmonic thin-film solar cells with broadband absorption enhancements. *Adv. Mater.* 21:3504–3509.

[19] Atwater, H.P., and A. Polman. 2009. Plasmonics for improved photovoltaic devices. *Nat. Mater.* 9(3):205–213.

[20] Cai, W.W., J.S. White, and M.L. Brongersma. 2009. Compact, high-speed and power-efficient electrooptic plasmonic modulators. *Nano Lett.* 9(12):4403–4411.

[21] Tang, L. S.E. Kocabas, S. Latif, A.K. Okyay, D.-S. Ly-Gagnon, K.C. Saraswat, and D.A.B. Miller. 2008. Nanometre-scale germanium photodetector enhanced by a near-infrared dipole antenna. *Nat. Photonics* 2:226–229.

[22] Akimov, A.M., A. Mukherjee, C.L. Yu, D.E. Chang, A.S. Zibrov, P.R. Hemmer, H. Park, and M.D. Lukin. 2007. Generation of single optical plasmons in metallic nanowired coupled to quantum dots. *Nature* 450:402–406.

[23] Hryciw, A.J., Y.C. Jun, M. L. Brongersma. 2010. Electrifying plasmonics on silicon. *Nat. Mater.* 9:3–4.

[24] Bergman, D.S., and M. I. Stockman. 2003. Surface plasmon amplification by stimulated emission of radiation. *Phys. Rev. Lett.* 90:027402.

[25] Hill, M.T., Y.-S. Oei, B. Smalbrugge, Y. Zhu, T. de Vries, P.J. van Veldhoven, F.W.M. van Otten, T.J. Eijkemans, J.P. Turkiewicz, H. de Waardt, E.J. Geluk, S.-H. Kwon, Y.-H. Lee, R. Nötzel, and M.K. Smit. 2007. Lasing in metallic-coated nanocavities. *Nat. Photonics* 1:589–594.

[26] Noginov, M.Z., G. Zhu, A.M. Belgrave, R. Bakker, V.M. Shalaev, E.E. Narimanov, S. Stout, E. Herz, T. Suteewong, and U. Wiesner. 2009. Demonstration of a spaser-based nanolaser. *Nature* 460.

[27] Oulton, R.F., V.J. Sorger, T. Zentgraf, R.-M. Ma, C. Gladden, L. Dai, G. Bartal, and X. Zhang. 2009. Plasmon lasers at deep subwavelength scale. *Nature* 461:629–632.

[28] P. Nordlander, C. Oubre, E. Prodan, K. Li, and M. I. Stockman. Plasmon Hybridization in Nanoparticle Dimers. *Nano Letters*, 4(5):899–903, May 2004

[29] Isabel Romero, Javier Aizpurua, Garnett W. Bryant, and F. Javier Garcia De Abajo. Plasmons in nearly touching metallic nanoparticles: singular response in the limit of touching dimers. *Optics Express*, 14(21):9988, October 2006

[30] Garnett W Bryant, F Javier Garcia de Abajo, and Javier Aizpurua. Mapping the plasmon resonances of metallic nanoantennas. *Nano letters*, 8(2):631–6, February 2008.

[31] J. P. Kottmann and O. J. F. Martin, Retardation-induced plasmon resonances in coupled nanoparticles, *Opt. Lett.*, vol. 26, no. 14, pp. 1096-1098, 2001.

[32] H. Tamaru, H. Kuwata, H. T. Miyazaki, and K. Miyano, Resonant light scattering from individual Ag nanoparticles and particle pairs, *Appl. Phys. Lett.*, vol. 80, p. 1826, 2002.

[33] W. Rechberger, A. Hohenau, *et al.*, Optical properties of two interacting gold nanoparticles, *Optics Communications*, vol. 220, no. 1-3, pp. 137-141, 2003

[34] K. H. Su, Q. H.Wei, *et al.*, Interparticle coupling effects on plasmon resonances of nanogold particles, *Nano Lett.*, vol. 3, no. 8, pp. 1087-1090, 2003.

[35] E. Cubukcu, E. A. Kort, K. B. Crozier, and F. Capasso, Plasmonic laser antenna, *Appl. Phys. Lett.*, vol. 89, p. 093120, 2006.

[36] R. M. Bakker, H. K. Yuan, *et al.*, Enhanced localized fluorescence in plasmonic nanoantennae, *Appl. Phys. Lett.*, vol. 92, p. 043101, 2008.

[37] P. K. Jain, W. Huang, M. A. El-Sayed, and P. Nordlander, On the universal

scaling behavior of the distance decay of plasmon coupling in metal nanoparticle pairs: a plasmon ruler equation, *Nano Lett.*, vol. 7, pp. 2080-2088, 2007.

[38] E. J. Smythe, E. Cubukcu, and F. Capasso, Optical properties of surface plasmon resonances of coupled metallic nanorods, *Opt. Express*, vol. 15, pp.7439-7447, 2007.

[39] H. Gai, J.Wang, and Q. Tian, Tuning the resonant wavelength of a nanometric bow-tie aperture by altering the relative permittivity of the dielectric substrate, *J. Nanophotonics*, vol. 1, no. 013555, p. 013555, 2007.

[40] J. Merlein, M. Kahl, *et al.*, Nanomechanical control of an optical antenna, *Nature Photonics*, vol. 2, no. 4, pp. 230-233, 2008.

[41] H. Fischer and O. J. F. Martin, Engineering the optical response of plasmonic nanoantennas, *Opt. Express*, vol. 16, pp. 9144-9154, 2008.

[42] A. J. Haes, S. Zou, G. C. Schatz, and R. P. V. Duyne, Nanoscale opticalbiosensor: short range distance dependence of the localized surface Plasmon resonance of noble metal nanoparticles, *J. Phys. Chem. B*, vol. 108, no. 22, pp. 6961-6968, 2004.

[43] W. H. Weber and G. W. Ford, Propagation of optical excitations by dipolar interactions in metal nanoparticle chains, *Phys. Rev. B*, vol. 70, no. 12, p. 125429, 2004.

[44] J. N. Farahani, D. W. Pohl, H. J. Eisler, and B. Hecht, Single quantum dot coupled to a scanning optical antenna: a tunable superemitter, *Phys. Rev. Lett.*, vol. 95, p. 017402, 2005.

[45] Pottier, P., Seassal, Ch., Letartre, X., Leclercq, J.-L., Viktorovitch, P., Cassagne, D. and Jouanin, Ch. (1999) *Journal of Lightwave Technology* 17, 2058.

[46] Joannopoulos J. D., Meade R. D., Winn J.N. (1995) *Photonic crystal: Molding the flow of light* Princeton University Press, New Jersey.

[47] Akahane, Y., Asano, T., Song, B. S., and Noda, S. (2003) High-Q photonic nanocavity in a two-dimensional photonic crystal, *Nature* 425, 944.

[48] Mc Nab, S., Moll, N. and Vlasov, Y. (2003) Ultra-low loss photonic integrated circuit with membrane-type photonic crystal waveguides, *Optics Express* 11, 2927.

[49] Painter, O., Lee, R.K., Scherer, A., Yariv, A., O'Brien, J.D., Dapkus P.D. and Kim, L. (1999) Two-dimensional photonic band-gap defect mode laser, *Science* 284, 1819.

[50] Fan, S., Villeneuve, P.R., Joannopoulos J.D. and Hauss, H. A. (1998) Channel drop tunnelling through localized states, *Phys. Rev. Lett.* **80**, 960.

[51] Letartre, X., Mouette, J., Seassal, C., Rojo-Romeo, P., Leclercq, J.-L., Viktorovitch, P.(2003) Switching devices with spatial and spectral resolution combining photonic crystal and MOEMS structures, *Journal of Lightwave Technology* **21**, 1691.

[52] Notomi, M., Susuki, H. and Tamamura, T. (2001) Directional lasing oscillation of twodimensional organic photonic crystal lasers at several photonic band gaps, *Appl. Phys. Lett.* **78**, 1325.

[53] Monat, C., Seassal, C., Letartre, X., Regreny, P., Rojo-Romeo, P., Viktorovitch, P., Le Vassor D'yerville, M., Cassagne, D., Albert, J.P., Jalaguier, E., Pocas S. and Aspar, B. (2002) InPbased two-dimensional photonic crystal on silicon: In-plane Bloch mode laser, *Appl. Phys. Lett.* 81, 5102.

[54] Notomi, M., Shinya, A., Kuramochi, E. and Ryu, H-Y. (2004) Waveguides, resonators and their coupled elements in photonic crystal slabs, *Optics Express* **12**, 1551.

[55] Drouard, E., Hattori, H. T., Grillet, C., Kazmierczak, A., Letartre, X., Rojo-Romeo P. and Viktorovitch, P. (2005) Directional channel-drop filter based on a slow Bloch mode photonic crystal waveguide section, *Optics Express* **13**, 3037.

[56] Belarouci, A.; Benyattou, T.; Letartre, X.; Viktorovitch, P.. (2010) '3D light harnessing based on coupling engineering between 1D-2D Photonic Crystal membranes and metallic nano-antenna', *Optics Express* 18, A381.

[57] L. Sherry, R. Jin, C. Mirkin, G. Schatz, and R. Van Duyne, "Localized surface plasmon resonance spectroscopy of single silver triangular nanoprisms," Nano Lett. **6**, 2060–2065 (2006).

[58]P. B. Johnson and R. W. Christy, "Optical Constants of the Noble Metals," Phys. Rev. B **6**, 4370–4379 (1972).

[59] Ditlbacher, H. et al. Silver nanowires as surface plasmon resonators. Phys. Rev.Lett. 95, 257403 (2005).

[60] Bozhevolnyi, S. I., Volkov, V. S., Devaux, E., Laluet, J.-Y. & Ebbesen, T. W. Channel plasmon subwavelength waveguide components including interferometers and ring resonators. Nature 440, 508–511 (2006).

[61] Weeber, J.-C., Bouhelier, A., Colas des Francs, G.,Markey, L. & Dereux, A.

Submicrometer in-plane integrated surface plasmon cavities. Nano Lett. 7, 1352–1359 (2007).

[62] D.P. Fromm, A. Sundaramurthy, P.J. Schuck, G. Kino, and W.E. Moerner. Gap-dependent optical coupling of single "Bowtie" nanoantennas resonant in the visible. *Nano Lett.*, 4:957, 2004.

[63] P.J. Schuck, D.P. Fromm, A. Sundaramurthy, G.S. Kino, and W.E. Moerner. Improving the mismatch between light and nanoscale objects with gold bowtie nanoantennas. *Phys. Rev. Lett.*, 94:017402, 2005.

[64] A. Sundaramurthy, K.B. Crozier, G.S. Kino, D.P. Fromm, P.J. Schuck, and W.E. Moerner. Field enhancement and gap-dependent resonance in a system of two opposing tip-to-tip Au nanotriangles. *Phys. Rev. B*, 72:165409, 2005.

[65] K.B. Crozier, A. Sundaramurthy, G.S. Kino, and C.F. Quate. Opticl antenna: Resonators for local ¯eld enhancement. *J. Appl. Phys.*, 94:4632,,2003.

[66] R.C. Compton, R.C. McPhedran, Z. Popovic, G. M. Rebeiz, P.P. Tong, and D.B. Rutledge. Antennas on a dielectric half-space: Theory and experiment. IEEE Transitions on Antennas and Propagation, AP-35:622-631, 1987.

[67] S. G. Johnson, J. D. Joannopoulos, and M. Soljačić, "MIT Photonic Bands," *http://ab-initio.mit.edu/wiki/index.php/MIT_Photonic_Bands*.

[68] Thanh-Phong Vo, Adel Rahmani, Ali Belarouci, Christian Seassal, Dusan Nedeljkovic, and Ségolène Callard, "Near-field and far-field analysis of an azimuthally polarized slow Bloch mode microlaser," Opt. Express **18**, 26879-26886 (2010)

[69] Y. Akahane, T. Asano, B.-S. Song, and S. Noda, "High-Q photonic nanocavity in a two-dimensional photonic crystal," Nature **425**(6961), 944–947 (2003).

# Chapter 2: Device Fabrication

**Introduction**

In this chapter the fabrication processes and steps that have been used to fabricate the hybrid plasmonic-photonic crystal devices will be reviewed. The challenge in the fabrication process includes highly accurate electron beam lithography, high quality dry etching techniques and ultimate alignment procedure between photonic and plasmonic devices. These devices have been fabricated in InP material systems and have been used in the experimental work demonstrated in this thesis. Photonic crystal cavities were fabricated in an active material that was grown by Dr Philippe Regreny (INL) through solid source molecular beam epitaxy, as described elsewhere [1]. It consists of four InAs compressively strained quantum wells separated by InP barriers. The fabrication process can be roughly divided into defining alignment marks, processing the photonic crystal laser and positioning the optical nano-antenna on the backbone of the photonic structure by another e-beam exposure followed by a lift-off process. The whole fabrication procedure is summarized in Figure 2.1.

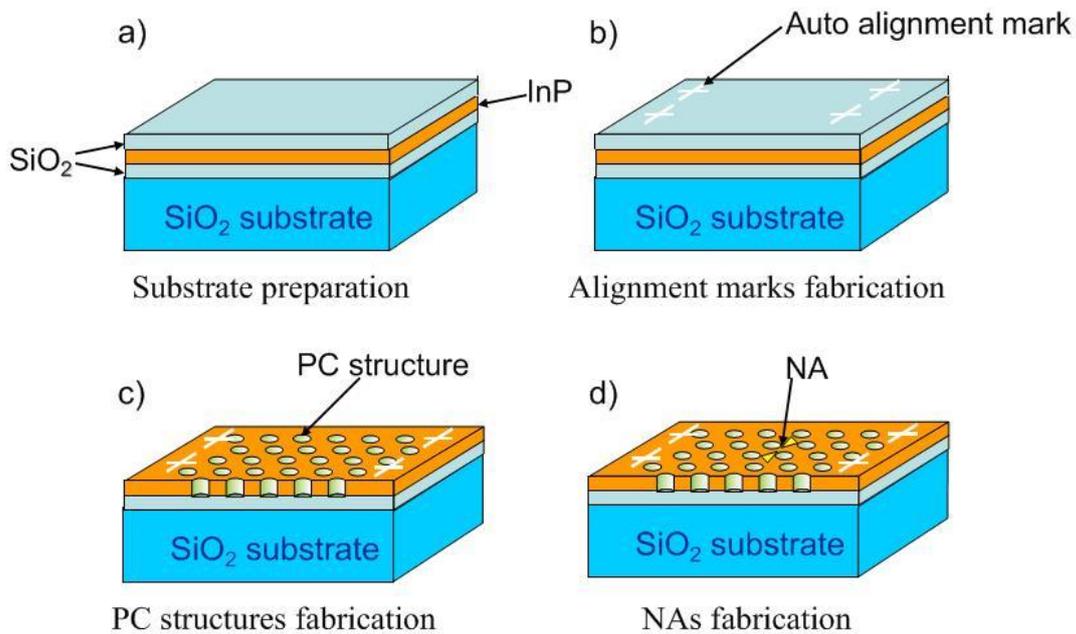

*Figure 2.1: Schematic diagram showing the fabrication steps*

## 2.1 Epitaxial layer structure

The InP slab is 250nm thick, has a refractive index of 3.17 at 1.55μm and supports only one guided mode around that wavelength. Four InAsP quantum wells (QW), separated by InP barrier layers, are grown at the center of the InP slab. The photoluminescence from the QWs occurs over a broad spectral range between 1250nm and 1650nm. This heterostructure is wafer-bonded onto a transparent SiO2 host substrate. Compared to photonic crystal cavities fabricated in free-standing membranes, this approach yields a better heat sinking, a higher mechanical stability of the structure, and enables front side and back side photo-excitation of the quantum wells. The 2D Photonic Crystal and Optical Nanoantenna structures are designed to exhibit modes in the wavelength range of the QWs emission.

The III-V heterostructure was grown by solid source molecular beam epitaxy (MBE) at INL. First a 300nm thick $In_{0.53}Ga_{0.47}As$ etch stop layer is grown on top of a 2-inch InP(001) wafer, followed by the growth of the active InP membrane [2, 3] (Figure 2.2). Then, a 10nm thick $SiO_2$ layer was deposited by Plasma Enhanced Chemical Vapour Deposition (PECVD) on the top of the InP slab. This step was assisted by Electron Cyclotron Resonance (ECR). The structure was transferred on the top of a fused silica host wafer using $SiO_2$-$SiO_2$ wafer bonding. This step was performed at CEA-LETI. The InP wafer and the each-stop layer were subsequently eliminated by selective wet chemical etching at INL. Finally, a layer of $SiO_2$ (100nm thick) was deposited on top of the active InP slab by plasma assisted sputtering, to act as a hard mask during the reactive ion etching process [2, 3].

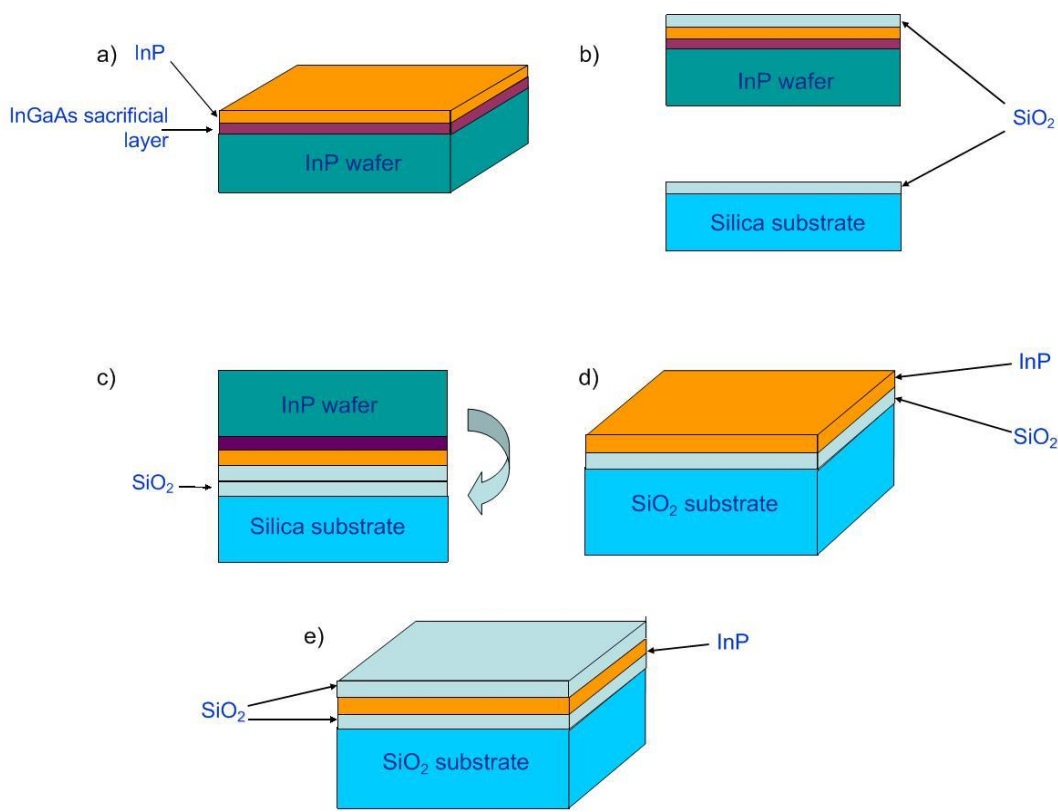

*Figure 2.2 Diagram of substrate prepareration: a) MBE grown of InGaAs layer and InP layer; b) PECVD deposition of SiO$_2$ layer; c) Active layers transfer via SiO$_2$-SiO$_2$ wafer bonding; d) Wet chemical etching; e) Hard mask layer deposition.*

## 2.2 Electron-beam lithography

### 2.2.1 General description

Electron-beam lithography was used to create any critical feature in the device masks because of its superior resolution and easily modified masks as compared to photolithography. A 30kV Inspect F FEI scanning electron microscope (SEM) was employed to raster expose the e-beam resist, with the aid of a Raith lithography software. An accelerating voltage of 30kV and working distance of 6mm was always used. The beam current (~10pA) was measured with a Faraday cup to maintain a constant dose during the beam write. A magnification of 1100×, corresponding to a calibrated field of view of 100$\mu$m on a side, was predominantly used. A proper selection of the electron beam resist is very crucial to obtain the resolution inherent to the e-beam lithography systems. Two types of electron beam resists were used in this thesis work: 950K molecular weight single layer PMMA resist for the definition of photonic crystals and 950K/495K bilayer PMMA resists to write alignment marks and NAs. PMMA is an abbreviation for poly-methylmethacrylate. The different sequences are shown schematically in Figure 2.3.

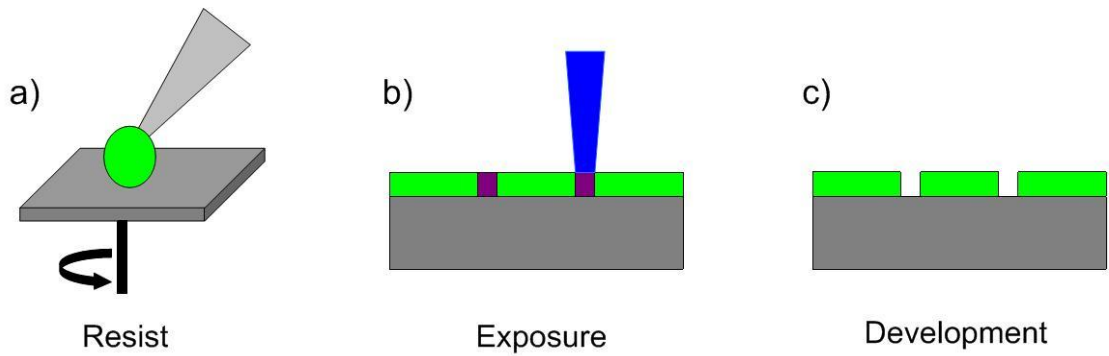

Fig. 2.3 Diagram of E-beam lithography: a) Resist coating; b) E-beam exposure; c) Development.

Of all the steps in the fabrication process, this step has presented the maximum degree of variability and, as a result, considerable effort was invested in identifying and addressing the issues. These are now highlighted.

**Proximity effect** - During lithography, an e-beam isn't contained within an area intended for exposure. Forward scattering (due to electron-electron interactions within the beam) and backward scattering (by the substrate below the resist) widens the e-beam, and as a result proximate features tend to get an additional e-beam exposure. In the worst case, this leads to the merging of relatively close features. This effect is compounded by an e-beam that is defocused. Proximity-effect-correction (PEC) is accomplished by employing Monte Carlo simulation methods that compute electron trajectories (for several million electrons at a time) for a given accelerating voltage and wafer stack, and can be used to adjust dosing for proper exposure. It is found that for photonic crystal presented in this work, it suffices to shrink mask features by a small amount to get optimal exposure.

**Charging effect** - Resist charging can be a significant problem in electron-beam lithography. The beam deflection caused by it leads to pattern displacement, pattern distortion, and registration error (see Figure 2.4). The typical solution is to add a conductive layer either above or below the resist. The choice of conductor (metal, polymer, etc.) will depend on the details of the application. In some cases, it may be difficult to determine a coating/removal process that is compatible with the entire device fabrication process. One simple solution to prevent pattern distortion from charging is to evaporate ~10nm of aluminium on top of the resist before writing and etch it off in NaOH or KOH after the writing and before developing.

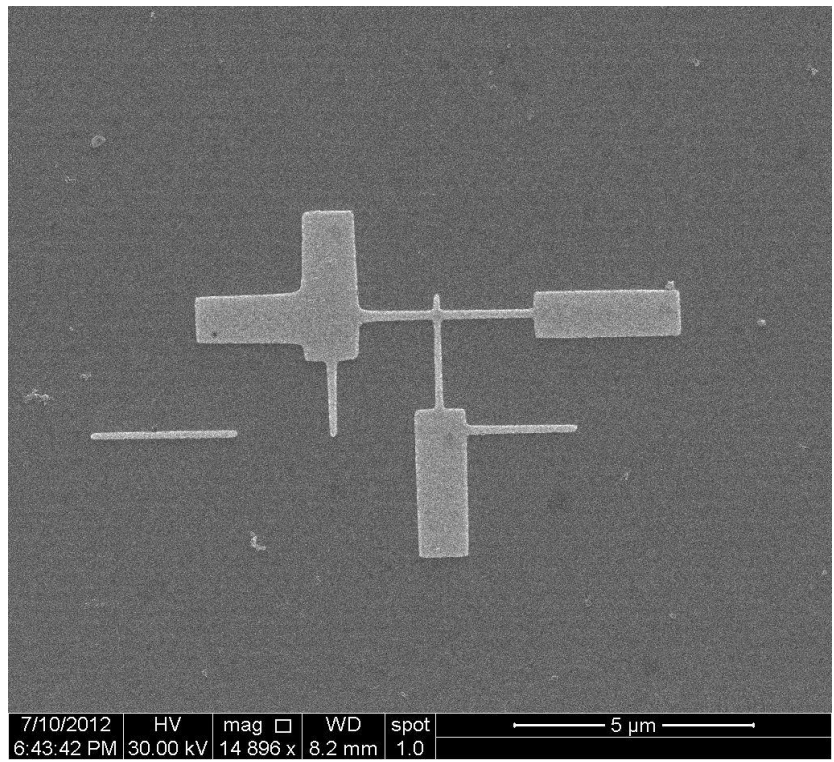

*Figure 2.4 An auto alignment mark fabricated with charging effect*

## 2.2.2 Alignment issues and technique

Several structures have been fabricated by a high-precision three-step electron beam lithography process and layer-by-layer stacking nanotechniques. Complete control over alignment mark definition provides the capability to align to a wide array. The markers define a local coordinate system which is then used for the next layers to achieve alignment on the nanoscale from layer to layer. The steps are repeated for the next layer using the metal markers to align the second layer with respect to the first layer in the electron beam system. The steps can be repeated in principle as long as the markers are detectable. Overlay accuracy down to 20nm measure by on-chip verniers can be routinely achieved. Before patterning the photonic and plasmonic structure, several robust gold marks have been processed on the substrate. The set of photonic crystal defined in the first layer was patterned in a positive tone resist and transferred into a silica hard mask and in the InP membrane. Finally, the set of plasmonic structures defined in the second layer was exposed, aligned with respect to the common coordinate system, followed by thermal evaporation of gold and lift-off.

### 2.2.3 Structure design

The full pattern can be designed in a working area of 500μm by 500μm, containing 25 squares of 100μm by 100μm. The size of each square is equal to the write field for each step of lithography. A set of 8 or 9 hybrid structures can be exposed in a 100μm by 100μm write field (Figure 2.5 and Figure 2.8). The location and dose of each pattern can be set, as well as the layer with the Raith software.

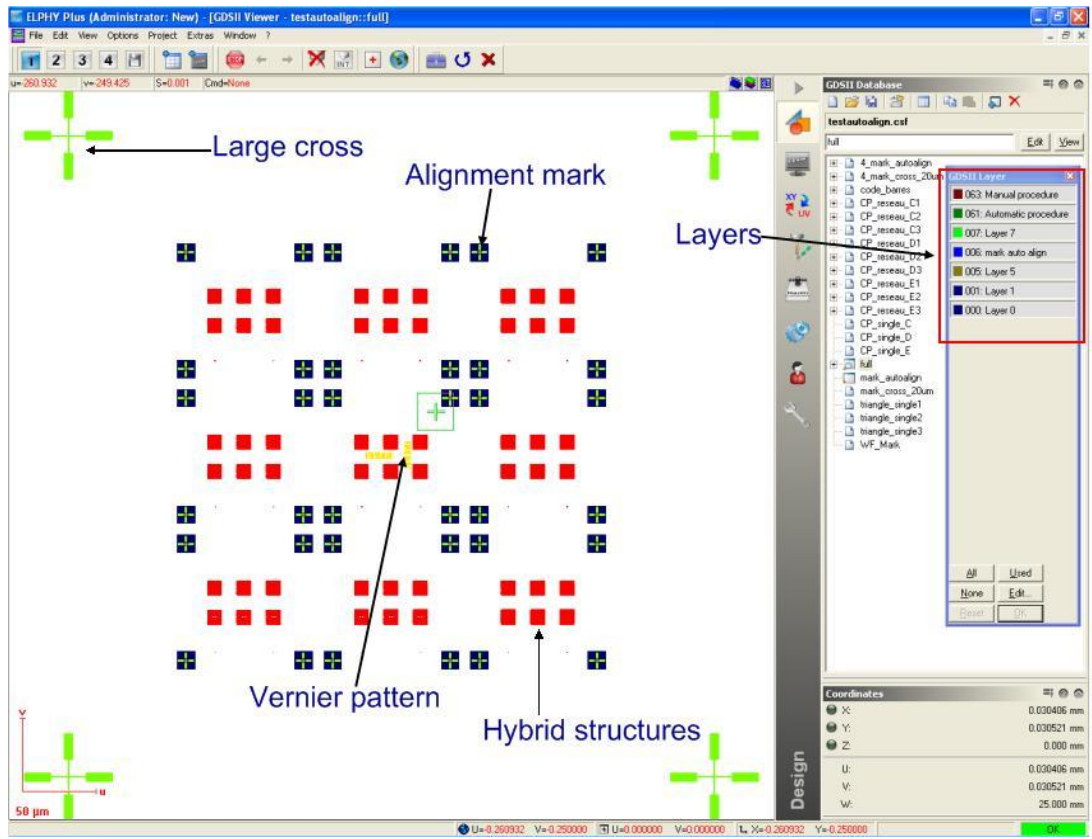

*Figure 2.5 A full pattern of CL7 or CL5 PC cavities with NAs. The different colours of the hybrid structures and large crosses show the different dose factors of these two structures.*

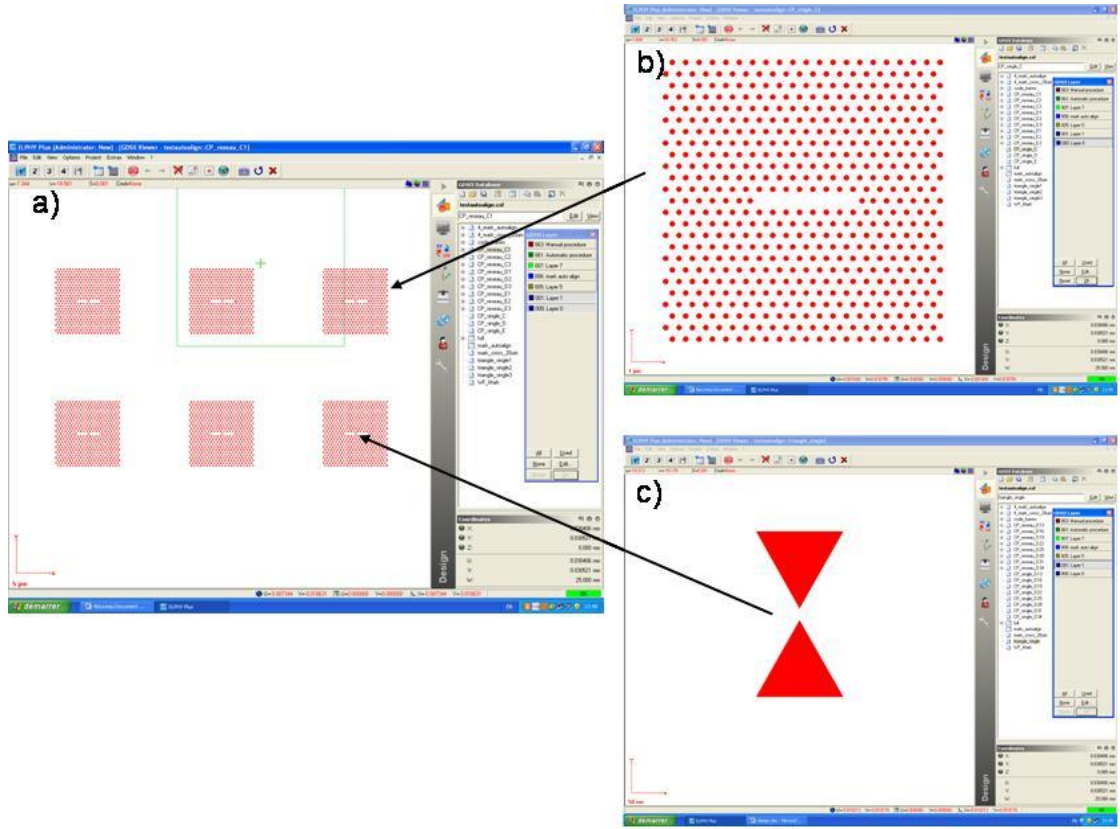

*Figure 2.6 One group of CL7 PC cavities with NAs: a) The whole pattern. b) Zoom in one PC structure. c) Detail of the NA.*

Each CL7 or CL5 PC cavity is made from a triangular array of cylindrical holes with a period of 420nm (Figure 2.6b). The structures are divided into 3 groups by radius of the holes. Holes radii have been set to 90nm, 100nm and 110nm for each group. To form a CL7 cavity, 7 holes are omitted at the centre of the array. For CL5 structure, 5 holes are omitted at the centre of the array as well. Each group is composed of 6 structures. The doses of each structure are set from 1.5 to 2.0. To increase the quality factor, two holes on either corner of the cavity are shifted by 80nm outward [4]. The bowtie NA consists of two isosceles triangles, head to head. The base length of the triangle is 140nm and its height is 125nm. The gap between the two arms is 20nm as reported in Figure 2.6c. The dose of the NAs is set to 3. They have been located at 3 strategic positions on the backbone of the PC cavities (Figure 2.7). One position is in the centre of the cavity, the direction of the NA is vertical to the cavity to reach a coupling with the optical field $Ey^2$ and confine the optical field within the antenna gap as detailed in Chapter 1. The second position is also in the centre, but the direction is horizontal to the cavity. This configuration allows us to investigate whether the NA can be coupled with the optical field at the corners of the triangles. Finally, the last position is near the upper edge or lower edge of the cavity, and close to the hole, where the optical NA can couple with the Ex component of the field. The direction of the NA is horizontal to the cavity.

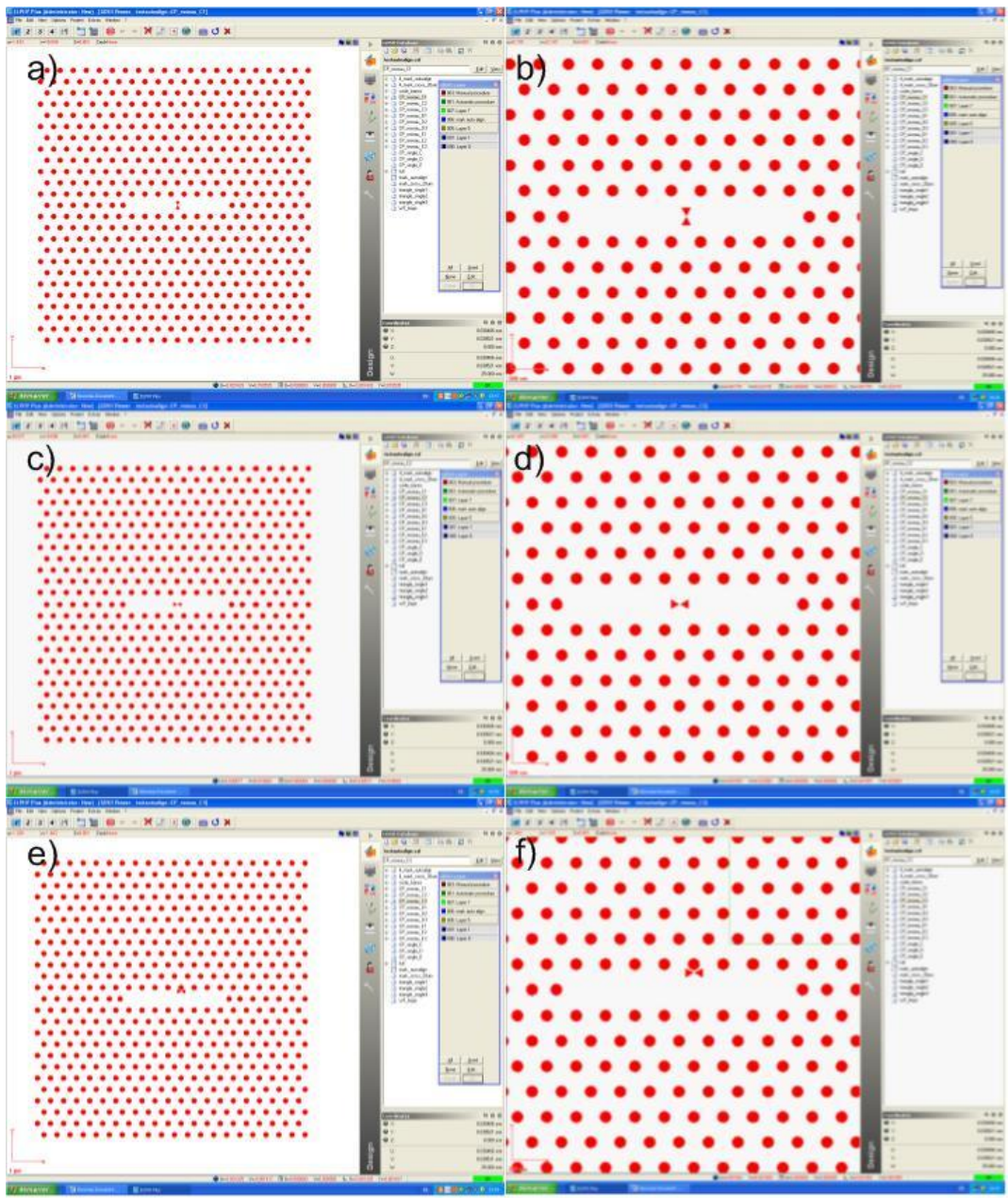

*Figure 2.7 The 3 positions where the NAs are positioned on the PC cavities. a), c) and e) are overview of the hybrid structures. b), d) and f) are the details of the cavities.*

Bloch mode based photonic crystal cavities have also been fabricated (Figure 2.8). The graphite-type PC has a size of 60µm x 60µm). 3 periods have been selected: 450nm, 460nm and 470nm, corresponding to the distance between the two nearest holes of the unit cell. The hole radius is 120nm. Each group is made of 8 structures with doses ranging from 1.3 to 3.4. The geometrical parameters of the NA have been set to display a plasmonic resonance at 1.5µm and are identical to the one grafted on the CL cavity. A group of 16 NAs has been uniformly positioned on the PC to study the optical coupling between the antenna and the monopolar mode the graphite structure as shown in Figure 2.9.

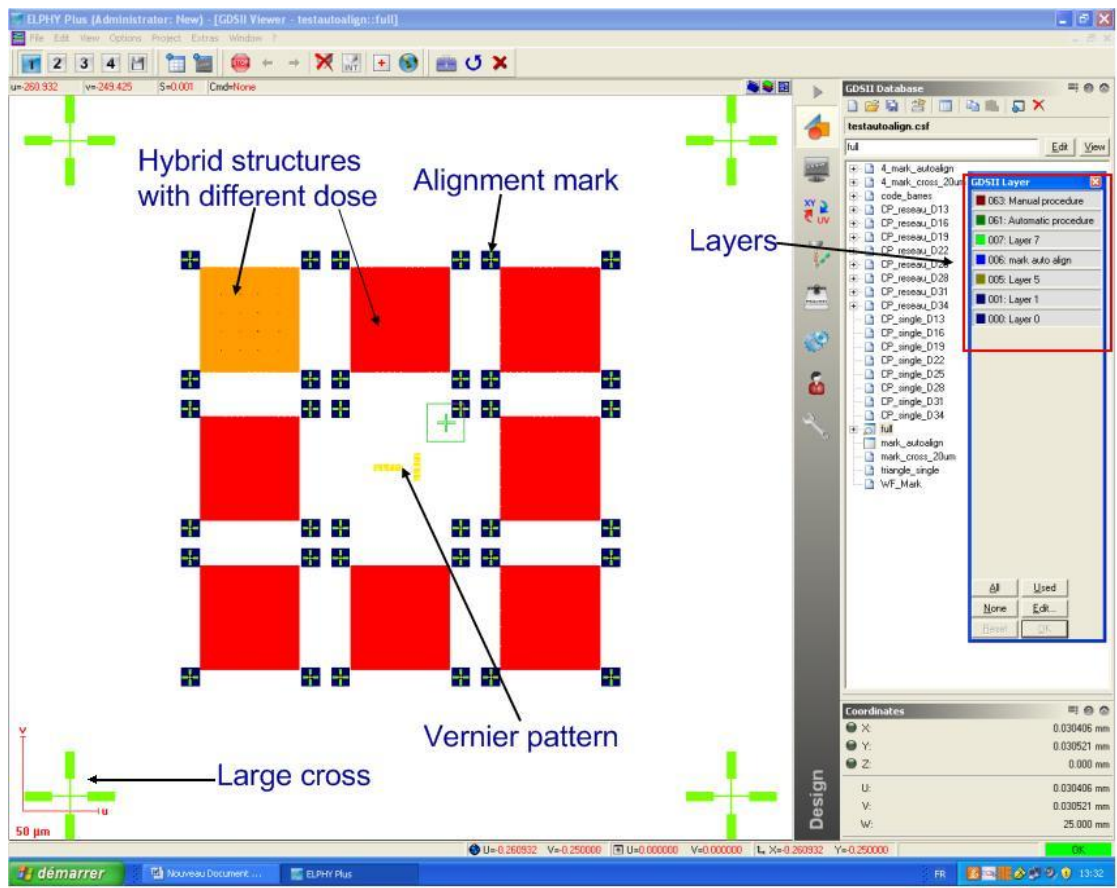

*Figure 2.8 A full pattern of graphite PC with NAs. The different colours of the hybrid structures denote the different dose factors that have been used.*

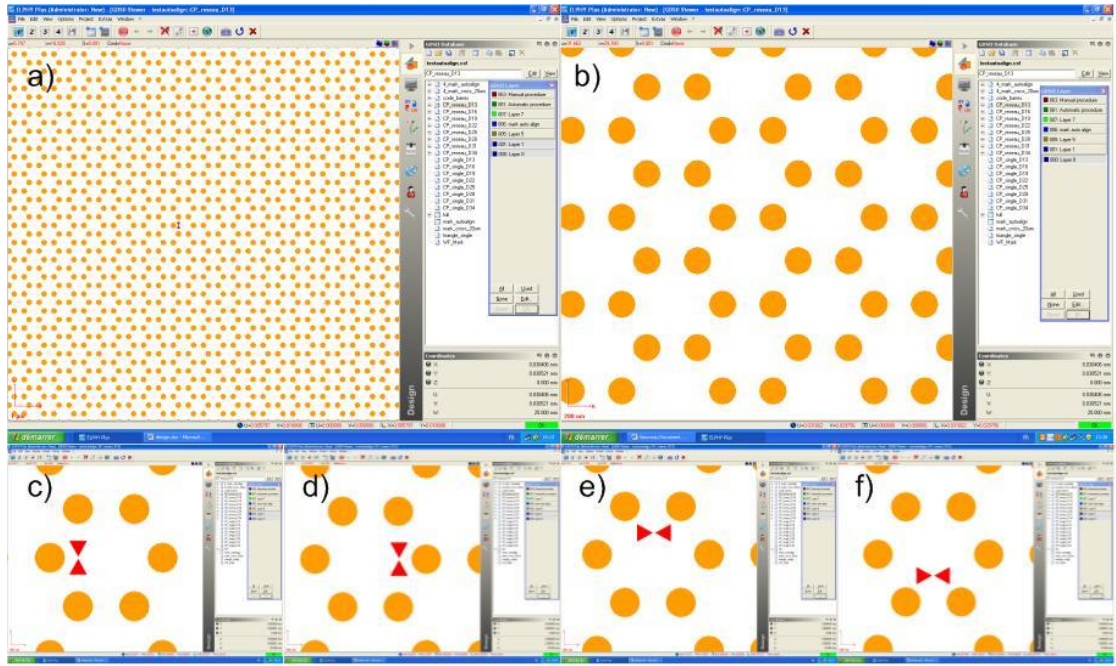

*Figure 2.9 Details of graphite PC with NAs: a) View of the PC with a NA; b) Zoom in of the PC; c) to f) Different positions of the NA.*

## 2.4 Alignment marks fabrication

Before Patterning the photonic crystal sheet, several metallic markers have been processed on the substrate using a double layer of positive resist PMMA. Using a double layer of PMMA with a more sensitive PMMA as a first layer yields an undercut after exposure and development and hence facilitates the lift-off process [5, 6]. The process flow is described as follow (Figure 2.10):

a- Spin on MMA (8.5) MAA EL 9 at 3000 rpm for 30s, which gives a 170nm-thick film. Bake for 90s at 150°C on a hot plate. Spin on 950 PMMA $C_2$ at 2000 rpm for 30s, which gives a 70nm-thick film. Bake for 90s at 180°C.

b- The sample is then exposed setting the diaphragm to 7, the voltage 30kV and the spot size to 1.0. A 10pA beam current intensity has been measured at the Faraday cup location and used to calculate the dwell time before scanning. The operations of "stage to sample adjustment" and "write field alignment" must be operated very carefully. The detail of the operations is shown in appendix B.

c- Develop with 2-Methyl-2-Pentanone (methyl isobutyl ketone, or MIBK) diluted with 2-Propanol as 1:1 for 60s. Rinse with 2-Propanol, blow-dry with nitrogen, and bake at 100°C on a hot plate for 60s.

d- Metals are deposited by electron gun evaporation at a rate of ~1Å/s. An adhesion layer of 10nm chromium is evaporated before the evaporation of 40nm gold and 10nm chromium.

e- The sample is put in a trichloroethylene bath at 60°C for 10min to dissolve the residual PMMA and lifts-off the excess metal, leaving the markers on the substrate.

f- The markers define a local coordinate system which is then used for the next layers to achieve alignment on the nanoscale from layer to layer.

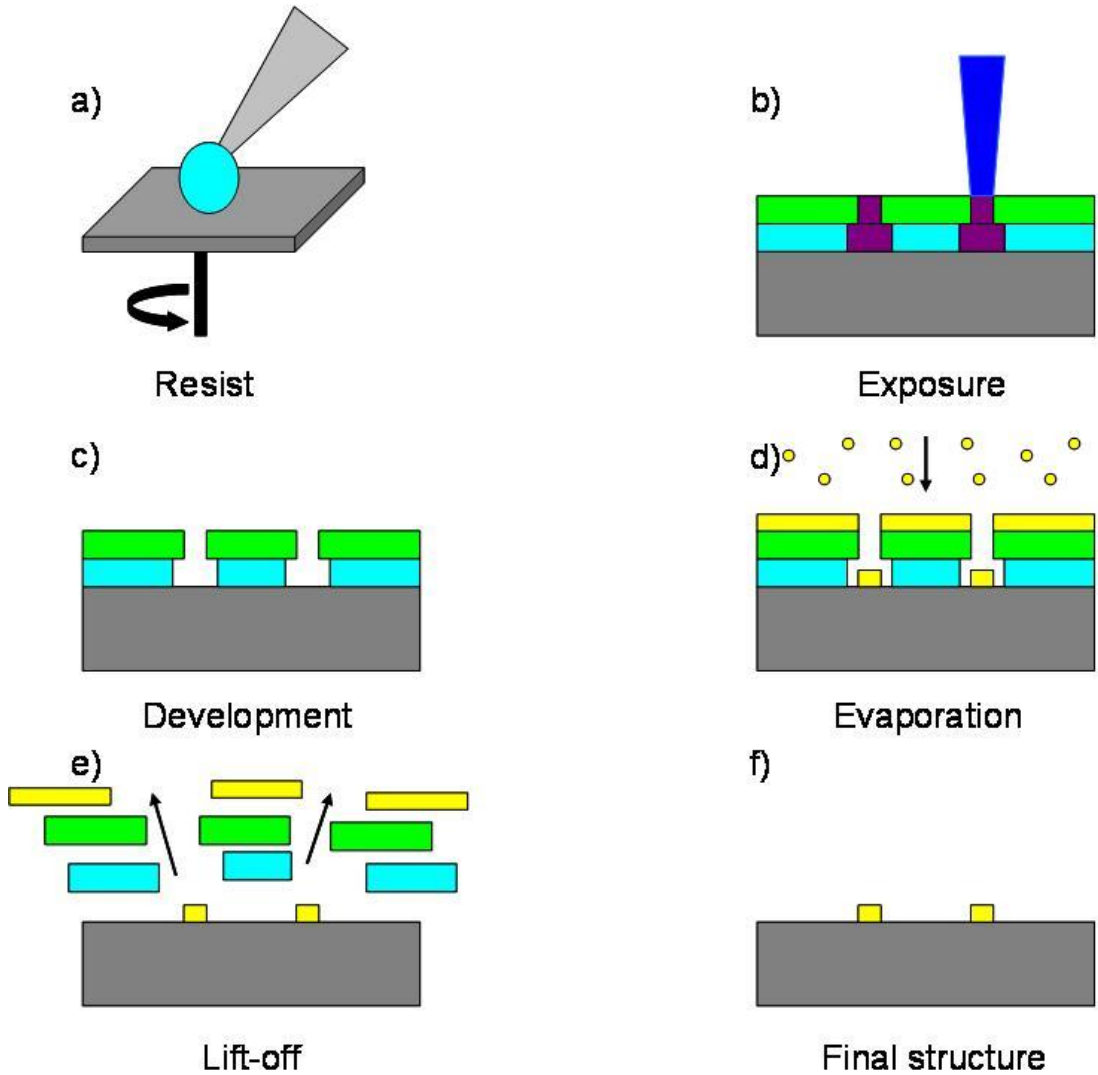

*Figure 2.10 Diagram of the lift-off process*

## 2.5 Fabrication of PC structures

The PC patterns are created through electron beam (E-beam) lithography of 70nm thick layer of PMMA A4 (5000 rpm for 30s, baked on a hot plate at 180°C for 90s). The diaphragm is set to 6, the voltage to 30kV, and the spot size to 2.0. A value of 40pA for the beam current has been measured and used to estimate the dwell time. Computer-controlled alignment down to 20nm using the gold alignment markers is applied to ensure the accurate positioning of the PC structure with respect to the common coordinate system. Each mark is scan horizontally and vertically over the arm of the cross (Figure 2.11). The software can record and display the intensity of the secondary electron as a function of the position [7, 8]. The intensity of the marks is much higher than the rest of the substrate (Figure 2.12). The alignment marks are scanned repeatedly until a satisfactory alignment is obtained (more details can be found in appendix C).

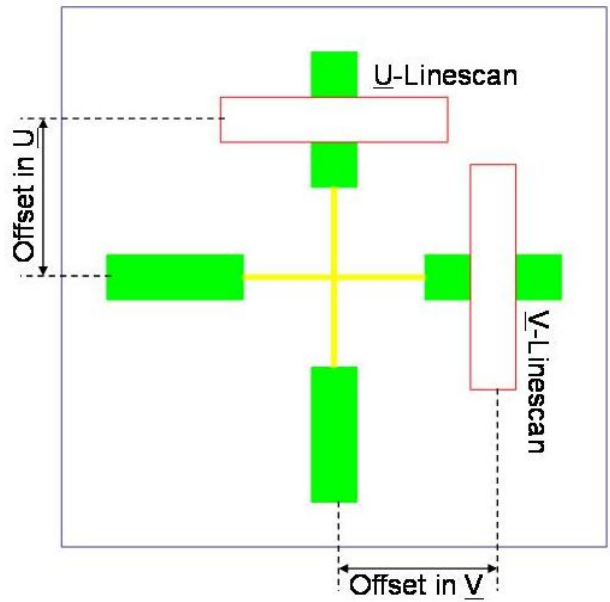

*Figure 2.11 Schema of the two line scans on the auto-alignment mark*

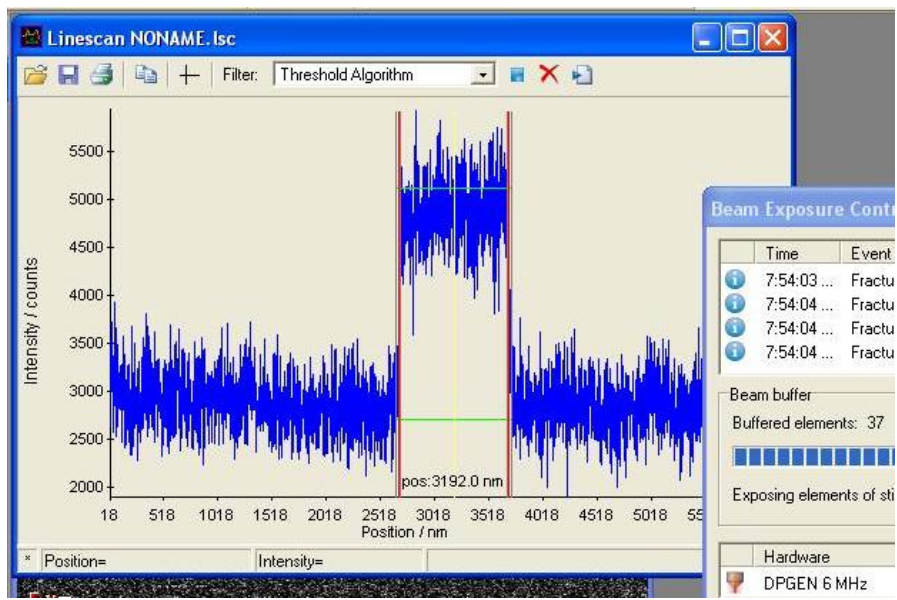

*Figure 2.12 The line scan window.*

After exposure, the resist has been developed in MBIK/IPA 1:1 during 45s, rinsed 2-isopropanol and finally baked at 100 ˚C for 60s. Reactive ion beam etching (RIBE) is used to open the holes in the SiO2 layer with a fluorine gas. The hard mask pattern has finally been transferred to the active InP membrane using a CH4/H2 gas mixture. These fabrication steps are described in detail in Table 2.1 and Figure 2.13.

| Step | Gas | Flow (sccm) | Pressure (mTorr) | $P_{RF}$ (W) | Time (s) |
|---|---|---|---|---|---|
| Etching $SiO_2$ | $CHF_3$ | 16 | 15 | 100 | 600 |
| Removing PMMA A4 | $O_2$ | 20 | 100 | 100 | 120 |
| Etching InP | $CH_4$ | 15 | 30 | 200 | 300 |
| | $H_2$ | 30 | | | |
| Removing C | $O_2$ | 20 | 100 | 100 | 120 |
| Removing $SiO_2$ | $CHF_3$ | 16 | 15 | 100 | 600 |

**Table 2.1 Gases and working conditions during each step of RIE**

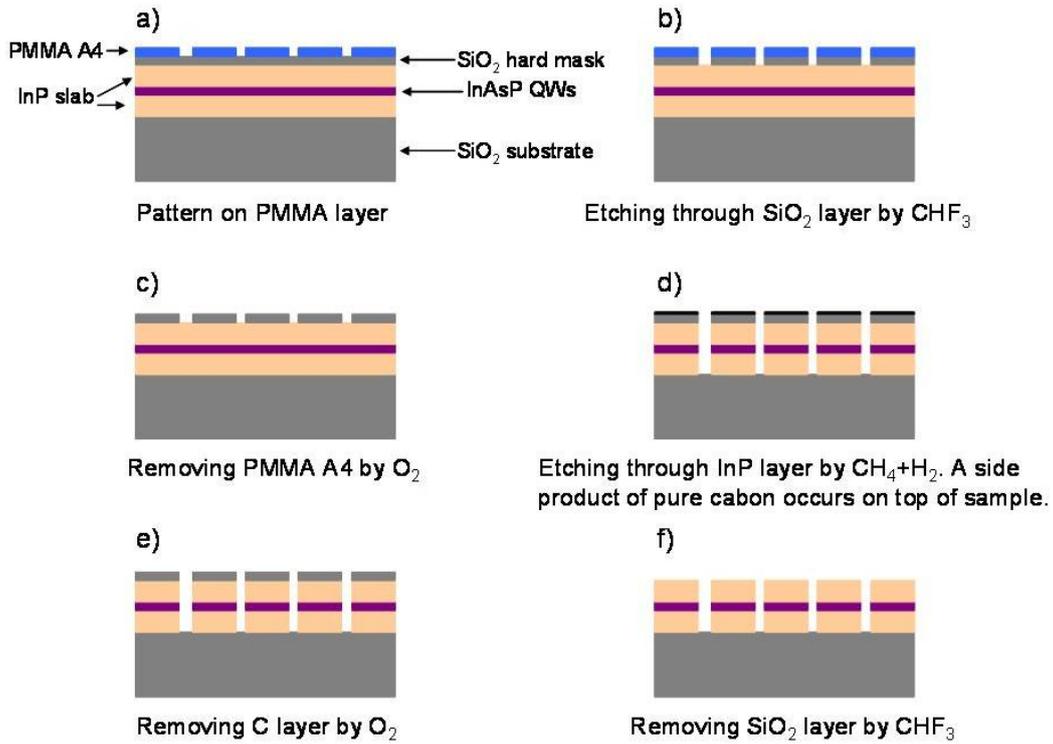

*Figure 2.13 Diagram of RIE steps to transfer PC patterns into the InP slab*

Figure 2.14 displays an alignment mark after the process of the photonic crystal layer. Despite the RIE procedure, the mark is still robust and exhibits sufficient contrast (measured by secondary electron line scan) to be used gain for multilevel E-beam lithography.

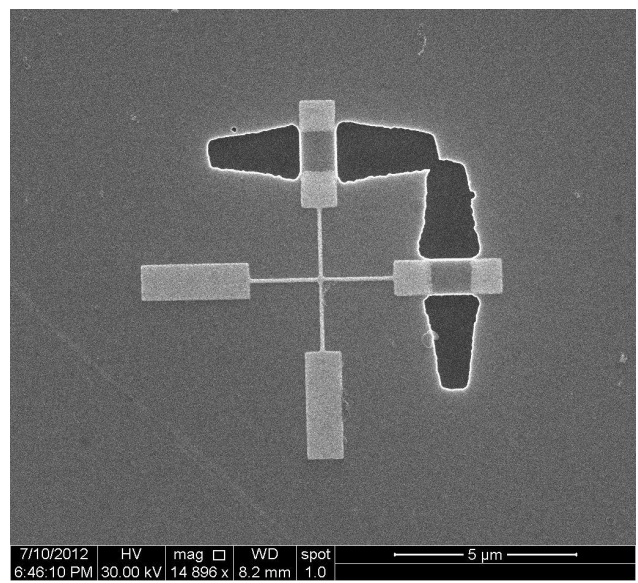

*Figure 2.14 Auto-alignment mark after PC structure fabrication*

## 2.6 Fabrication of NAs

The steps are repeated for the plasmonic sheet using metal markers to align the second layer with respect to the first one in the E-beam system. Accurate stacking alignment can be repeated in principle as long as the markers are detectable with the electron beam. Individual metallic NAs are thus deterministically positioned on the backbone of the PC cavity by a second aligned e-beam exposure followed by thermal evaporation of 4nm Ti and 40nm gold, and a lift-off process. Optical antennas with reproducible gap sizes down to 10nm have been successfully produced

An alignment mark after the process of the plasmonic layer is reported on Figure 2.15. The metal deposition over the horizontal and vertical line scan is clearly visible.

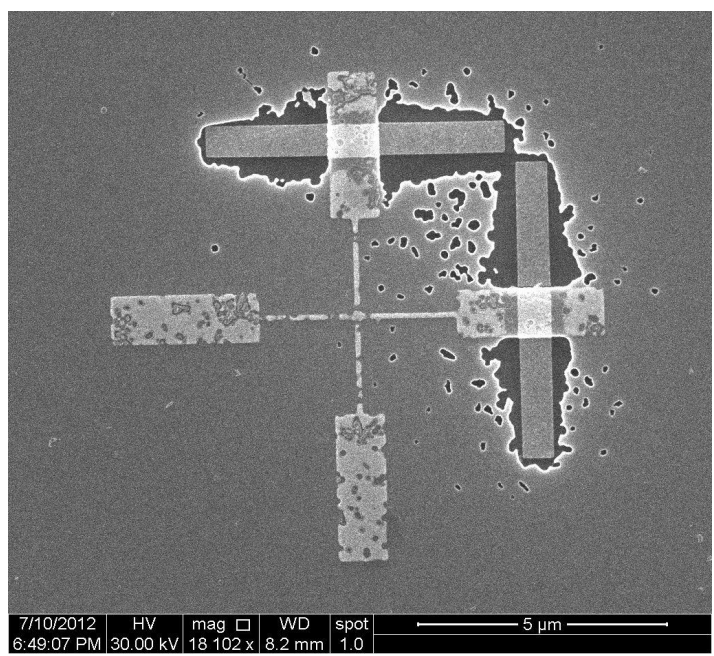

*Figure 2.15 Auto-alignment mark after NAs fabrication*

## 2.7 Fabrication results and discussion

The fabrication quality of the hybrid devices has been investigated by SEM, scanning near-field optical microscope (SNOM) and atomic force microscopy (AFM). For SEM, 5kV Inspect F FEI SEM and 5kV Tescan SEM were employed. They were used to investigate the feature of the devices. The SNOM system equipped with ND-MDT shear force head has been used to investigate the near-field optical properties of the structures. The distance between the optical near-field probe and the surface of sample is kept constant by a shear-force feed-back loop [9]. The position of the probe is collected and recorded during the raster movement. This enables the analysis of the sample topography as well [3]. The sample topography has been also studied with Veeco di CP-II AFM.

### 2.7.1 Alignment accuracy

Micro-vernier scales for subsequent alignment inspection have also been designed at the center of the work field and fabricated together with the mark (metallic part of the vernier) and the photonic crystal (etch part of the vernier) as shown in Figure 2.16. Overlay measurements demonstrate that the deviation in the alignment error could be as small as 10nm. Figure 2.16 displays both the middle and end regions of the overlapped vernier which confirm the alignment accuracy to be sub-10nm (resolution of the vernier) in the x direction and 10nm in the y direction. The auto-alignment resolution is confirmed by another group of vernier structures (Figure 2.17)

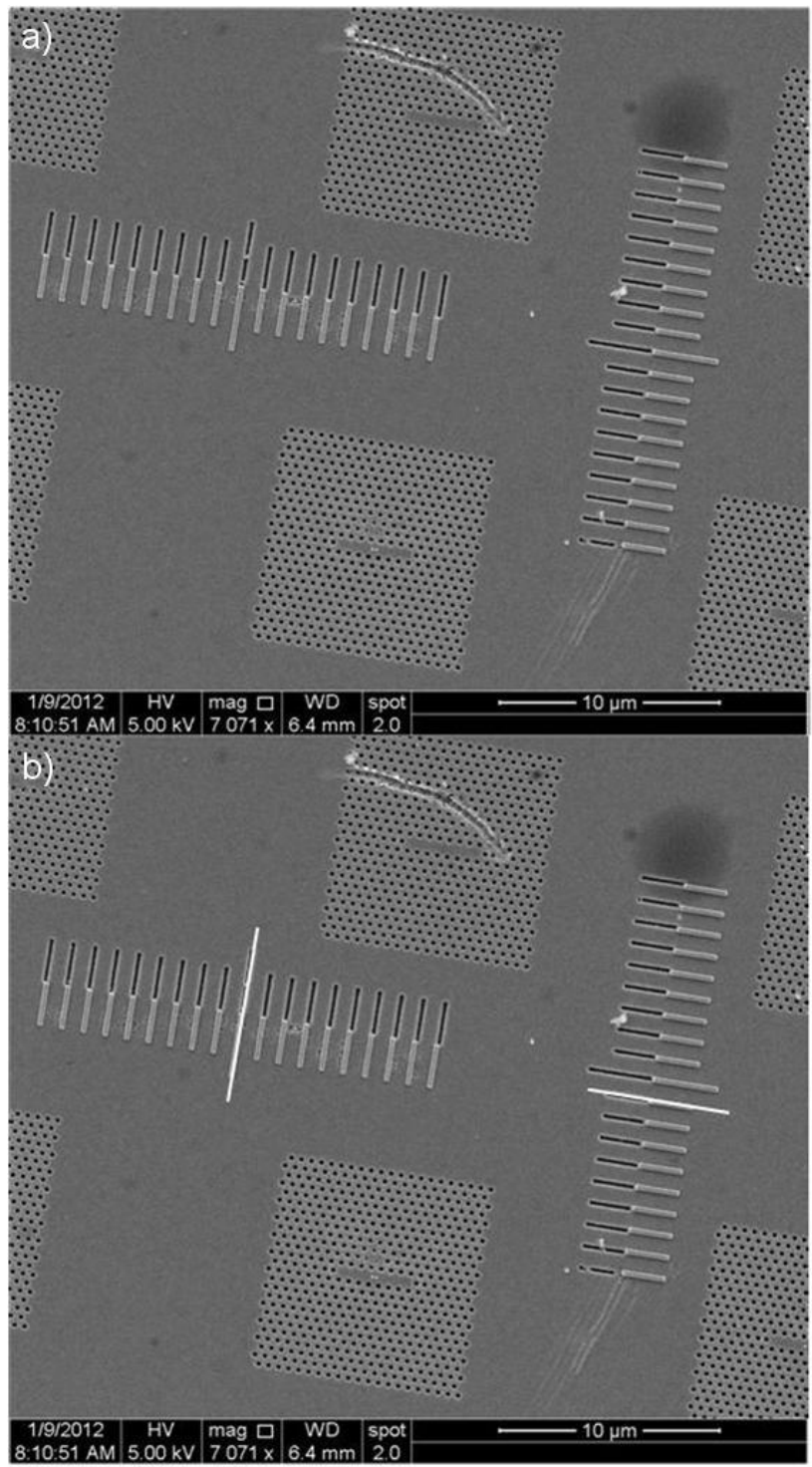

*Figure 2.16 Measurement of auto-alignment of a group of hybrid devices. a) SEM image of the vernier pattern; b) measurement result of the mismatch, the white line is used to label the matched rods.*

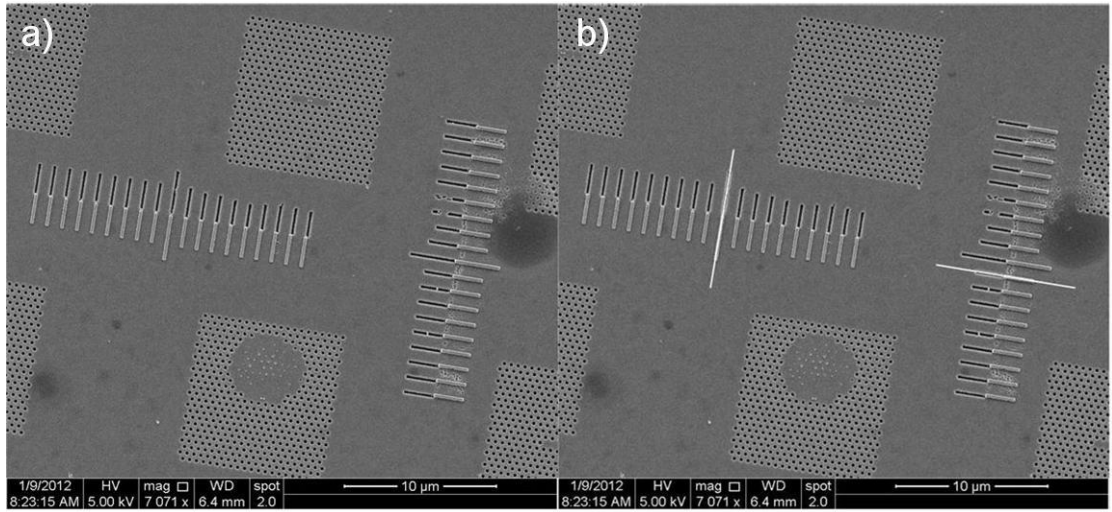

*Figure 2.17 Measurement of auto-alignment for another group of hybrid devices. a) SEM image of the vernier pattern; b) measurement result of the mismatch, the white line is used to label the matched rods.*

The alignment between the PC structures and NAs is also very accurate as can be seen in Figure 2.18. Both images show high quality CL5 and CL7 PC fabrication. The position, shape, and direction of the NAs are accurate, indicating that the locations of the NAs match the design very well.

The locations of NAs on graphite PC structures also match the design very well (Figure 2.19). The NAs are located accurately at 4 different positions to overlap the monopolar mode of the PC. Additionally, the quality of PC structure is good also too. The holes are very round with smooth edges, indicating that the beam astigmatism correction was excellent.

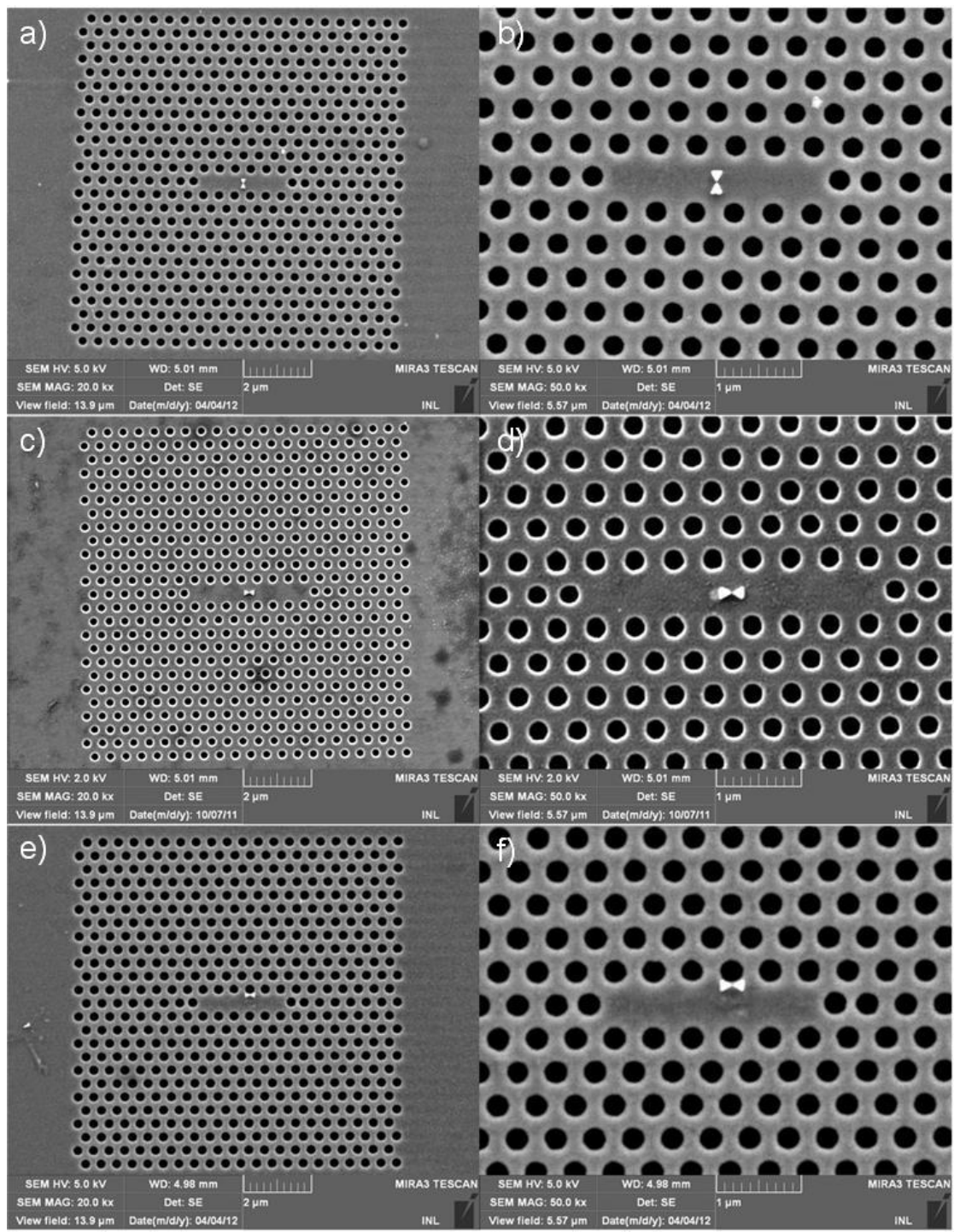

*Figure 2.18 SEM images of fully fabricated CL5 or CL7 plasmonic-photonic devices: a), c) and e) are overview of the hybrid devices; b), d) and f) are zoom in the cavities.*

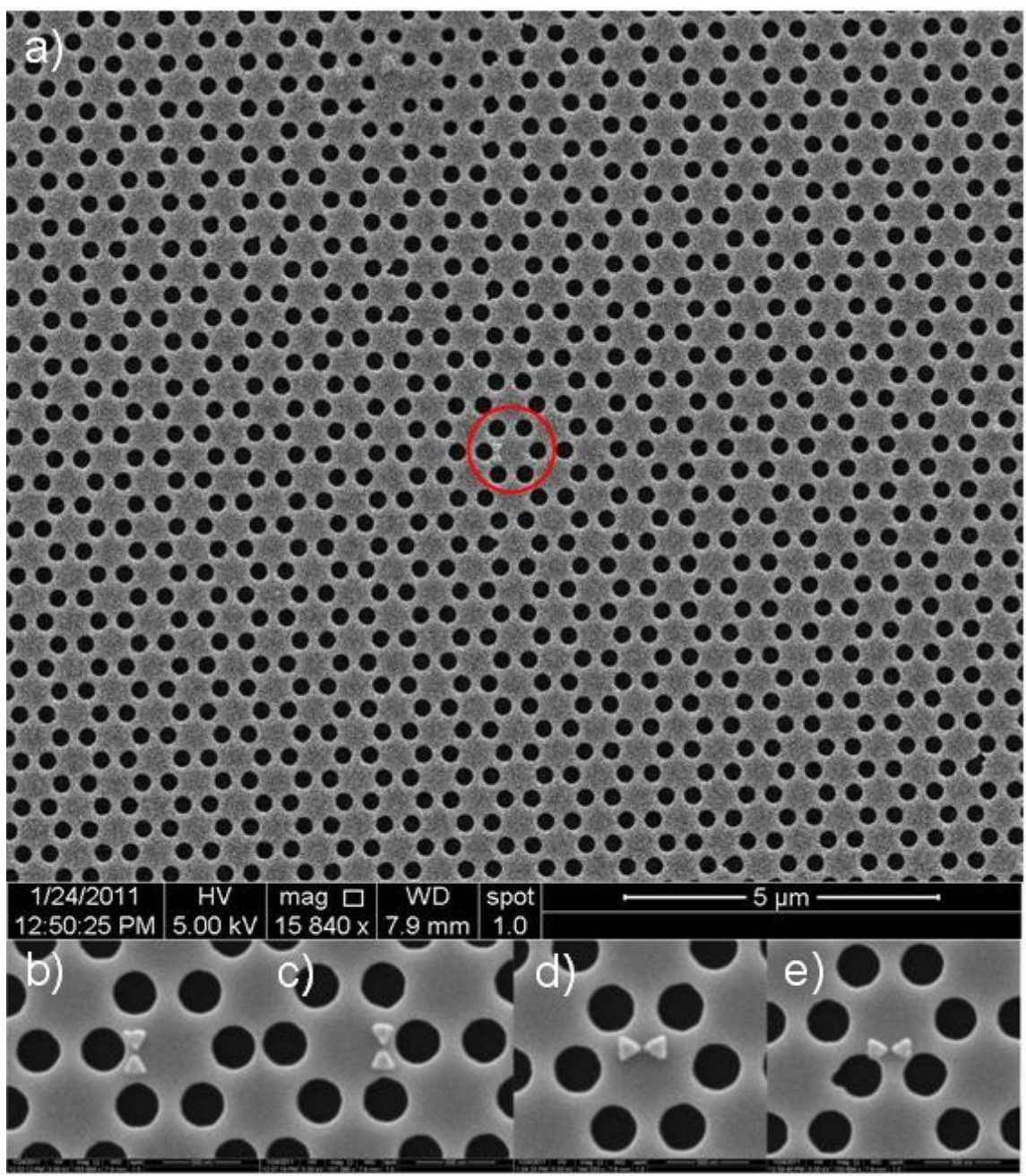

*Figure 2.19 SEM images of graphite PC hybrid devices: a) overview of the hybrid devices and the red circle marks the location of the NA; b), c), d) and e) are images of 4 different positions of NAs.*

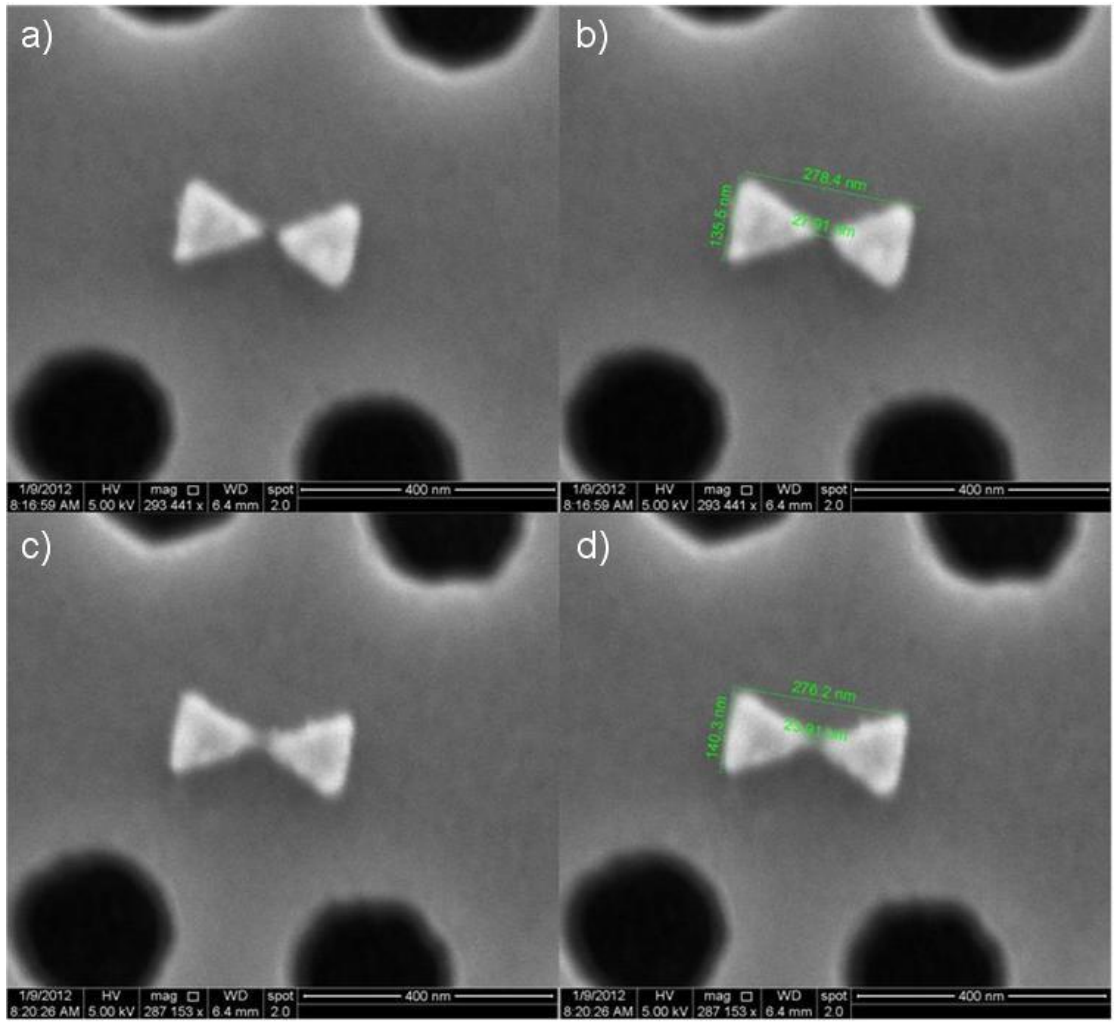

*Figure 2.20 Details of NAs: a) and c) are SEM images of two NAs; b) and d) are measurement results corresponding to a) and c).*

The bowtie NAs were observed in detail (Figure 2.20). The triangle base is estimated to be in the in range of 135nm to 140nm while the length of the NAs spans from 270nm to 280nm. The gap sizes are in range of 20 to 30nm. The geometrical parameters almost match the designed values (base triangle: 140nm; length of NA: 270nm; gap size: 20nm). However, the surface of the NAs seems not smooth, which is common for thin gold layer deposition.

**2.7.2 The effects of SNOM tip**

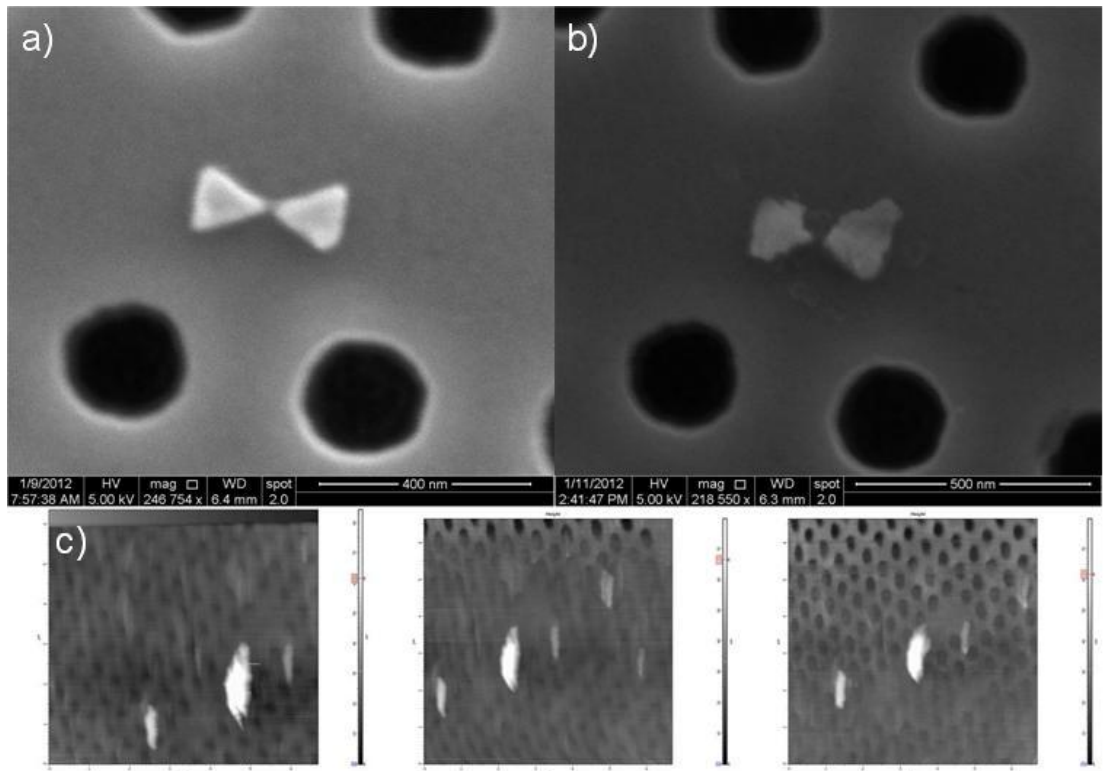

*Figure 2.21 Effect of the fiber tip during a SNOM scan: a) SEM image before the scan; b) SEM images after the scan; c) topography images recorded during 3 consecutive scan.*

SNOM is used to investigate the near-field optical properties of the structures. In the SNOM system, a near-field probe equipped with the shear force head scans the surface of the sample, the distance between the surface and the probe is controlled by a shear-force feedback system [9]. However, the NAs cannot support the initial value (set point) of shear force as much as former researches [3]. In a test, one hybrid structure was scanned by a SNOM dielectric tip in near-field. The set point was as same as former researches. The structure was not pumping by laser source during the test. Hence the affect of pump laser was excluded. After 3 times of scan, the structure was observed by SEM again (Figure 2.21 b)). As shown in Figure 2.21, the shape of the NA is changed a lot. The triangles are deformed and the gap size is larger than before. In some other devices, the triangles are moved to other positions by the shear force (Figure 2.22).

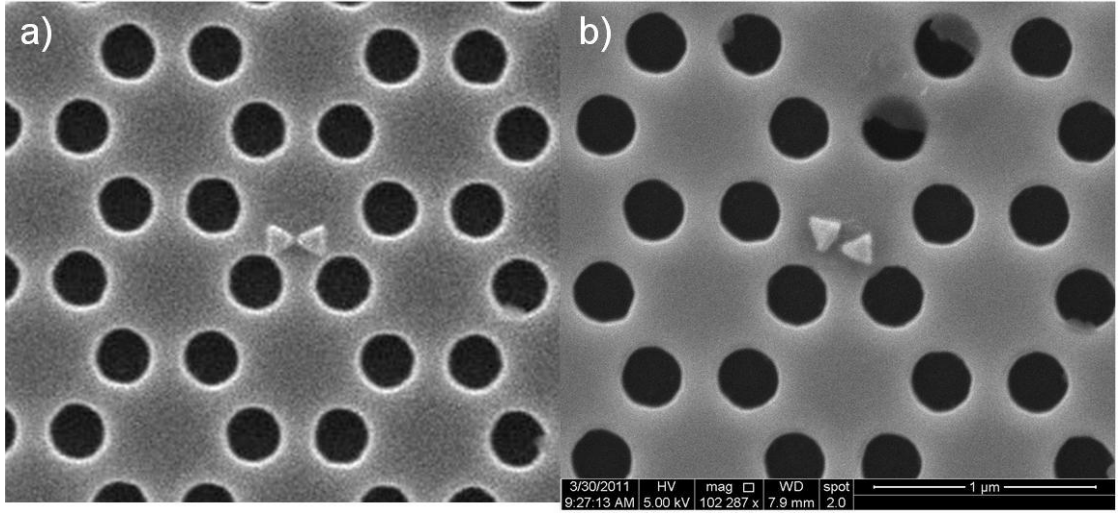

*Figure 2.22 A NA moved by SNOM probe: a) SEM image before SNOM scan; b) SEM image after SNOM scan.*

The AFM measurement shows similar result: the NAs cannot support a very high atomic force from the AFM probe. Two hybrid structures were characterized by AFM and the results are shown in Figure 2.23 and Figure 2.24. The PC cavities and NAs can be clearly seen in an overview image (Figure 2.23). And the surface of the sample is not smooth, there are some little projections be observed not far from the hybrid structure. In the AFM images of localized characterization on NAs (Figure 2.24 b) and e)), the triangles are not very clear. After the characterization, the structures were observed by SEM again. Compare with the SEM images of the two NAs which were observed before AFM, for one structure, the NA is not be deformed a lot (Figure 2.24 c)), while for the other, the shapes of the triangles and the gap size change a lot (Figure 2.24 f)). Since the AFM is depending on the atomic-force between the sample surface and the AFM probe, it means not every NA can support a very high shear-force of the SNOM, or atomic-force of the AFM.

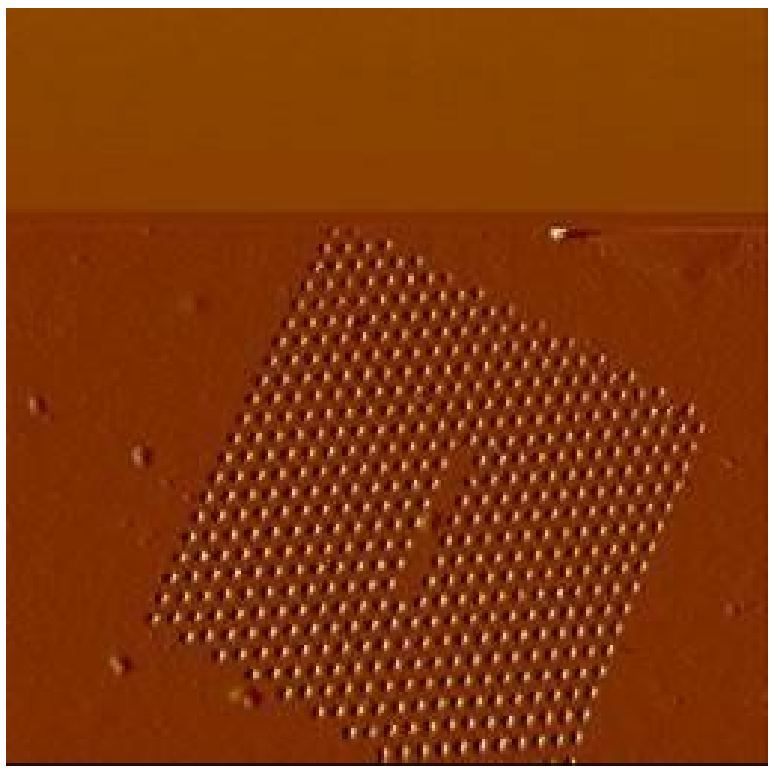

*Figure 2.23 AFM image of a hybrid device.*

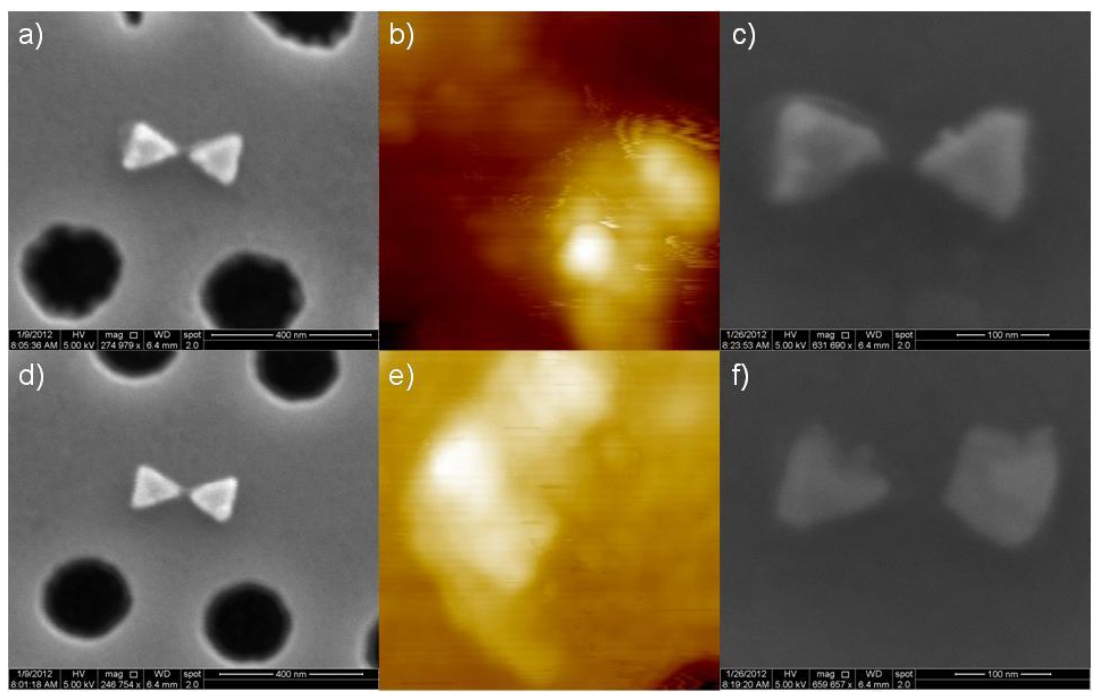

*Figure 2.24 Measurement process of the AFM for two NAs: a) and d) SEM image before the AFM measurement; b) and e) AFM images of the two NAs; c) and f) SEM images after the AFM.*

The reason of these events is maybe the Au film is soft and the metal layer does not adhere to the substrate very well [10]. Therefore, to avoid destroy the NAs, during SNOM measurement, the initial value of the shear-force feedback between the probe and surface of the sample should be set lower than before. In this way, the shear-force exerted on the NAs will be smaller than before and the NAs can support it. The set-up of the initial value depends on the fabrication quality of the samples.

**Conclusion**

To demonstrate the fundamental concepts of plasmonic-photonic hybrid structure, a complex e-beam lithography multilayer process has been developed. Each of the above steps was optimized over a two-year period in a basically serial manner of processing samples. Overlay measurements revealed the deviation in the alignment error could be as small as 10 nm. This type of hybrid structure can exhibit both high quality factors and pronounced hot spot of the electromagnetic field, potentially enhancing the interaction of the cavity mode with emitters or other types of active materials. This novel system may open the route to applications in integrated optoplasmonic devices for quantum information processing, as efficient single photon sources or nanolasers, or as sensing elements for bio-chemical species.

**Bibliography**

[1] M. Hopkinson, J.P.R. David, P.A. Claxton and P. Kightley. Growth of strained InAs/InP quantum wells by molecular beam epitaxy, Appl. Phys. Lett., 60 : 841-843 (1992).

[2] T. Zhang, A. Belarouci, S. Callard, P.R. Romeo, X. Letartre and P. Viktorovitch. *Plasmonic-photonic hybrid nanodevice,* International Journal of Nanoscience, 11(4) : 1240019 (2012).

[3] T-P. Vo. *Optical Near-Field Characterization of Slow-Bloch Mode Based Photonic Crystal Devices,* doctoral dissertation, Ecole Centrale de Lyon (2011).

[4] G. Le Gac. *Etude de l'impact d'une pointe SNOM sur les propriétés des modes optiques d'une cavité à base de cristaux photoniques,* doctoral dissertation, Ecole Centrale de Lyon (2009).

[5] K.E. Docherty. Improvements to the Alignment Porcess in Electron-Beam Lithography, doctoral dissertation, University of Glasgow (2009).

[6] W. Hu. Ultrahigh Resolution Electron Beam Lithography for Molecular Electronics, doctoral dissertation, University of Notre Dame (2004).

[7] *Software Manual for Elphy Plus System Version 3.00,* (Raith GmbH).

[8] *ELPHY Plus Software Reference Manual Version 5.0,* (Raith GmbH).

[9] *Instruction Manual Smena Shear-Force,* (NT-MDT).

[10] *Thin Film Evaporation Guide,* (Vacuum Engineering & Marteials Co., Inc.)

# Chapter 3: Optical Characterization in Far-Field for CL5 and CL7 Hybrid Structures

## Introduction

In this chapter, we present the results of far-field optical characterization of the hybrid nanodevices. Two kinds of investigation have been conducted on the structures: far-field characterization, which enables analyses of the global optical properties of structure; and near-field optical characterization, which yield local information on the light intensity. These investigations have been performed with two different set-ups. The far-field set-up is particularly efficient to obtain information on the lasing properties of the structures, including the spectral characteristics, the lasing threshold and the lasing curve. The near-field set-up allows getting complementary information concerning the spatial distribution of the lasing mode. As it is not diffraction limited, it has been used to study directly the effect of the NA on the PC mode. After a brief description of the labels used to identify the different structures, the optical far-field characterization experiments will be described and the results presented. The near-field optical characterization will be presented in next chapter.

## 3.1 CL5 and CL7 hybrid devices presentation

To indentify the different structures, we define here a specific nomenclature for each sample. Three set of samples have been fabricated with the same fabrication parameters. Apart structural fluctuation due to the fabrication process, these three samples should have the same geometrical parameters (period a, radius of the hole r or filling factor, NA dimension): they will be referred as sample S1, S2 and S3. S1 and S2 have 144 structures for each. S3 has 36 structures since it has only one kind of hybrid structure.

The different structures consist of two types of linear cavities, CL5 (where 5 holes have been removed from the crystal) and CL7 (where 7 holes have been removed). For each cavity type, 3 different sizes of holes are used: 90nm, 100nm and 110nm. The holes radiuses directly influence the filling factor of the PC and therefore the emission wavelength of the cavities. We use letter c, d and e to differentiate the 3 groups (Table 3.1).

| Letter | c | d | e |
|---|---|---|---|
| Radius | 90 nm | 100 nm | 110 nm |

**Table 3.1 Radiuses of PC cavity groups**

For each group, 6 structures are fabricated with different doses of e-beam lithography, ranging from 1.5 to 2.0. The dose influences the real holes radiuses of the PC cavities which can be, at the end of the fabrication slightly different from the nominal radius.

As was described before in Chapter 1 and Chapter 2, for the hybrid structures, NAs are put at different strategic positions in the cavity with different orientations (Figure 3.1). To obtain a coupling with the component Ey of the field (NA perpendicular to the long axis of the cavity), one position is chosen: in the centre of the cavity (Figure 3.1 (a)). To obtain a coupling with the component Ex of the field (NA parallel to the long axis of the cavity), two positions are chosen: in the center and at an edge of the cavity. We use the extension "Hyc", "Hxc" and "Hxs" to represent the different cases. One group of PC cavities without NAs was also fabricated as a reference group. We use "NO" to label them.

Therefore, each structure can be named in form of:

**Sample -Cavity type-Dose number - group-NA position**

For instance, a CL5 hybrid device on S1 with holes radius of 90nm and fabricated with the E-beam dose of 1.5, on which the NA is placed in the centre of the cavity along the short axis is named:

**S1-CL5-1.5c-Hyc.**

**Important**: some of the samples were observed by SEM prior to optical characterizations. It appears that these observations might have contaminated or modified the surface of those samples. These effects will be discussed.

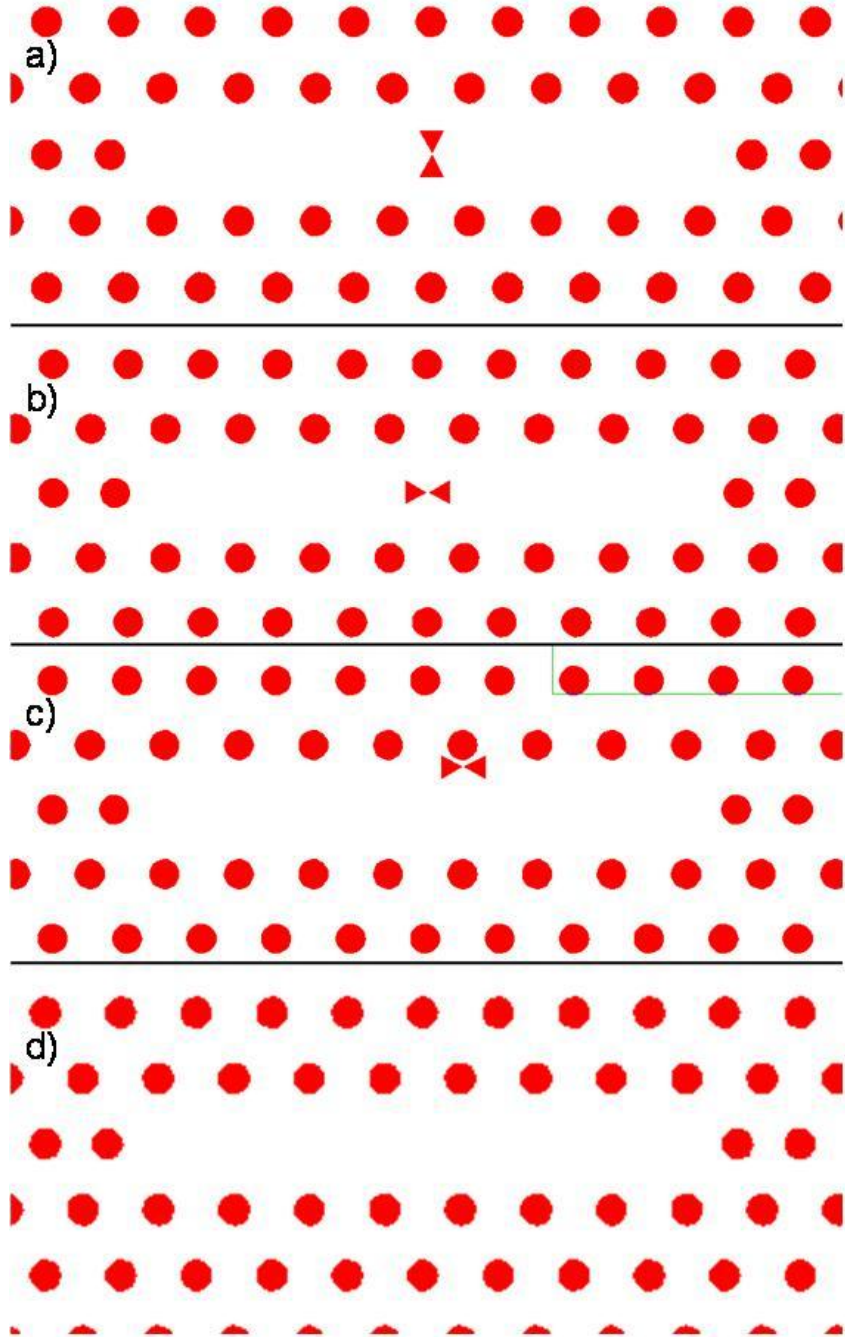

*Figure 3.1 Positions of NA in the CL7 PC cavity: (a) structure "Hyc": NA couples to Ey; (b) structure "Hxc": NA couples to Ex; (c) structure "Hxs": NA couples to Ex; (d) structure "NO": No NA.*

## 3.2 Equipment set-up

Far-field optical set-up that has been developed in INL is dedicated to the measurement of the photoluminescence emitted by a structure. Our far-field set-up enables global analyses of the optical responses of the fabricated structures and is used generally used prior to SNOM measurements to get the spectral characteristics of the structures [1].

The optical characterization is performed by micro-photoluminescence spectroscopy at room temperature (Fig. 3.2). The samples are optically pumped at room temperature, using a pulsed laser diode (LD, Wavelength = 808nm). The pump beam is focused under normal incidence with an achromatic objective lens, onto the structures. The light emitted in free space by the structure is then collected through the same objective lens. Emission spectra are recorded using a spectrometer (0.15nm resolution) and an InGaAs photo detector array after passing an 1100nm-long pass filter. A dichroic mirror, placed in the optical path enables to partially reflect the emitted light to an infra-red camera for in-line observation [1, 2, 3].

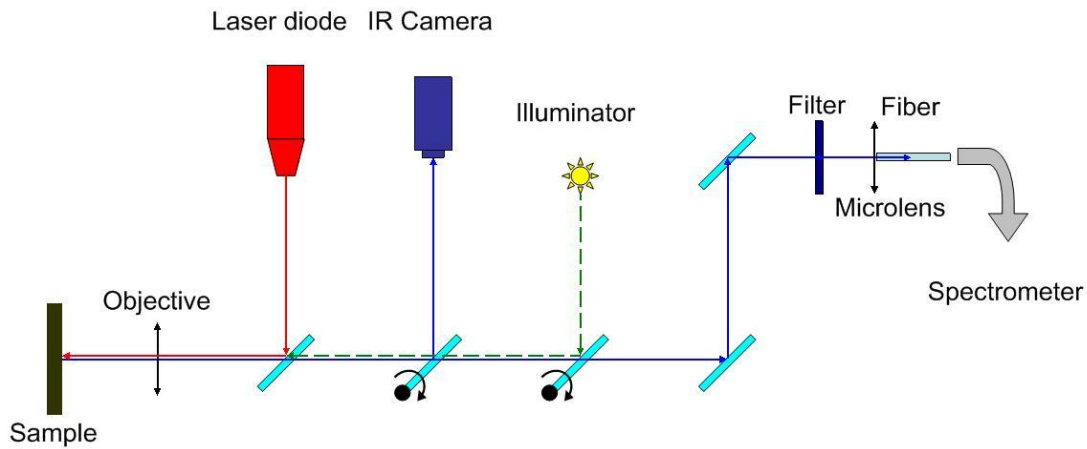

*Figure 3.2 Photoluminescence set-up for Far-field characterization. Pump laser (red line) was focused on the sample by an objective. The reflected light (blue line) was collected by the same objective and then directed to spectrometer through fiber for spectral analysis.*

In order to measure the lasing threshold, it is important to know the incident power on the sample. This latteris controlled by the alimentation current of the pump LD and was calibrated as a function of the laser diode monitoring current. Fig. 3.3 shows that the relationship is linear. In our case, the duty cycle of the pulse LD is about 10% (repetition rate 233 ns, pulse width 22.3 ns), , the collection objective is 20X which yields the following relationship between the incident power of the sample and the monitoring current of the LD $P_{sample}$ (mW) = 0.06518 × $P_{source}$ (μA). This calibration curve allows us to infer directly the effective incident power on the sample from the laser diode monitoring current. The detailed procedure of the calibration is described in the appendix C.

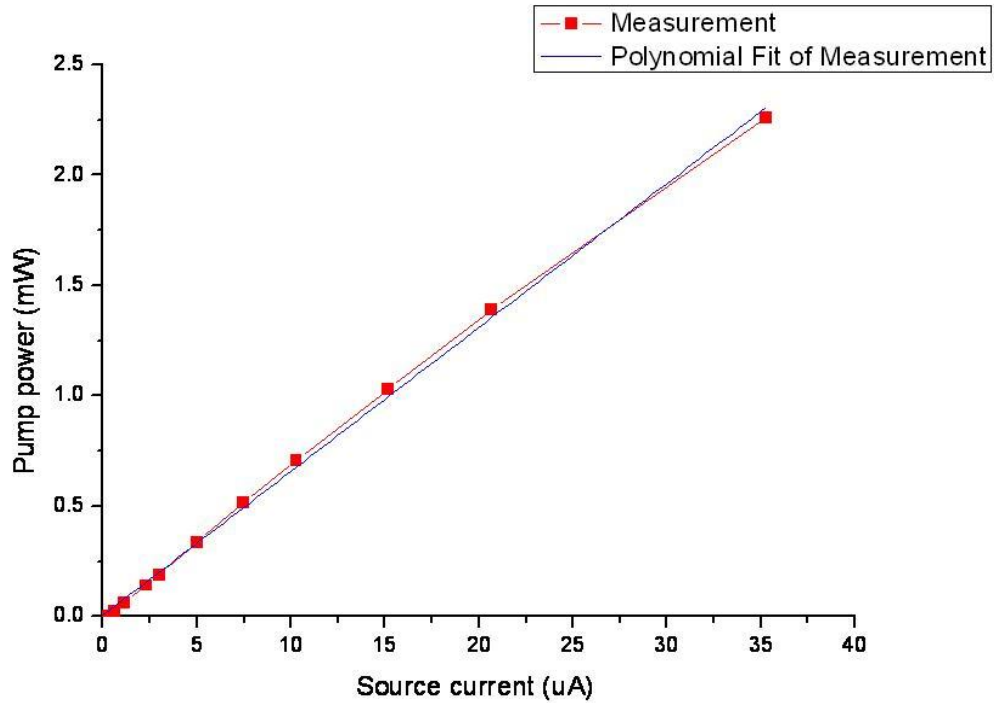

*Figure 3.3 The relationship of the incident power on the sample and the laser diode monitoring current. The red curve is the measurement values and the blue curve is the fitting result via Origin software.*

## 3.3 Threshold and linewidth of the lasing emission measurement

In this section, we investigate the effect of the NA (position and orientation) on the lasing effect, the spectral properties (lasing wavelength) and the lasing threshold of the PC cavity modes.

The far-field optical characterizations of 40 structures were performed. Here we will present the results concerning structures (CL5 and CL7 without NA, hybrid CL5 and hybrid CL7) fabricated in the same operating conditions to compare their optical properties.

### 3.3.1 Threshold of the lasing emission of CL7 and CL5 cavities without NA

The far-field optical characters of 15 structures of pure CL7 and CL5 PC cavities were measured as references for the hybrid structures. Here, the spectrum results of S1-CL7-1.5e-NO and S2-CL5-1.8c-NO are present as representatives.

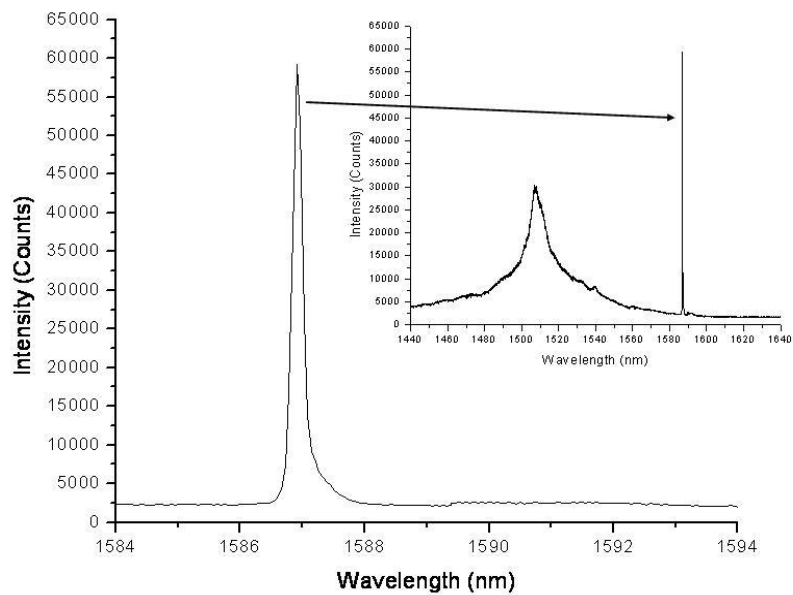

*Figure 3.4 Spectrum of laser emission of structure S1-CL7-1.5e-NO. The whole spectrum is inserting. The lasing wavelength is 1586.91nm.*

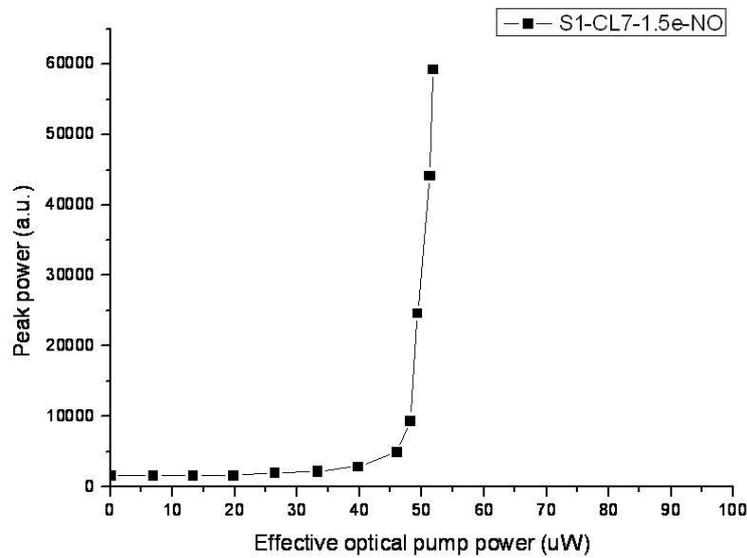

*Figure 3.5 Laser peak intensity plotted against the effective pump power for structure S1-CL7-1.5e-NO. The laser threshold is 45µW.*

**a. CL7 cavity**

Figure 3.4 shows the spectrum of laser line of S1-CL7-1.5e-NO just above the threshold. For this sample, the far-field measurement was taken before the SEM observation. The lasing wavelength is 1586.91nm. The whole spectrum is shown inset. The broad shorter wavelength peak at 1507.08nm corresponds to emission from the quantum wells emission with the modes of greater energy of the CL7 (which does not present a lasing effect). Figure 3.5 shows the variation of the laser peak intensity versus the effective incident pump power. The laser threshold of S1-CL7-1.5e-NO is 45µW.

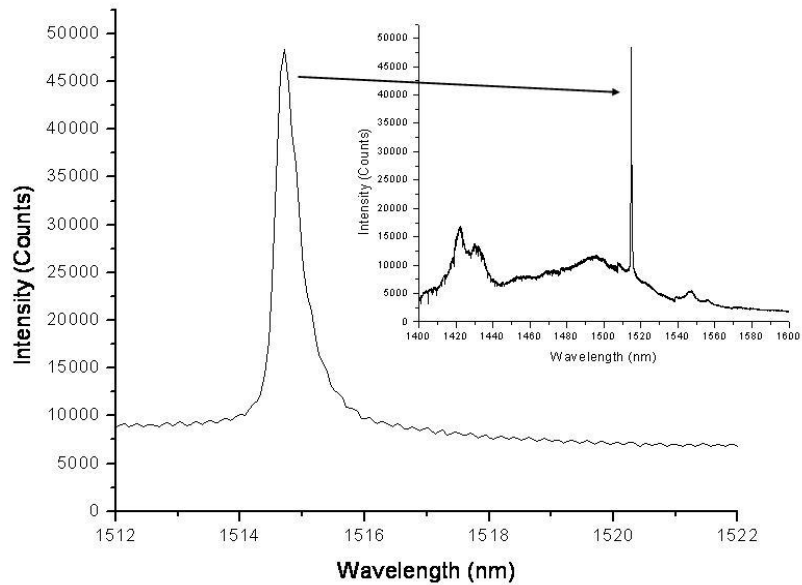

*Figure 3.6 Spectrum of laser emission of structure S2-CL5-1.8e-NO. The whole spectrum is inserting. The lasing wavelength is 1514.71nm.*

**b. CL5 cavity**

For CL5 PC cavities, Figure 3.6 shows the spectrum of laser line of S2-CL5-1.8e-NO just above the threshold. The lasing wavelength is 1514.71nm. The whole spectrum is shown inset. The spectrum of the QW emission by the CL5 is also visible: the other peaks (1495 nm, 1422.2nm) correspond to the modes of greater energy that do not present lasing effect. Figure 3.7 shows the variation of the laser peak intensity versus the effective incident pump power. The laser threshold of S2-CL5-1.8e-NO is 20µW.

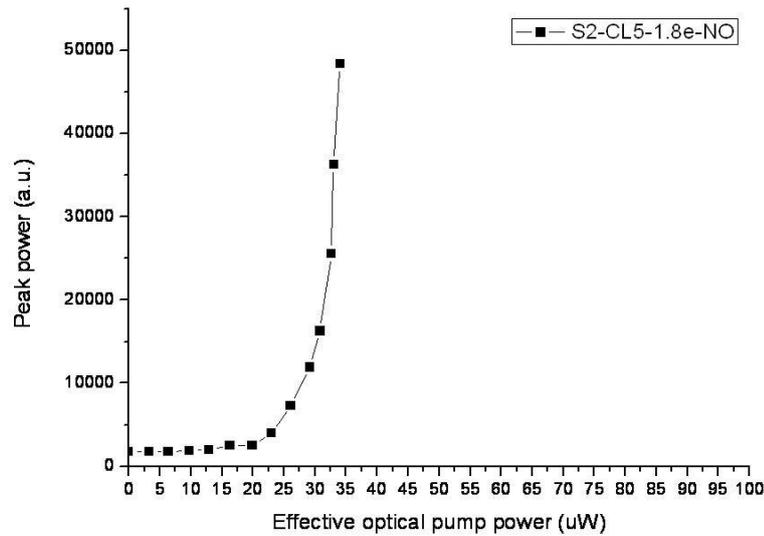

*Figure 3.7 Laser peak intensity plotted against the effective pump power for structure S2-CL5-1.8e-NO. The laser threshold is 20μW.*

## 3.3.2 Threshold of the lasing emission CL7 and CL5 cavities with NA

For the hybrid structures, the spectrum results of S1-CL7-1.5e structures and S2-CL5-1.8c structures are present as representatives. Since their fabrication parameters of the PC cavities are same as the pure PC cavities which are shown before. Hence the variation of the spectrum result is due to the NA. Therefore, we can evaluate the impact of the NA to the spectrum results of the structures.

### 3.3.2.1 S1-CL7-1.5e structures

The presence of NA does not kill the laser mode of the PC cavity. Figure 3.8 shows the spectrum of S1-CL7-1.5e-Hyc just above the threshold. The far-field measurement was taken before the SEM observation. The lasing wavelength is 1590.81nm. The whole spectrum is shown inset. The peaks of higher energy correspond to the higher modes of the CL7.

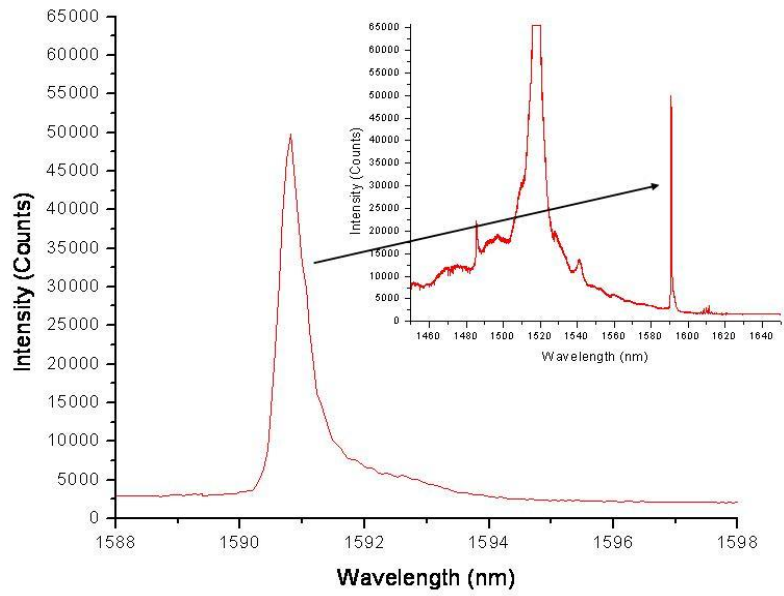

*Figure 3.8 Spectrum of laser emission of structure S1-CL7-1.5e-NO. The whole spectrum is inserting. The lasing wavelength is 1590.81nm.*

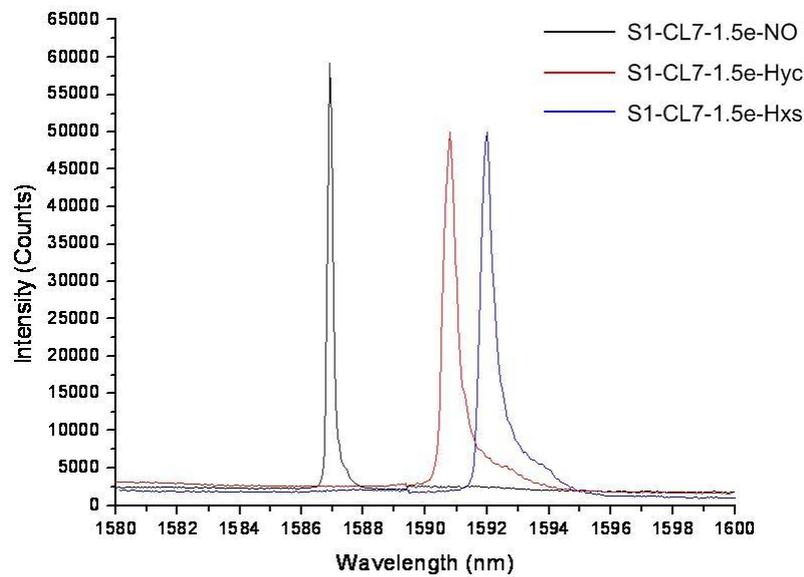

*Fig. 3.9 Spectrum of laser emission of S1-CL7-1.5e structures. The laser wavelength of S1-CL7-1.5e-NO is 1586.91nm, the laser wavelength of S1-CL7-1.5e-Hyc is 1590.81, the laser wavelength of S1-CL7-1.5e-hxs is 1591.92nm.*

The laser spectra of the structures of hybrid structures are shown in Figure 3.9. They are shown together with the spectrum of pure PC cavity without NA. Spectra show that the lasing wavelength seems to be red-shifted in the case of structure with NA. The red-shift is respectively 3.9 nm and 5 nm for S1-CL7-1.5e-Hyc and S1-CL7-1.5e-Hxs. The presence of a red-shift was not predicted by the theory. For these structures, the spectra measurements have been performed on 23 structures. Depending on the structures, both red-shifts and blue-shifts were observed. The blue-shift is expected in theory, the reason of the presence of a red-shift is not fully understand: another effect may be in play for this structure. This issue will be discussed later in section 3.3.3.

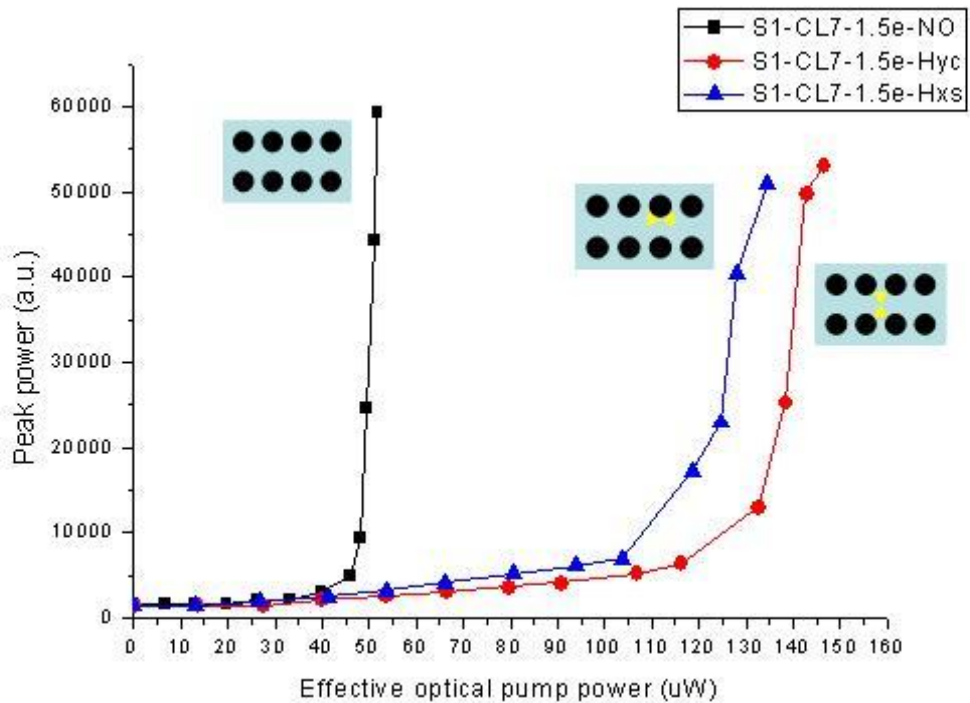

*Figure 3.10 Threshold of the S1-CL7-1.5e structures. The threshold of S1-CL7-1.5e-NO is 45µW, the threshold of S1-CL7-1.5e-Hyc is 130µW, the threshold of S1-CL7-1.5e-NO is 100µW.*

The variation of the laser peak intensity versus the effective incident pump power was also measured to evaluate the effect of the NA on the laser threshold. The results are shown in Figure 3.10. It shows that the presence of the NA has an impact on the laser threshold. For the hybrid structures, the laser thresholds are 130μW and 100μW for S1-CL7-1.5e-Hyc and S1-CL7-1.5e-Hxs respectively, compared to 45μW for the structure wihout NA. This results, that have been observed for different CL7 structures, show that the laser thresholds of hybrid structures are higher than the pure PC cavity without NA, indicating the influence of the latter on the optical losses (quality factor) of the lasing mode. More specifically this effect can be explained in two ways:

- A Q factor decrease can be due to additional diffraction losses induced by the presence of the NA. In that case, the NA can be seen by the PC mode as a diffusion center.

- If a coupling occurs between the PC and NA modes, the resulting hybrid mode is a linear combination of both modes. Therefore its Q factor lies in between the values of the NA and PC modes (see chapter 1). As QNA<QPC, then Qhybrid <QPC.

Discriminating between these two explanations cannot be inferred from far field measurements and near field experiments are needed to get more information on the spatial properties of the lasing mode.

Note that in the case of this set of structures, the lasing threshold is higher for the Hyx structures than for the Hxs structures but this behavior is not recurrent: for other structures, the Hxs threshold is higher than he Hyc.

### 3.3.2.2 S2-CL5-1.8e structures

As for CL7 structures, the Figure 3.11 shows the spectrum of laser line just above the threshold of structure S2-CL5-1.8e-Hyc. The lasing wavelength is 1505.47nm. The whole spectrum is shown in inset.

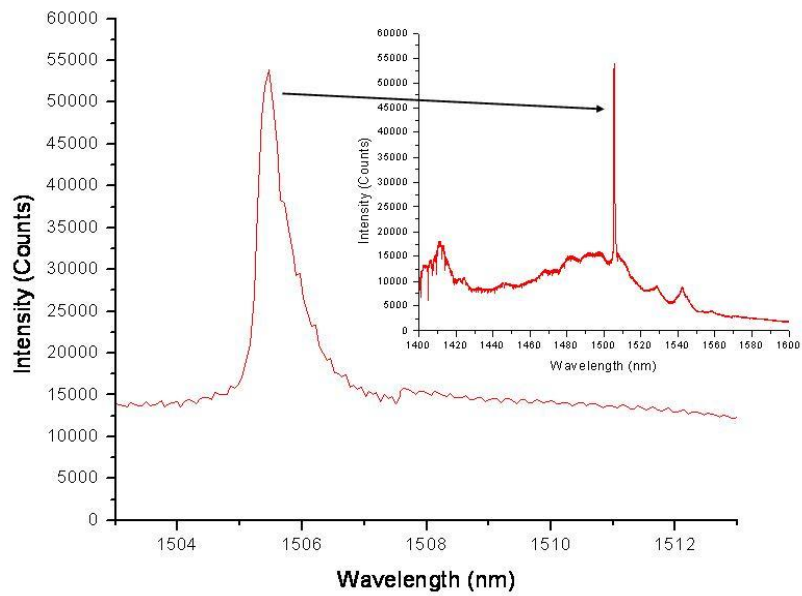

*Figure 3.11 Spectrum of laser emission of structure S2-CL5-1.8e-Hyc. The whole spectrum is in insert. The lasing wavelength is 1505.47nm.*

The laser spectra of the structures S2-CL5-1.8e-Hyc and S2-CL5-1.8e-Hxs are shown in Figure 3.12. They are shown together with the spectrum of S2-CL5-1.8e-NO. The laser lines are also just above the threshold. The lasing wavelength are 1514.71nm (S2-CL5-1.8e-NO), 1505.47nm (S2-CL5-1.8e-Hyc) and 1512.52nm (S2-CL5-1.8e -Hxs). It shows that for this group of structures, the presence of the NA creates a blue shift of the lasing wavelength, in agreement with theoretical simulations (see. Chapter 1).

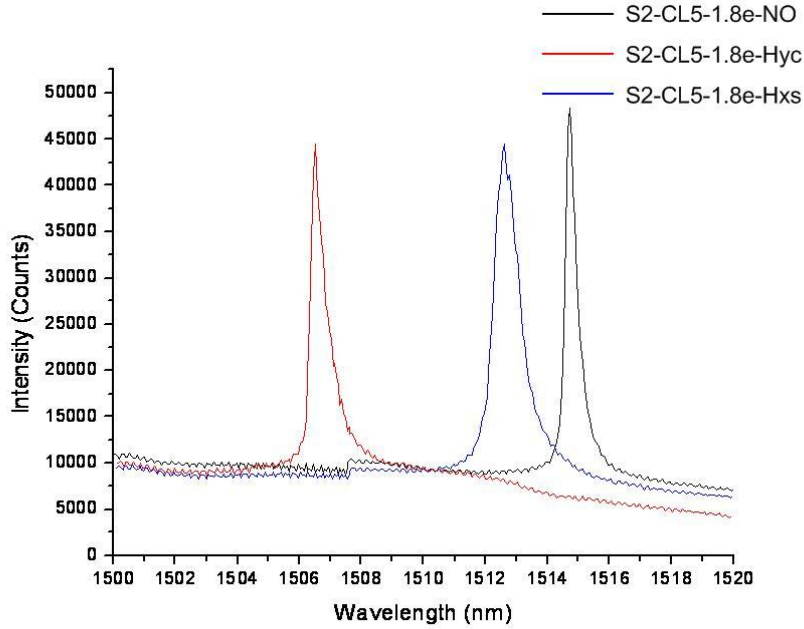

*Figure 3.12 Spectrum of laser emission of S2-CL5-1.8e structures. The laser wavelength of S2-CL5-1.8e-NO is 1514.71nm, the laser wavelength of S1-CL7-1.5e-Hyc is 1505.47, the laser wavelength of S1-CL7-1.5e-hxs is 1512.52nm.*

The variation of the laser peak intensity versus the effective incident pump power of the group of S2-CL5-1.8e structures, including pure PC cavity and hybrid structures are shown in Figure 3.13. The measurement condition is same as the group of S1-CL7-1.5e structures. Note that conversely to the previous sample, this sample (S2) was observed by MEB Tescan before far-field measurement. Figure 3.13 shows that the variations of the laser threshold of S2-CL5-1.8e structures are similar to those of the CL7 structures. The thresholds of PC cavities with NA are higher than without NA. The laser threshold of the S2-CL5-1.8e-NO is about 20µW. For the hybrid structures, the laser threshold is about 27µW for S2-CL5-1.8e-Hyc, 22µW for S2-CL5-1.8e-Hxs. The threshold values are much smaller than those of CL7 structures. It is probably due to a better quality of the fabrication of sample S2.

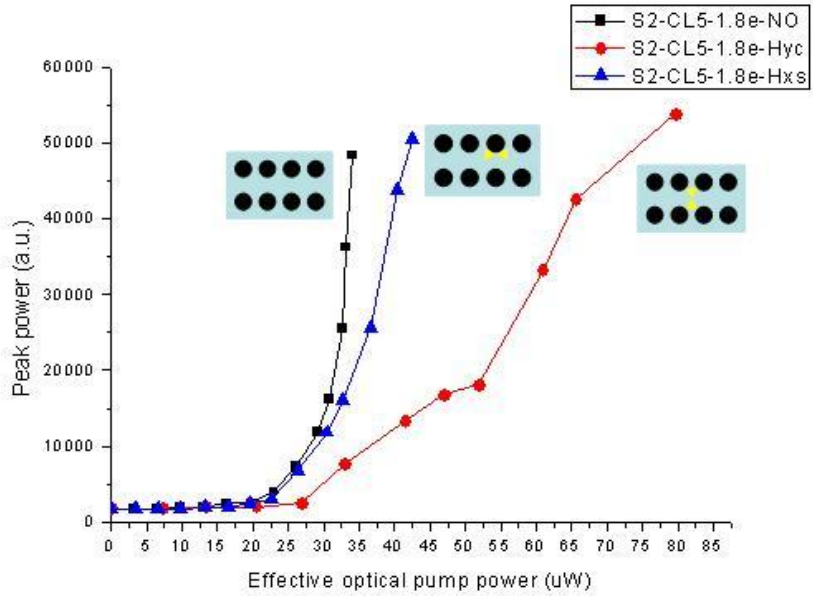

*Figure 3.13 Threshold of the S2-CL5-1.8e structures. The threshold of S2-CL5-1.8e-NO is 20µW, the threshold of S2-CL5-1.8e-Hyc is 27µW, the threshold of S2-CL5-1.8e-Hxs is 22µW.*

### 3.3.3 Influence of the pumping power on the lasing wavelength

In the previous section, we observed that the presence of the NA in the CL7 structures produced a red-shift, which is not in agreement with the theoretical predictions. Conversely, an expected blue-shift was observed for the CL5. To explain these facts, we investigate another cause of wavelength variation, which is the pumping power : the mode wavelength of an active structure depends also of the pumping parameters. For photonic crystal cavities, without NA, the behaviour of the wavelength mode with the pumping power has been studied previously in former PhD thesis (Gaelle Legac, Lydie Ferrier) [2, 4]. A blue shift is observed at low pumping power, before the pumping threshold. This blue-shift is attributed to the modification of the optical index by carriers injection due to pumping [5]. After the threshold, the carrier population is clamped by the laser feed-back mechanism. If the pumping increases after the threshold, others effects have to be taken into account, in particular thermal effects as structure heating. This effect modifies the optical index as well and produces a red shift (PhD Gaelle Legac) [2]. To evaluate quantitatively these effects on the present structures, with and without NA, we studied the evolution of the mode wavelength with the pumping power for CL5 and CL7 structures.

### 3.3.3.1 S1-CL7-1.5e structures

First, we studied the evolution of the wavelength of the lasing mode of the S1-CL7-1.5e structures (with and without NA) presented in the previous section. During the measurement, the pump power is increased slowly, from 0 to about 0.45mW. When the pump power is below the threshold (0.045mW), only spontaneous emission occurs. Above the threshold, lasing occurs. The emission wavelengths measured for different values of pump power are shown in Figure 3.14. As expected, it shows that below the threshold, the emission wavelength undergoes a blue shift of with the increasing pump power: the blue-shift occurs for all the structure, with and without NA and is about 3 - 4 nm. When the threshold is reached and after, the emission wavelength presents a moderate red shift with the increase of the pump (2.3 nm/mW). In Figure 3.14, we see clearly that whatever the pumping power, there is always a clear red-shift between the structure without NA and with NA. We also seethat the three structures of the group show the same variations, and particularly, the identical variation rate for the red-shift (2.3 nm/mW). It indicates that this behavior is not affected by the presence or not of NA and can no explain the observed red-shift (it could have been the case if the slope value was changed by the presence of the NA). However, we also see that the value of the shift depend on the pumping powe. The same measurements have been performed on other CL7 structures and it shows the same behavior (Figure 3.15).The case of S1-CL-1.7e is interesting because, here we see that depending on the pumping conditions, the shift may be different. These measurements show that to compare the wavelengths of different structure, it is important to be clearly above the threshold.

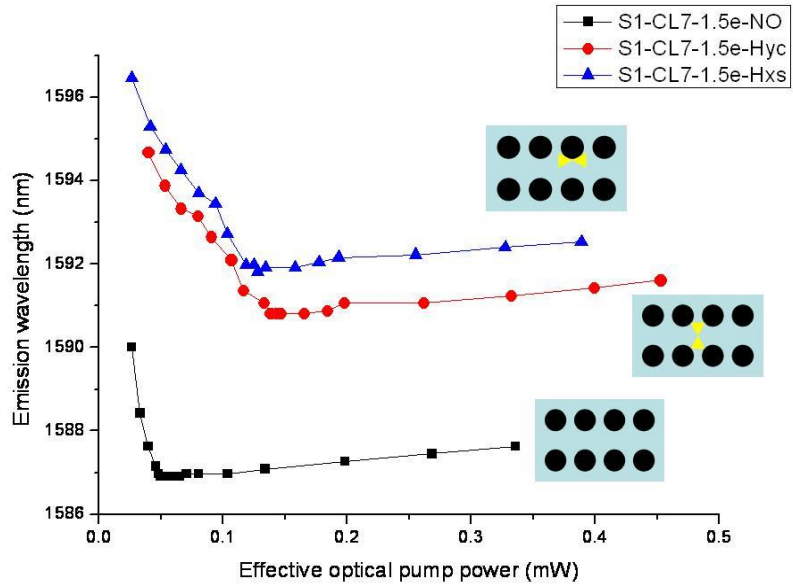

*Figure 3.14 Wavelength variations of S1-CL7-1.5e structures along the variations of pump power.*

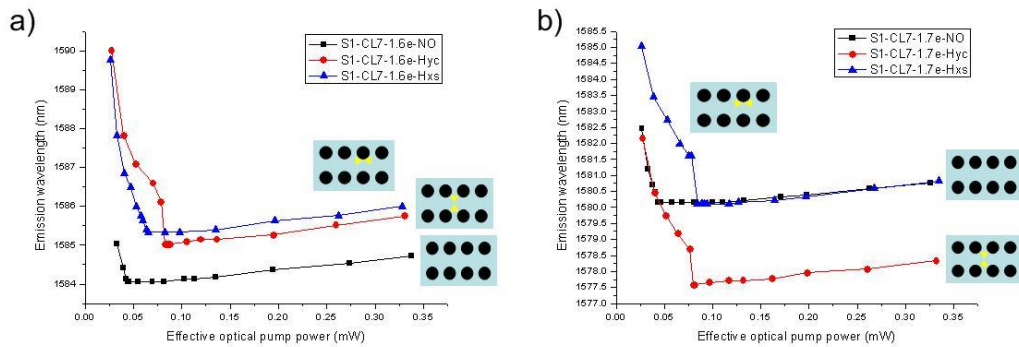

*Figure 3.15 Wavelength variations of S1-CL7-1.6e (a)) and S1-CL7-1.7e (b)) structures along the variations of pump power.*

### 3.3.3.2 S2-CL5-1.8c structures

The same study has been performed on other CL5 structures. However, the experimental conditions were a bit different because the CL5 structures were observed with SEM (MEB Tescan) before measurements.

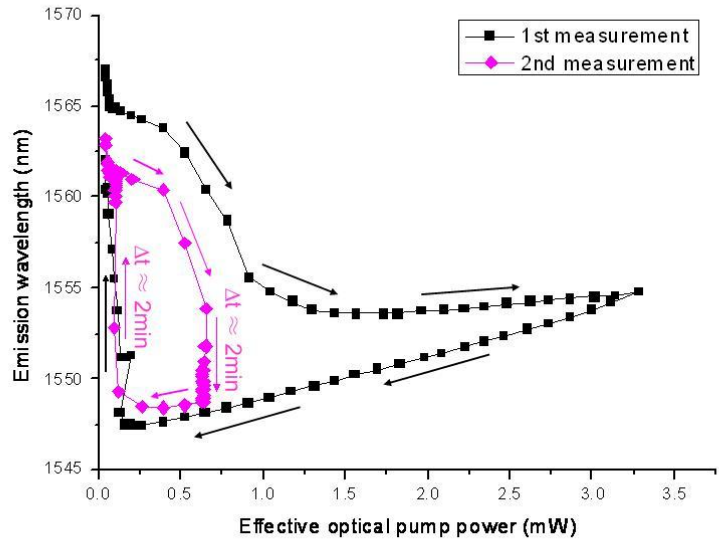

*Figure 3.16 Wavelength variations of S2-CL5-1.8c-NO along the variations of pump power.*

The evolution of the lasing wavelength has been investigated on two CL5 (S2-CL5-1.8c-NO, S2-CL5-1.8c-Hxs) by performing a cycle up-and-down for the pumping power.

For structure S2-CL5-1.8c-NO, the measurements have been performed two times. Results are shown in the Figure 3.16. In the first measurement, the pump power is increased slowly from 0 to 3.26mW, and then decreased slowly from the maximum to 0. The wavelengths measured during this process are shown in Figure 3.16. It shows that below the threshold (0.013mW), the wavelength blue-shifts from 1566.98 nm to 1565 nm. However, above the threshold, the wavelength keeps on blue-shifting from 1565 nm to 1552 nm (at 1.5 mW). Above 1.5mW, the wavelength red shifts moderately (0.75 nm/mW). The pump power is then decreased and the emission wavelength blue-shifts again moderately (-2.3 nm/mW) until the pump power decreases to 0.13mW. Below this value, the wavelength increases again with the decreasing pumping power. The final emission wavelength is 1562.03 nm at the end, which is blue-shifted with respect to its initial value of 1566.8 nm.

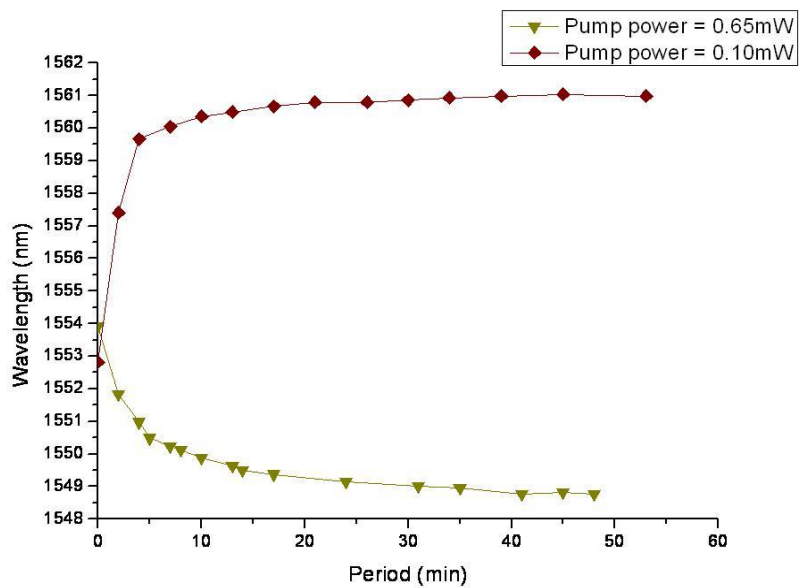

*Figure 3.17 Variation of laser wavelength along period at fixed pump power of: 0.65mW (dark yellow) and 0.10 (wine).*

A second measurement was performed on the same structure two days later. The evolution of the wavelength was investigated during a new cycle. The goal was to investigate the effect of the pumping duration on the wavelength shift. In this measurement, the pump power was increased slowly on a smallest range (0 - 0.65mW). The starting mode wavelength was 1563.21nm. By increasing the pump power, the emission wavelength blue shifts as observed previously. At 0.65mW, the pump power was maintained and several spectra were recorded every 2 minutes. During this time we observed a regular blue shift from 1553.92nm to 1548.77nm. The evolution of the wavelength shift is presented in Figure 3.17. After 40 minutes the emission wavelength kept constant and stable (1548.77nm). To complete the cycle, the pump power was then decreased slowly, until about 0.1mW. In this process, the emission wavelength exhibited a blue-shift and then a red-shift as shown in Figure 3.16. The pump power was kept at about 0.013mW and several spectra were recorded every 2 minutes for 40min. We observed that the emission kept on red-shifting, from 1552.82nm to 1560.98nm, before stabilizing (Figure 3.17). Finally the pump power was decreased to 0: the emission wavelength red-shifted until the emission disappeared at 1562.9nm, very close to its initial value. We only did observe this behavior for structures that have been observed by SEM. These experiments show that the time of exposition to the pumping power is important and that a physical and reversible phenomenon participates to the wavelength shift. This will be discussed later.

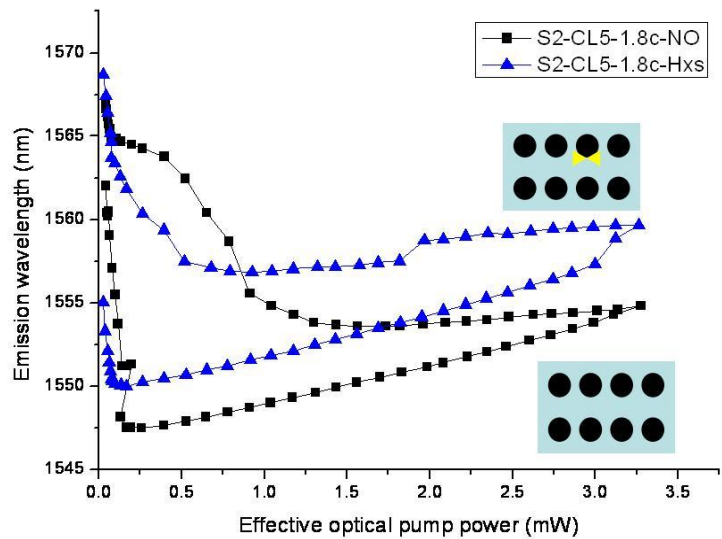

*Figure 3.18 Wavelength variations of S2-CL5-1.8c-NO and S2-CL5-1.8e-Hxs along the variations of pump power.*

The same type of measurements was performed on S2-CL5-1.8c-Hxs. Figure 3.18 shows the results for the structure S2-CL5-1.8c-NO and S2-CL5-1.8c-Hxs during one cycle. It shows that the trends of emission wavelength variation of both structures are similar. It also shows that the hysteresis cycle of this structure with NA is globally red-shifted with respect to the cycle of the structure without NA.

**Discussion:**

For structures that have not been observed by SEM, the influence of the pumping power on the wavelength variation is perfectly classical: it shows that above the threshold, the wavelength is moderately red-shifted with a variation rate of 2.3 nm/mW, which does not depend on the presence or not of the NA. It does not explain why, in certain cases (S1-CL7-1.5e) a red-shift is observed and in other cases (S1-CL7-1.7e) a blue-shift is observed.

For structures that have been observed by SEM prior far-field measurement, the behaviour is different: the reason of the cycle variation of laser wavelength with the pump power is not clear and a cross-disciplinary analysis should be performed to understand this phenomenon. However, a possible explanation is proposed here. Recently, another group shows that the presence of water can impact the refractive index of a PC nanocavity. When water fills the holes of the PC, the resonance wavelength can be red-shifted of more than 10 nm [6]. In our case, after the fabrication process, there may be some organic residue left on the surface of the sample or in the holes of the PC structures. During the long SEM observation by MEB Tescan, some chemical change due to the electric beam scan may have occurred. After the chemical change, the organics would become hydrophilic and adsorb water strongly. When exposed to a pumping laser diode, the water may evaporate: when the pump power is low, the thermal effect is weak and the water remains inducing a red-shift of the laser wavelength of the PC cavity (compared to the cavity without water). With the pump power increasing, the thermal effect became stronger and the water evaporates, inducing a variation of the effective index and an extraordinary blue-shift until the water totally evaporates. Without water, the laser wavelength red shifts normally because of the thermal effect. When the pump is decreased, the laser wavelength blue shifts due to the thermal effect decrease (with the same slope as the CL7 structures). When the thermal effect is weak enough, the water comes back again on the surface and the laser wavelength has an extraordinary red-shift. This effect would be compatible with the time constant that have been observed for this phenomena (about few 10 minutes)

**Conclusion**

In summary, several conclusions can be drawn from the far-field optical characterisations of CL7 and CL5 PC cavity.

- First, the presence of NA increases the threshold of the laser emission. We show that the laser thresholds of hybrid structures are higher than the pure PC cavity without NA, indicating the influence of the latter on the optical losses (quality factor) of the lasing mode. This can be due to additional diffraction losses induced by the presence of the NA or by the coupling effect between the cavity mode and the PC mode.

-Second, the NA also impacts the wavelength of the laser emission. The theory predicts blue-shift but experiments show blue- and red-shifts. The NA does not change the trend of the variation of the laser wavelength along the pump power variation.

**Bibliography**

[1] T-P. Vo. *Optical Near-Field Characterization of Slow-Bloch Mode Based Photonic Crystal Devices, doctoral dissertation,* Ecole Centrale de Lyon (2009).

[2] G. Le Gac. *Etude de l'impact d'une pointe SNOM sur les propriétés des modes optiques d'une cavité à base de cristaux photoniques,* doctoral dissertation, Ecole Centrale de Lyon (2009).

[3] T. Zhang, A. Belarouci, S. Callard, P.R. Romeo, X. Letartre and P. Viktorovitch. *Plasmonic-photonic hybrid nanodevice,* International Journal of Nanoscience, 11(4) : 1240019 (2012).

[4] L. Ferrier. *Micro-nanostructures à base de cristaux photonique pour le contrôle 3D de la lumière,* doctoral dissertation, Ecole Centrale de Lyon (2008).

[5] F. Raineri, C. Cojocaru, R. Raj, P. Monnier, A. Levenson, C. Seassal, X. Letartre and P. Viktorovitch. *Tuning a two-dimensional photonic crystal resonance via optical carrier injection,* Optical Letters, 30(1) : 64-66 (2005)

[6] B. Cluzel, L. Lalouat, P. Velha, E. Picard, E. Hadji, D. Peyrade, F. de Fornel. *Extraordinary tuning of a nanocavity by a near-field probe,* Photonics and Nanostructures - Fundamentals and Applications, 9 : 269-275 (2011)

# Chapter 4: Optical Characterization in Near-Field for CL5 and CL7 Hybrid Structures

## Introduction

In the hybrid structures, the interaction between the PC cavity mode and the NA is expected to produce a radical change of the spatial distribution of the light intensity of the structure mode. Particularly, a strong concentration of the light should be observed in the gap of the NA. To investigate the spatial properties of the mode, we use Scanning Near-field Optical Microscopy (SNOM): this characterization method is adapted to the observation of light localization phenomenon as it not limited by diffraction. However, this technique is not non-invasive as it requires placing an optical nano-probe in the near-field of the sample, which modifies the electromagnetic environment of the structure. However, if this perturbation remains weak, it can be neglected. In this section, we will first describe the experimental set-up that have been used, the results obtained on PC cavities will also be presented as reference for the cavity mode without NA. Finally, the results obtained on hybrid structures will be presented and discussed.

## 4.1 SNOM principle and set-up description

Basically, the electromagnetic field present at the surface of an object can be described in two terms: propagating light and evanescent light. The former concerns the components that can propagate away from surface at a minimum distance comparable to the light wavelength, in a zone labeled far-field. The latter are confined to the surface within a distance that can be well under the wavelength, in a zone labeled near-field. Propagating light can be detected in the far-field by using conventional far-field microscopes; it gives precious information about the spectral properties of the object, about its radiation pattern or the distribution of light at the object surface as far as low spatial frequencies are concerned. However, propagation acts as a low pass filter for spatial frequencies and in far-field, the optical resolution is limited, according to Rayleigh criterion, to l/2. In order to overcome these limitations and access to small light spatial modulation, specific devices that allow unveiling the information in near-field have been developed. Scanning Near-field Optical Microscopy (SNOM) has been developed for this demand. Since its development, SNOM related technologies have been utilized in many research areas and has become an important part in nano-optics research. SNOM operating principle relies on a fundamental idea proposed in 1928 by Synge, an Irish scientist. In order to extend the optical resolution into the nanometer regime, he proposed to use a very small object (<100nm), used as a probe that scatters evanescent light which can turn into propagating light detectable by a sensitive photo detector. In typical SNOM systems, this nano-object is the apex of a tapered optical fiber, which can be used to illuminate the structure, or to collect the optical signal. The latter is recorded point by point, as a function of the probe position above the surface of the structure and the SNOM images are constructed by a computer during the scan. Meanwhile, in order to maintain sub-diffraction spatial resolution, the distance between the tapered SNOM tip and surface must be kept as low as few tenth of wavelength and regulate using force microscopy, as shear-force or atomic force microscopy. Shear force is particularly used in SNOM because it is a short range damping force exerted on a laterally vibrating tapered SNOM tip. The vibration amplitude of the tip is damped due to the shear force interaction between tip and the sample surface. This damping depends of the distance from the tip to surface, which is regulated by a closed feedback loop to remain constant around 10nm.

A SNOM configuration (collection or illumination) and the type of probes are chosen to fit a specific experiment purpose. In our case, we use a SNOM configuration designed to work in transmission and collection mode with transparent samples. This SNOM set-up is able operate either with tapered silica fiber probe or nano-aperture probe milled at the apex of metalized polymer probe. The SNOM in transmission/collection mode is advantageous in experiments that need to control precisely the position and the size of pumping excitation spot with respect to the structures on sample. In order to characterize optical properties of the hybrid structures at $\lambda \approx 1.5\mu m$, a commercial SNOM (NT-MDT SMENA) has been employed. The set-up was developed during the PhD thesis of T-P. Vo [1] and is presented in Figure 4.1.

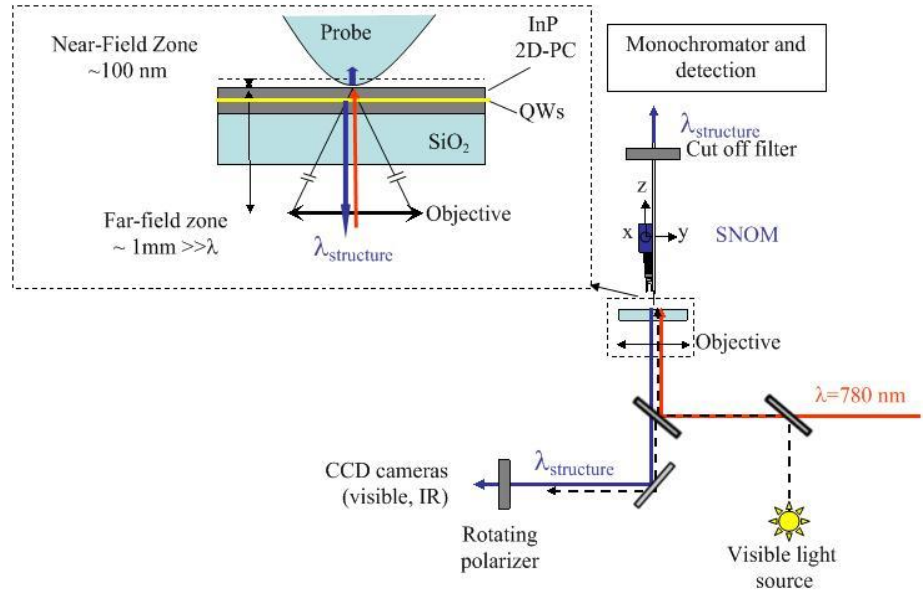

*Figure 4.1 SNOM in transmission set-up [2].*

Since the sample under investigate is transparent. A SNOM in transmission set-up is built based on a commercial Axio Observer.D1. It is dedicated to transparent sample. The experimental set-up has four main sub-systems: Illumination, SNOM head and detection chain [1, 2]. The schematic diagram of SNOM set-up is shown in Figure 3.19.

### 4.1.1 Illumination

The structure was pumped by a 50mW laser diode with a centre wavelength of 780 nm. The diode is pulsed (6.25% of duty cycle) to avoid overheating the structure. The laser diode is also modulated at low frequency (2kHZ) to perform a lock-in detection and ameliorate the signal to noise ratio. The DL light is coupled into a monomode fiber (9μm core diameter) through Thorlab's coupler (objective 20X) and is sent into the optical system of the microscope after passing a dichroic mirror. For looking for the structure, a white light emitted by a halogen bulb (HAL100) is sent into the microscope to illuminate the sample [1]. As was already said in Chapter 2, the photoluminescence of the structure occurs in the range of 1250-1650 nm. At this wavelength region, InP material is transparent.

### 4.1.2 SNOM scaning system

#### 4.1.2.1 Shear force regulation

The distance between the optical near-field probe and the surface of sample is kept constant by shear-force feed-back loop based on a quartz tuning fork system. The probe vibrates parallel to the surface of sample at the quartz resonance frequency, about 32-36 kHZ, with quality factor Q = 500 - 1000. The raster movement of the probe over the surface of the sample is controlled by the piezoelectric scanner located in the shear-force head. The position of the probe is collected and recorded during the raster movement. This enables the investigation of the sample topography [1].

#### 4.1.2.2 Collection of the optical signal

The optical signal in near-field is collected by an optical probe. When the probe is very close to the surface, the interaction between the optical signal and the probe converts the evanescent fields into propagating fields. The optical signal is guided into a monomode optical fiber and is passed to the detection system [1].

### 4.1.3 Optical detection system

The optical signal is directed into a diffraction grating monochromater (Jobin Yvon micro-HR,). This system is equipped with micro-slits to allow analyzing spectra with spectral resolution up to 0.6nm [1]. The micro-slits also have another function. During the near-field optical field distribution measurement, when the light intensity saturate the detector, instead of decrease the pump power of the laser source, one can close the micro-slits to a lower value to decrease the amount of light collected by the detector. This avoids to mofify the pumping power, which has an influence on the lasing wavelength.

The SNOM set-up has also been conceived to allow to record the far-field image of the optical signal emitted by the sample. The infra-red light emitted by the sample is collected from the back side of the sample by the same objective which has been used for the excitation. The light is then sent to another microscope output ports equipped with a IR-camera. Therefore, the image of the radiating light at the surface of the sample can be obtained [1]. It should be mentioned here that the Far-field images are not filtrated and the image may contain optical signals at different wavelengths.

### 4.1.4 Near-field probes

In SNOM, the probe diameter affects the spatial resolution of the distributed evanescent field. Therefore the probe has to be small enough, comparable to the size of objects at least, for resolving the nanometric-scale structure of sample. In this research two types of probe have been used: homemade dielectric probe (or silica probe) and commercial metallised probe (Al-coating) [1]. For the purely evanescent fields, the dielectric probe can get good resolution. However, when the optical field radiates a lot, the dielectric probe is not adapted. In this case the commercial metallised probe is employed. For this research, dielectric probes are used in investigation of the CL5 and CL7 structures and metallised probes for the graphite structures.

### 4.1.4.1 Silica probes

The silica probes are made via chemical etching (Tuner's method) using hydrofluoric acid HF [3, 4]. The optical glass fiber is F-SMF-28 which supports well wavelength at 1550nm. First, the outer protective polymer layer of a single mode fiber is stripped out. Then the fiber is cleaved and dipped into HF (40%) solution at 30 ℃ during 60 minutes in a Teflon vessel. An organic cover-layer (iso-octane) is put on the surface of the solution to protect the fiber mounts from the corrosive HF vapour. Up to eight probes can be fabricated in one batch of fabrication. Due to decreasing meniscus height between hydrofluoric acid and organic over-layer as the fiber diameter is reduced by the etchant, tapered silica probe is formed. With this method, probes with conical angles typically 30° and diameter 200nm can be got [1].

### 4.1.4.2 Metal-coated probe

The fabrication of commercial Al-coating circular probe is conducted by Lovalite SAS and the principle of manufacturing process is reported elsewhere [5, 6, 7, 8]. The spatial resolution of SNOM optical images is controlled by the light-transmitting aperture defined at the probe apex by the metallic coating [9]. The aperture size is about 100nm [2].

## 4.2 Near-field characterizations of S2-CL5 structures

A group of CL5 PC cavities of sample 2, with and without NAs, have been investigated. The sample was observed by SEM MEB-Tescan before these SNOM investigations. The SEM observation for all structures kept a long time. A homemade dielectric tip fabricated from silica optical fiber was employed for near-field measurements. Prior to any near-field measurements, the tip is placed at several microns above the structure to allow spectral investigation in far-field. The light emitted by the sample is collected by the tip, in the far-field, and send in the monochromator for detection. This allows a direct comparison between spectra performed in far-field and in near-field.

### 4.2.1 Spectroscopic results

With the SNOM equipment set-up, one can investigate the spectrum information of far-field and near-field. It is realized by target the near-field probe to the emission light, the probe can collects the emission light and gets the spectrum information. With these measurements, one can select the fixed wavelength for the optical distribution investigation. And it also helps to place the near-field probe more accurate above the cavity where the emission is the strongest.

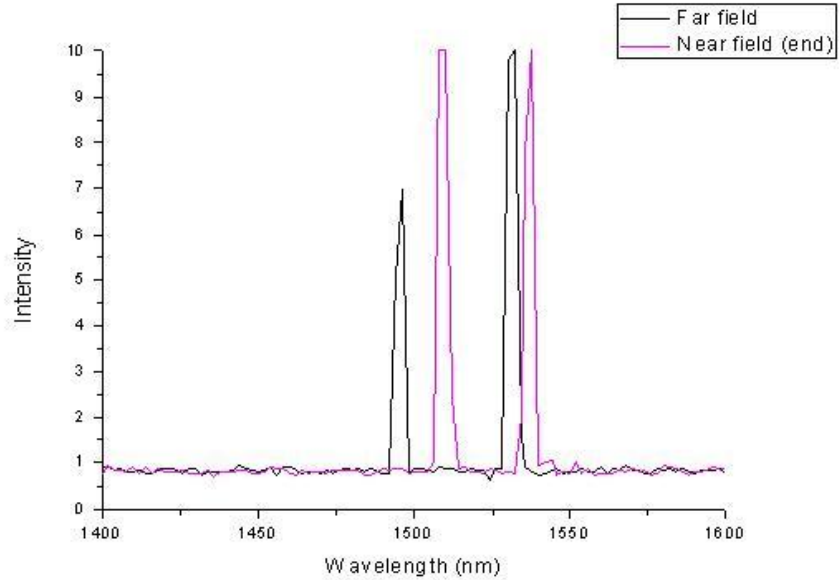

*Figure 4.2 Laser emission spctra of S2-CL5-1.8e-NO in far-field and near-field.*

Figure 4.2 shows the photoluminescence spectrum of structure S2-CL5-1.8e-NO recorded by SNOM, in far-field and in near-field at room temperature. To optimize the signal, the micro-slits are full open and the resolution of the monochromator is low (10nm). The Figure 4.2 reveals that two modes are lasing (1496nm and 1532nm), both in far-field and near-field measurements. From the near-field optical field distribution results which are shown in next section, we indentified the mode at 1496nm as the fundamental cavity mode. The mode at 1532nm is not a cavity mode: it is a delocalized Slow Bloch Mode (SBM) presents at the K-point of the dispersion diagram of the photonic crystal triangular lattice. In far-field, the lasing wavelengths for both modes are remains constant and very stable in time.

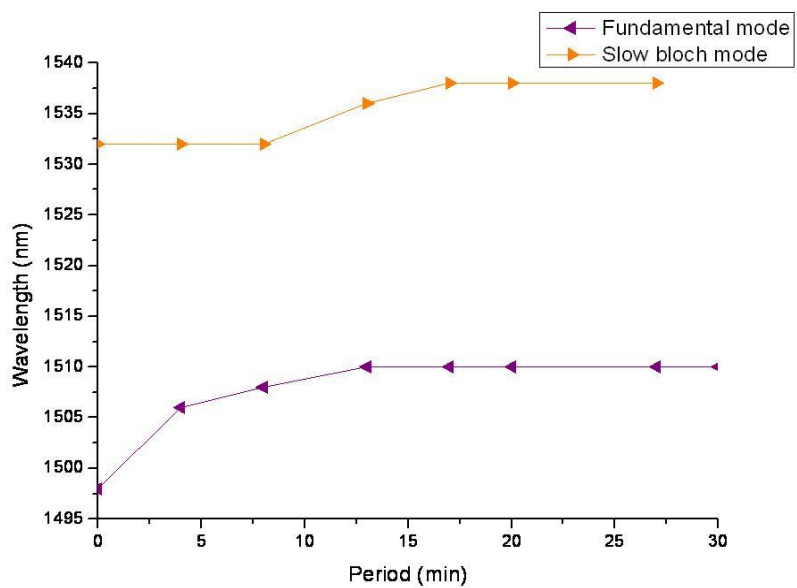

*Figure 4.3 Variation of laser wavelength with time in near-field for S2-CL5-1.8e-NO. Purple curve is the fundamental cavity Mode and orange curve is Slow Bloch Mode.*

However, when the probe approached the surface into the near-field, both wavelengths underwent a red-shift, which magnitude depends on the mode. However, the shifts are different. The fundamental cavity mode under went a larger shift ($\Delta\lambda \sim$ 14nm) than the SBM ($\Delta\lambda \sim$ 6nm). Note that the shifts are not instantaneous: once in the near-field, the wavelength shifts slowly (~15 min) to its final value (Figure 4.3 shows the evolution of the wavelength with the time). After 15min, laser wavelength move to the final value, the wavelength is remained very stable. It will not move during all the near-field measurement for a whole day. This red-shift cannot be explained by the effect of probe-tuning only due to the local modification of the refractive index of the mode environment by the tip. This effect is instantaneous and induced shifts of one or less nanometer. This latter effect will be discussed later in this section.

When the probe is withdrawn from near-field to far-field, the laser wavelength of the PC cavity had a very slow blue-shift. The wavelength variation of the two modes is shown in Figure 4.4. After one night, the laser wavelength back to the original (about

1496nm).

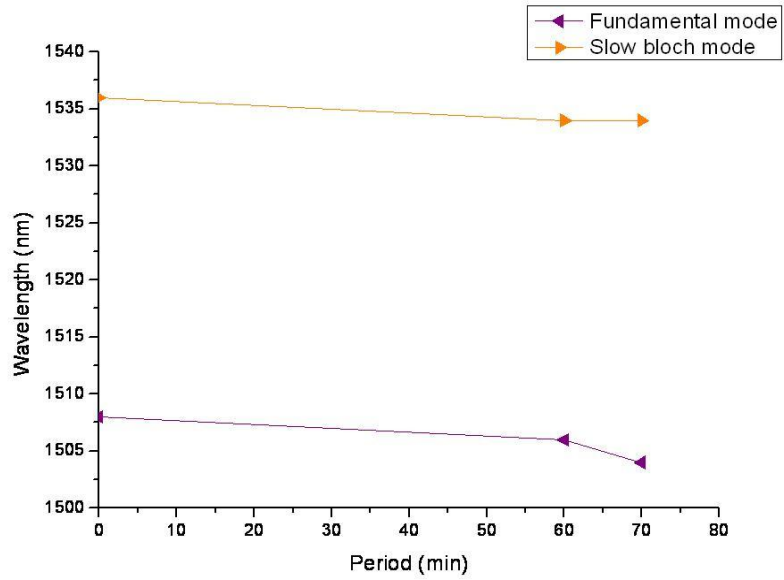

*Figure 4.4 Variation of laser wavelength along period in far-field zone after near-field measurement. Purple curve is Fundamental cavity Mode and orange curve is Slow Bloch Mode.*

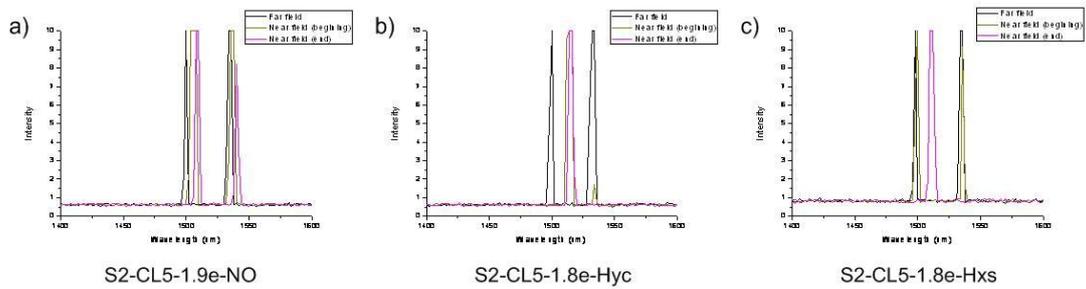

*Figure 4.5 Laser emission spctra of S2-CL5-1.9e-NO, S2-CL5-1.8e-Hyc and S2-CL5-1.8e-Hxs in far-field and near-field.*

The same phenomenon was observed on various structures, with and without NAs, with different designs (location of NA) and fabrication parameters (dose of E-beam lithography). This is illustrated in Figure 4.5 for three different structures. It indicates that the wavelength shift does not depend on the presence of NA. We can assume that the shift is mainly driven by the interaction between the probe and the sample.

The fact that wavelength shift is not instantaneous but takes a certain amount of time when the probe comes into the near-field is not fully understand yet. But, considering the previous result obtained in far-field, some hypothesis can be done. Figure 4.6 shows the measurement results for wavelength variations of structure S2-CL5-1.8e-NO, measured by far-field set-up. The photoluminescence spectra at pump power 1.64mW and 0.14mW are shown as well. The emission wavelength variation shows that with a high pump power, the emission has a blue shift, compared with the low pump power. In our previous hypothesis, this is due to the evaporation of water on the surface of the structure or in the holes of PC structures. In our near-field set-up, the pumping power is well above the threshold, it means that the remaining water evaporates quickly and that, when the probe is far from the sample, we are already on the second branch of the hysterisis cycle (red point in Fig 4.6 (a)). To explain the slow red-shift when the probe comes in the sample near-field, we had to take into account the properties of the silica probe: the silica probes are hydrophilic [10] and a thin layer of water may exits at its surface. When the probe comes into near-field, the water may cover the surface of the structure or fill the holes of the PC cavity again. It could explain a strong red-shift. However, according to our hypothesis, the wavelength should blue-shift again with the water evaporation. We do not observe any blue-shift. In conclusion, on this sample, the tip induce a strong red shift of the modes, this shift presents a time constant in agreement with thermal effect. Most probably, an interaction takes place between the tip and the surface involving some contamination brought by the tip.

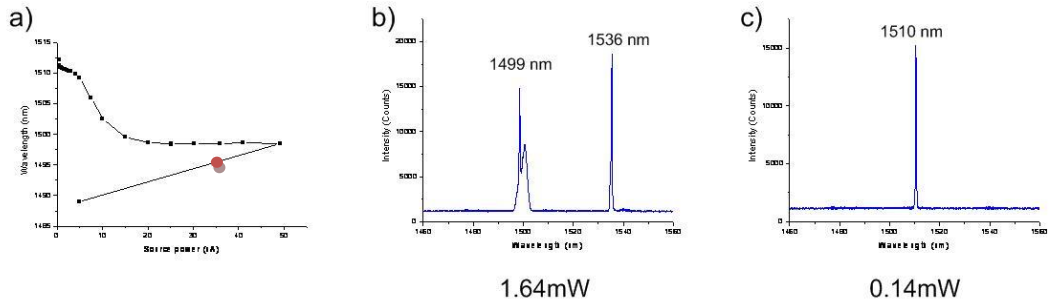

*Figure 4.6 Wavelength variations of S2-CL5-1.8e-NO along the variations of pump power and two specra of laser emission.*

## 4.2.2 Far-field images of the laser mode

As described in section 4.1.3, with the IR-camera of the SNOM system [1], one can get the far-field image of the radiating modes. The far-field image is formed via collected the emission light by the IR-camera from the backside of the sample. The resolution cannot exceed the diffraction limit. The collection wavelength is not fixed. Therefore the far-field image is the pattern of the entire laser mode.

Figure 4.7 shows the far-field images of structure S2-CL5-1.8e-NO, S2-CL5-1.8c-Hyc and S2-CL5-1.8d-Hxs. One can see that the present of the NA modifies the far-field pattern of the PC cavity. The far-field pattern of S2-CL5-1.8e-NO is large and emanative. It seems have two bright spots linked to each other. While the far-field patterns of S2-CL5-1.8c-Hyc and S2-CL5-1.8d-Hxs are concentrated and smaller. And each of them has only one bright spot.

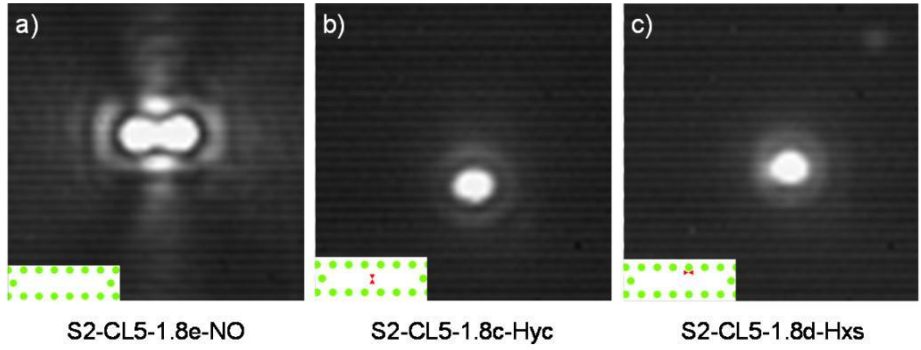

*Figure 4.7 Far-field images of structure S2-CL5-1.8e-NO (a)), S2-CL5-1.8c-Hyc(b)) and S2-CL5-1.8d-Hxs(c)).*

### 4.2.3 Near-field optical field distribution

The optical near-field maps are recorded at fixed wavelength. Note that for the same design, the difference of holes radius of design and dose of the E-beam lithography only change the laser wavelength, but the patterns of the optical field distribution are conserved.

### 4.2.3.1 Pure PC cavity without NA

First, structure S2-CL5-1.9e-NO was investigated. The emission wavelength was measured at different 4 locations noted as position 1 (P1) to position 4 (P4). They are shown in Figure 4.8. The wavelengths measured at these location are $\lambda_{P1} = 1514.39$nm, $\lambda_{P2} = \lambda_{P4} = 1514.21$nm and $\lambda_{P3} = 1513.62$nm.

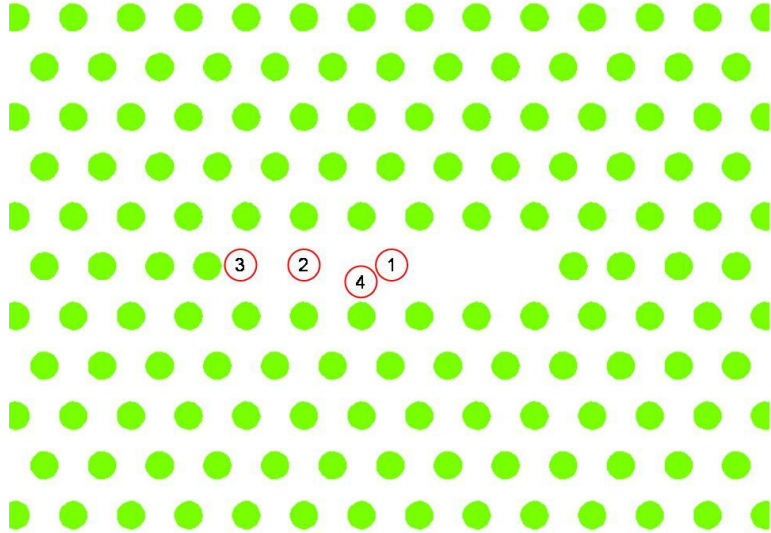

*Figure 4.8 Different positions of fixed wavelength for S2-CL5-1.9e-NO.*

Figure 4.9 shows the SNOM field intensity mapping measured at $\lambda_{P1}$, $\lambda_{P2}$ and $\lambda_{P3}$. The micro-slits value is same for all the results. The topography image shows the cavity clearly. The cavity profile is same as the SEM image. The optical field distributions at the different fixed wavelengths are shown in figures d), e) and f). The superposition of the optical pattern with the topography are also shown in figures g), h) and i). This measurement is in agreement with the simulation pattern of the fundamental mode. We find that the near-field intensity distributions are different and depend on the fixed wavelength. At $\lambda_{P1}$ = 1514.39nm (d)), the region in the centre of the cavity is very bright. At $\lambda_{P2}$ = 1514.21nm, the bright area extend to the corners of the cavity. And the intensity of centre region is weaker than before. At $\lambda_{P3}$ = 1513.62nm, the optical field at the edge of the cavity is very strong. Meanwhile, the optical field in the centre is grows weaker. Note that the strong optical signal out of the cavity comes from the holes losses.

**Discussion:**

This small wavelength tuning is due to the interaction between the photonic crystal and the probe and has been observed for other cavity [10, 11, 12]. The presence of the probe modifies the resonance condition of the cavity: at a given wavelength, only the resonant mode will exist only for a given position of the probe (in the centre of the cavity for $\lambda_{P1}$, at the edge of the cavity for $\lambda_{P3}$). Note that the maximum wavelength shift due to the tip position inside the cavity does not exceed 0.6nm.

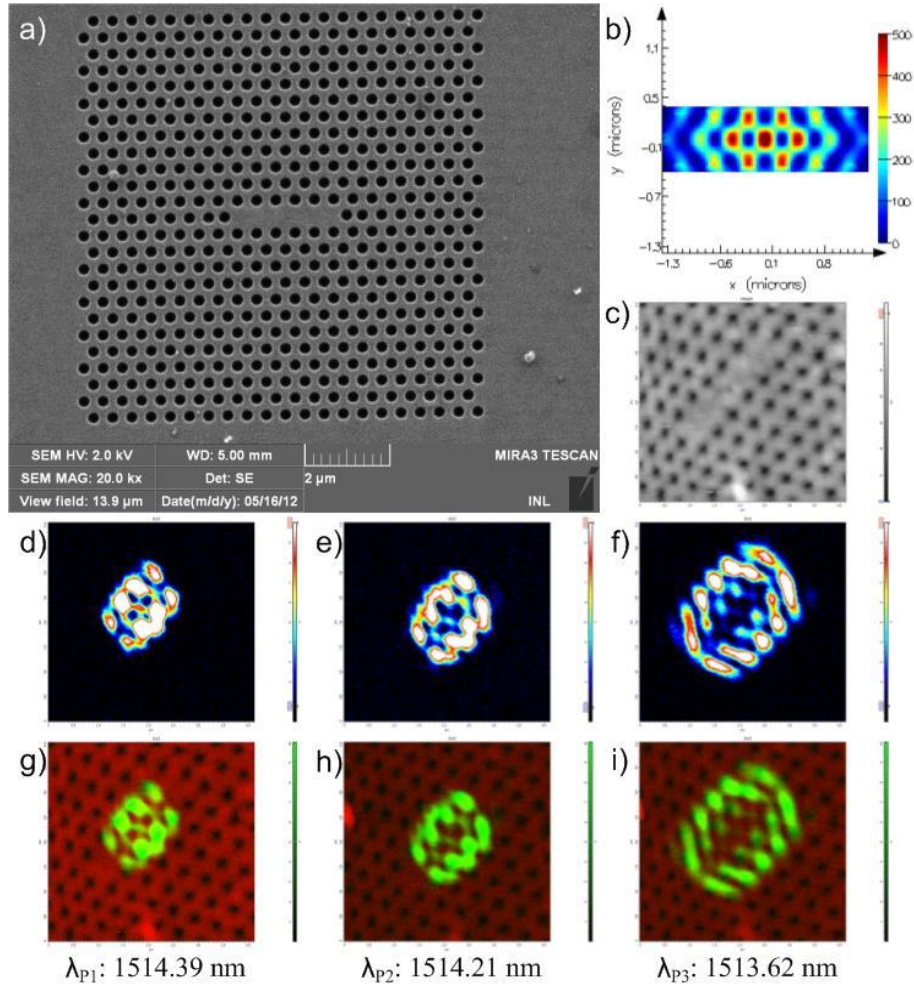

$\lambda_{P1}$: 1514.39 nm $\qquad$ $\lambda_{P2}$: 1514.21 nm $\qquad$ $\lambda_{P3}$: 1513.62 nm

*Figure 4.9 SNOM measurement result of S2-CL5-1.9e-NO. a), SEM image of the PC cavity; b), numerical simulation of the optical field map; c), topography of the PC cavity; d), e), f), optical field maps at fixed wavelength of $\lambda_{P1}$ = 1514.39nm, $\lambda_{P2}$ = 1514.21nm, $\lambda_{P3}$ = 1513.62nm; g), h), i), superpositions of the optical pattern with the topography correspond to d), e), f).*

Cross sections of the intensity distributions can give the quantity information of the optical field intensity. As we had already said in Chapter 1, for this mode, the field in the centre of the cavity along central axis is mainly due to the contribution of $E_y^2$ polarization. And the field intensity measured along the side of the cavity is mainly due to Ex$^2$ polarization. Figure 4.10 shows the cross sections of optical fields Ey$^2$ along long axis cavity at different wavelengths. The cross sections are the blue lines in tiles a), b) and c). And tiles d), e) and f) recorded the intensities correspond to tiles a), b) and c). The slit are opened in same value. At $\lambda_{P1}$ = 1514.39nm, the intensities of the optical field Ey$^2$ decreases in order from centre to out side. At $\lambda_{P2}$ = 1514.21nm, the intensity in the centre decreases. And at $\lambda_{P3}$= 1514.21nm, the intensity in the centre is much weaker than before. The intensities outside become strong on the other hand.

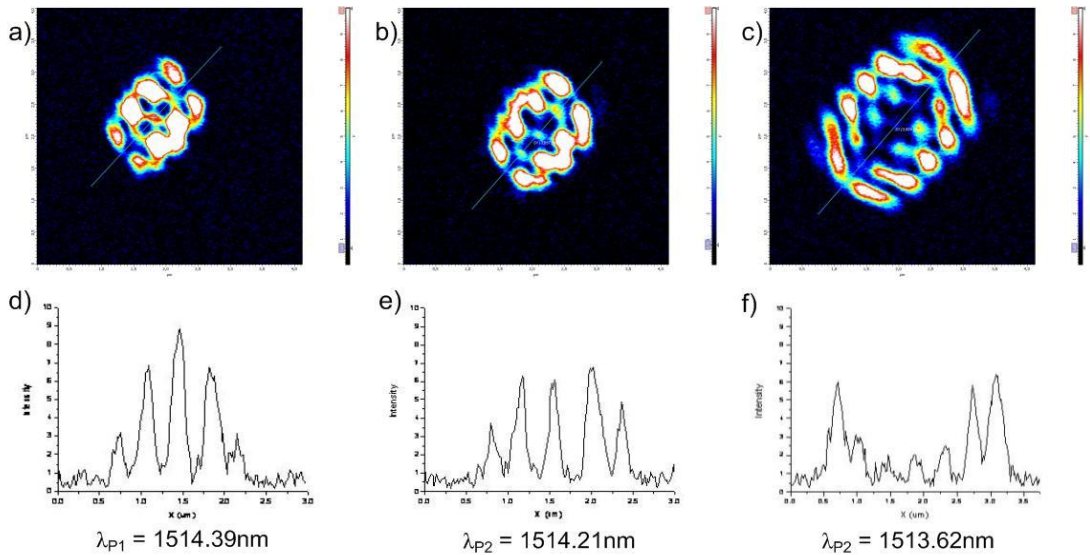

*Figure 4.10 Cross sections along the long axis S2-CL5-1.9e-NO. a), b), c), the schemas of the cross sections; d), e), f), the light intensities of the cross sections.*

The other structures of CL5 PC cavities without NA show the same features. Figure 4.11 is tiles of recorded images of structure S2-CL5-1.8e-NO. With various fixed wavelengths, the near-field intensity distribution shows the same frequency tuning phenomenon. For this structure, the range of wavelength variation is larger and the optical field intensity changes more. This probably indicates that the interaction between the tip and the photonic cavity is stronger. The maximum wavelength shift due to the tip position inside the cavity is 1.59nm.

In conclusion, the interaction between the near-field probe and the PC cavity influence the record result of the near-field optical distribution of the PC cavity. This phenomenon complicates the SNOM image interpretation and has to be considered in the study of the hybrid structure with NA.

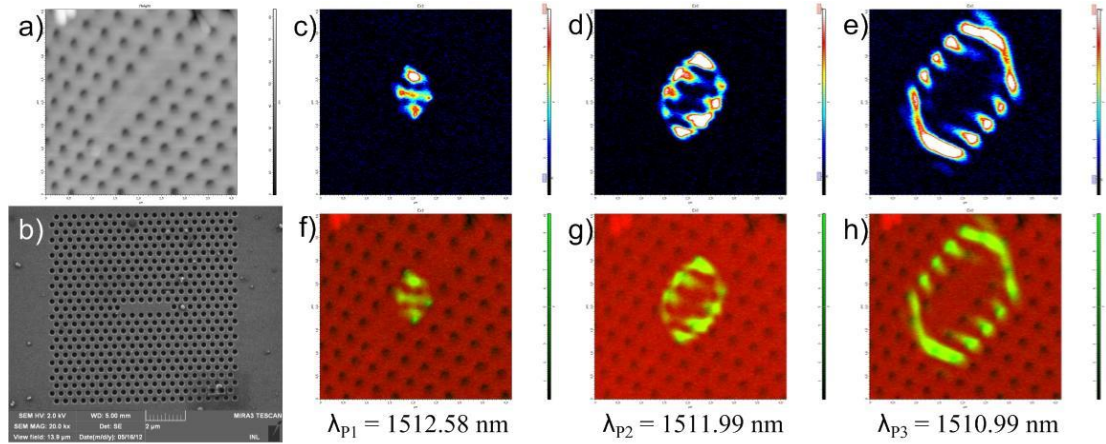

*Figure 4.11 SNOM measurement result of S2-CL5-1.8e-NO. a), topography of the PC cavity; b), SEM image of the PC cavity; c), d), e), optical field maps at fixed wavelength of $\lambda_{P1} = 1512.58nm$, $\lambda_{P2} = 1511.99nm$, $\lambda_{P3} = 1510.99nm$; f), g), h), superpositions of the optical pattern with the topography correspond to c), d), e).*

### 4.2.3.2 Hybrid structure with NA

### a. NA in the direction of Ey

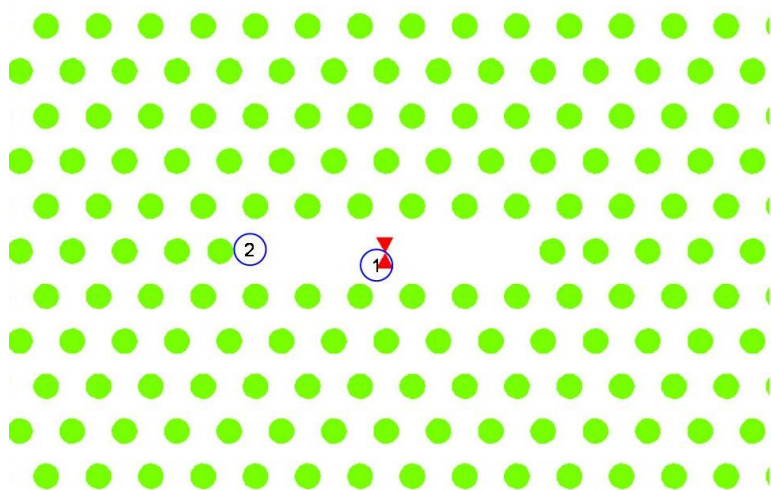

*Figure 4.12 Different positions of fixed wavelength for S2-CL5-1.8e-Hyc.*

In this section, CL5 PC cavity with NA are investigated. Figure 4.12 shows S2-CL5-1.8e-Hyc structure. The emission wavelength was measured at two different positions of the tip, noted as position 1 (P1) and position 2 (P2). P1 is near the centre of the cavity, to measure the lasing wavelength around the NA. P2 is near the edge of the cavity, to measure the lasing wavelength at the edge of the cavity. The wavelengths measured at these location are $\lambda_{P1}$ = 1517.38nm and $\lambda_{P2}$ = 1515.21nm.

Figure 4.13 shows the near-field optical maps recorded at $\lambda_{P1}$ = 1517.38nm (d) and $\lambda_{P2}$ = 1515.21nm (e). The patterns of the intensity distributions are different from the PC cavity without NA, which indicates that the NA modifies the mode. The topography shows the cavity clearly and one can see the feature of the NA. The structure profile is the same as the SEM image. The optical field distributions at different fixed wavelengths are shown in figures d) and e). The superposition of the topography and the optical maps are shown in figures f) and g). As observed before, the two near-field intensity distributions depend also on the fixed wavelength. This indicates that the wavelength tuning phenomenon still exists for hybrid structures and even seems stronger ($\Delta\lambda\sim$2nm). At $\lambda_{P1}$ = 1517.38 nm (d)), the region in the centre of the cavity is bright and there is a strong hot spot on the NA. The size of the hot spot is smaller than the size of NA. Most of the regions around the NA are dark. The regions of optical field far from the NA are bright. At $\lambda_{P2}$ = 1515.21nm, the optical fields at the edges is very strong. Meanwhile, the optical field in the centre is very weak. However, the position of NA is still bright. It is brighter than the area around. Compareed to the numerical simulation pattern, one can find that the enhancement is not as much as the simulation. This may be due to the losses which are yielded by the coupling between the PC cavity and NA. This decreases the efficiency of the enhancement [13].

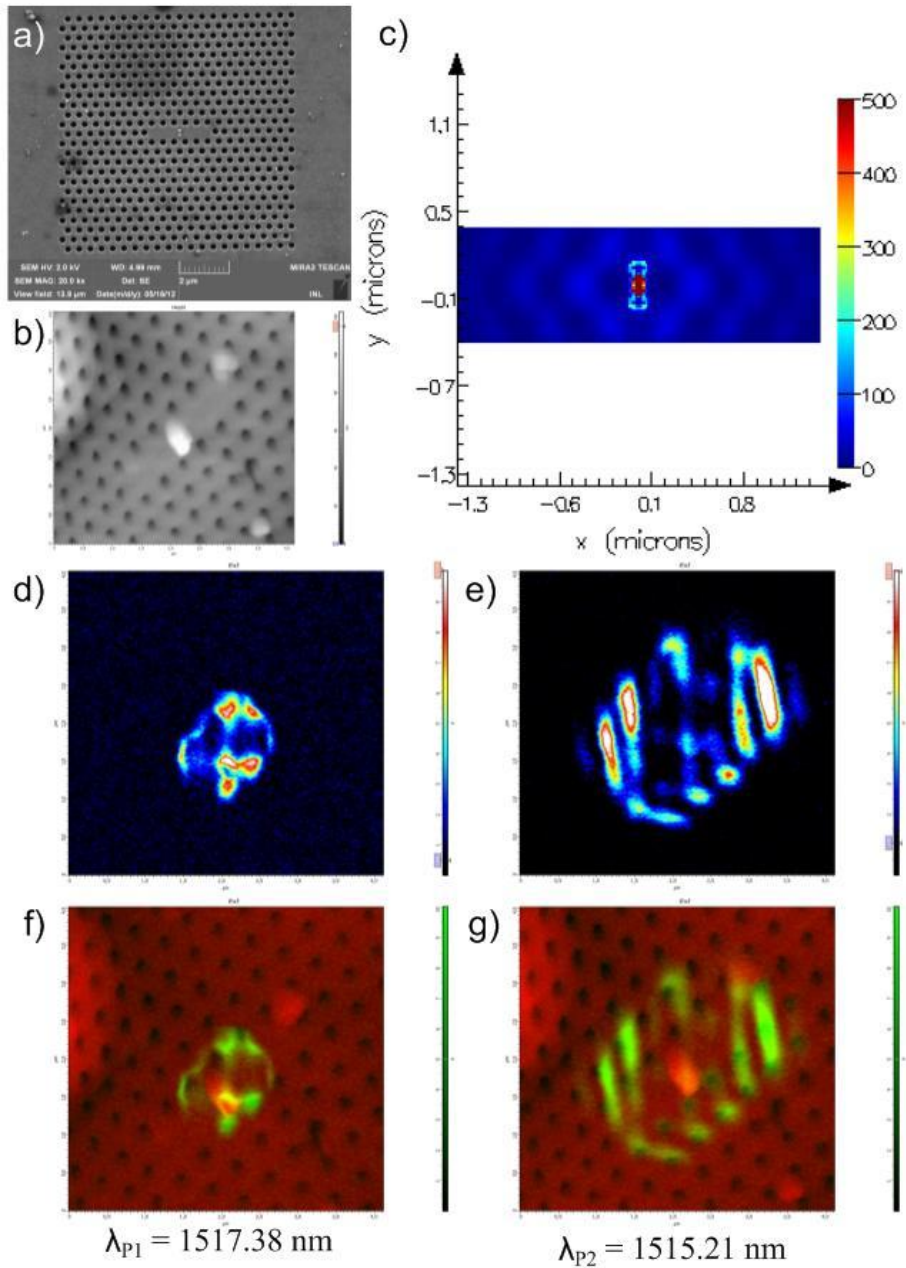

*Figure 4.13 SNOM measurement result of S2-CL5-1.8e-Hyc. a), SEM image of the PC cavity; b), topography of the PC cavity; c), numerical simulation of the optical field map; d), e), optical field maps at fixed wavelength of $\lambda_{P1} = 1517.38$nm, $\lambda_{P2} = 1515.21$nm; f), g), superpositions of the optical pattern with the topography correspond to d), e).*

To evaluate the impact of the NA on the mode distribution, we compare the optical maps obtained on S2-CL5-1.8e-Hyc to S2-CL5-1.9e-NO. Figure 4.14 shows the cross sections of optical field at $\lambda = 1517.38$nm for the structure with NA to the map recorded at 1514.4nm for the structure without NA. The presence of the NA seems to concentrate the light in a very tiny zone where the light intensity is much higher. It is however, very difficult to estimate the enhancement of the field with respect to the case without NA.

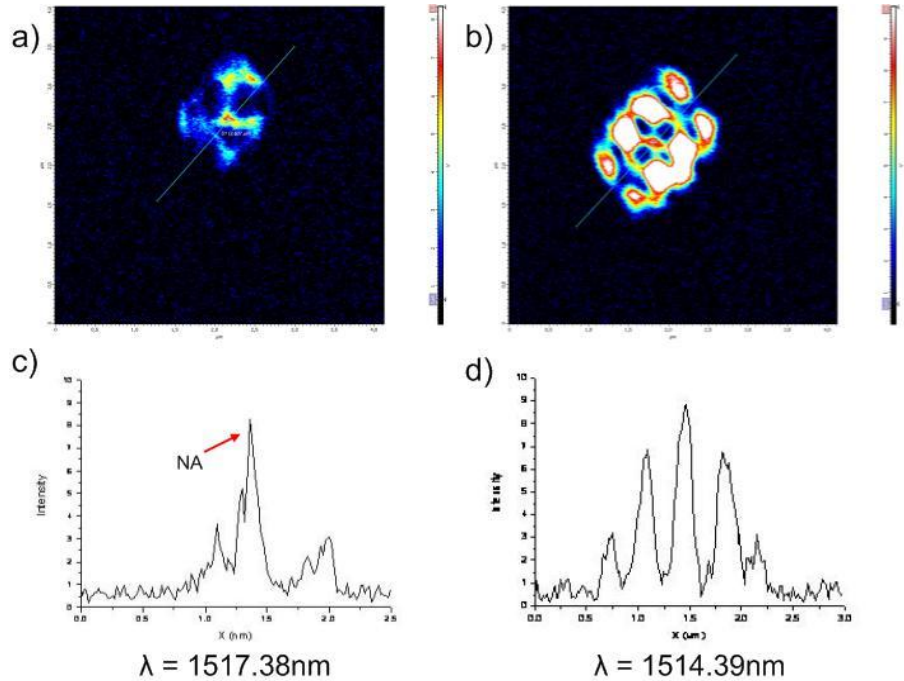

*Figure 4.14 Cross sections of S2-CL5-1.8e-Hyc at λ = 1517.38nm, compare with S2-CL5-1.9e-NO at λ = 1514.39nm. a), b), the schemas of the cross sections; c), d), the light intensities of the cross sections. The red arrows show the NA.*

The same comparison was performed for the two other wavelengths and is shown in Figure 4.15. Here again, the intensity of optical signal of NA is still much higher than the overall background signal

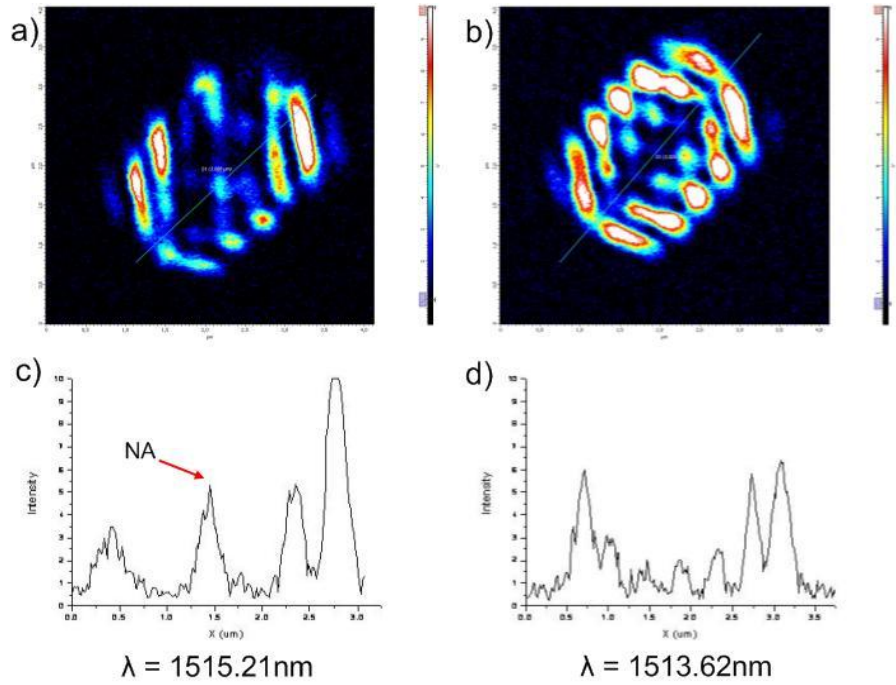

*Figure 4.15 Cross sections of S2-CL5-1.8e-Hyc at λ = 1515.21nm, compare with S2-CL5-1.9e-NO at λ = 1513.62nm. a), b), the schemas of the cross sections; c), d), the light intensities of the cross sections. The red arrows show the NA.*

## b. NA in the direction of Ex

In the structure S2-CL5-1.8e-Hxs (Figure 4.16), the NA is design to couple with the optical field $Ex^2$. The emission wavelength was measured for 4 different positions of the tip. The locations (cf. Figure 4.16) are noted as position 1 (P1) to position 4 (P4). The measured wavelengths at these location are $\lambda_{P1} = \lambda_{P3} = \lambda_{P4} = 1514.8nm$ and $\lambda_{P2} = 1514.4nm$.

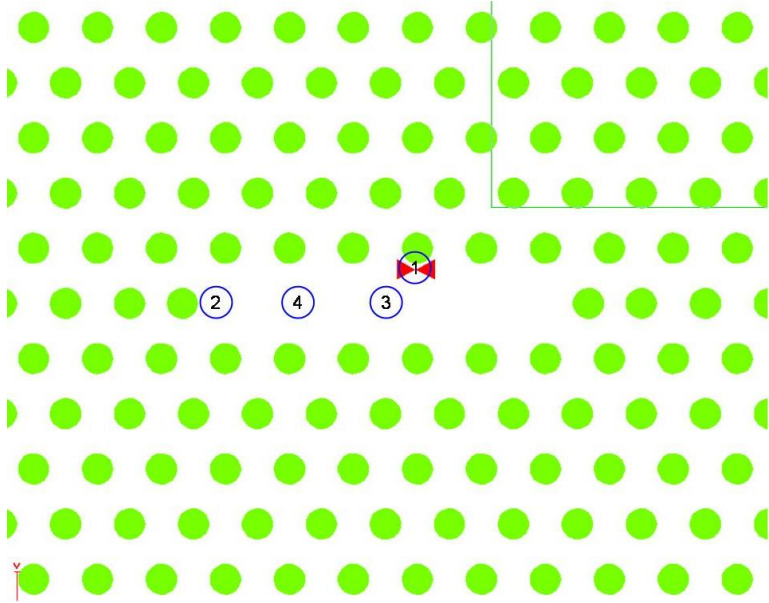

*Figure 4.16 Different positions of fixed wavelength of S2-CL5-1.8e-Hxs.*

Figure 4.17 shows the near-field cartographies performed with the wavelengths $\lambda_{P1}$ and $\lambda_{P2}$. The topography shows clearly the cavity and the feature of the NA. The structure profile is same as the SEM image. The optical field distributions at different fixed wavelengths are shown in figures d) and e). The superposition of the topography and the optical maps are shown in figure f) and g). As before, we observe the tip-position dependence of the resonance wavelength of the mode. The patterns of the intensity distributions are also different from the PC cavity without NA, which confirms that the NA modifies the mode. At $\lambda = 1514.8$nm (d)), The NA is very bright. The regions around the NA are very dark. The optical fields far from the NA are bright, but they are not as bright as the NA. At $\lambda = 1514.4$nm, the optical fields at the edge of the cavity arise and the intensity of them is strong. Meanwhile, the NA still shows the brightest signal.

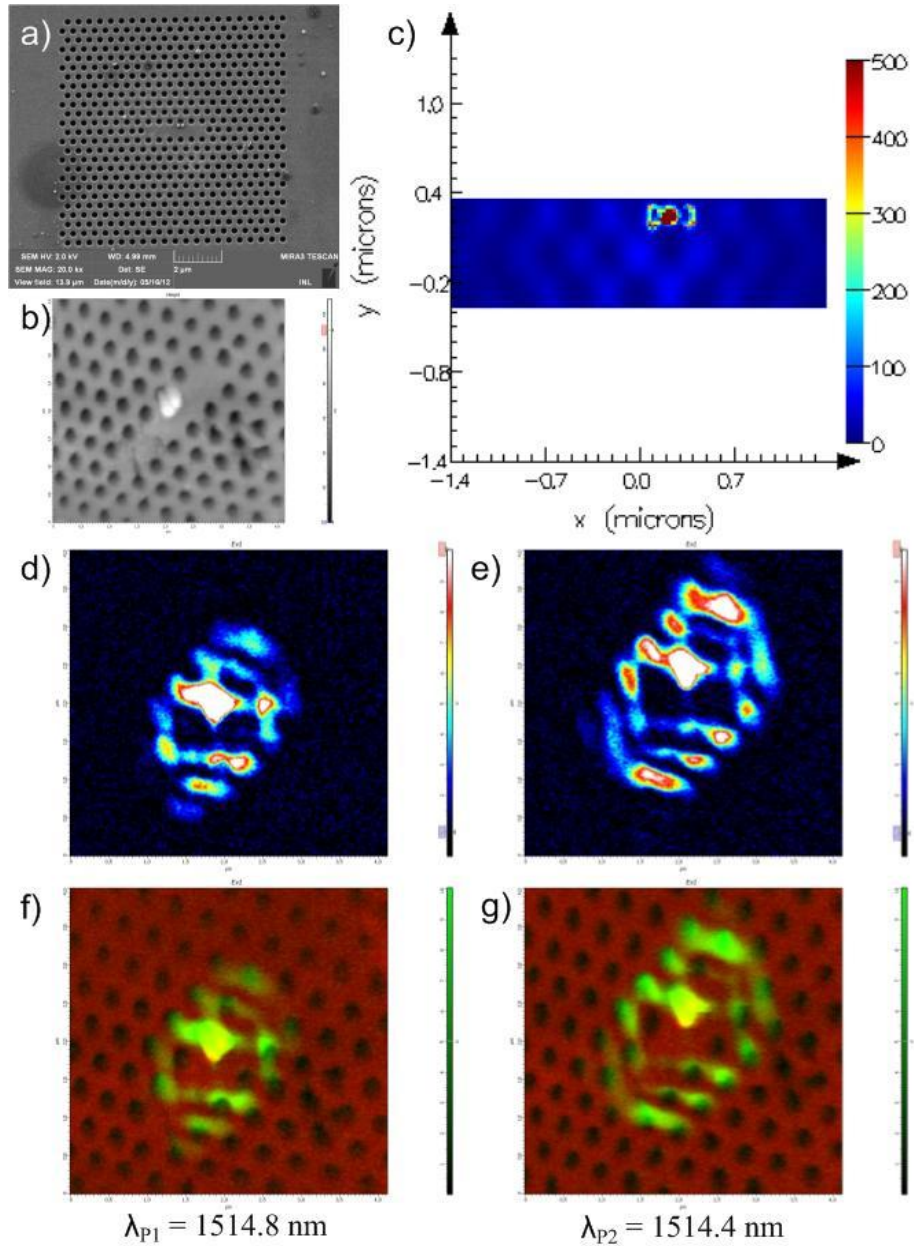

*Figure 4.17 SNOM measurement result of S2-CL5-1.8e-Hxs. a), SEM image of the PC cavity; b), topography of the PC cavity; c), numerical simulation of the optical field map; d), e), optical field maps at fixed wavelength of $\lambda_{P1} = 1517.38nm$, $\lambda_{P2} = 1515.21nm$; f), g), superpositions of the optical pattern with the topography correspond to d), e).*

Figure 4.18 shows the cross section of the optical signal at the position of the NA at $\lambda = 1514.8$ nm of S2-CL5-1.8e-Hxs. From the intensity records, one can see, for this structure, that the light intensity at the position of the NA is much stronger than the signal from other location. In Figure 4.18, the signal from NA saturates the detector.

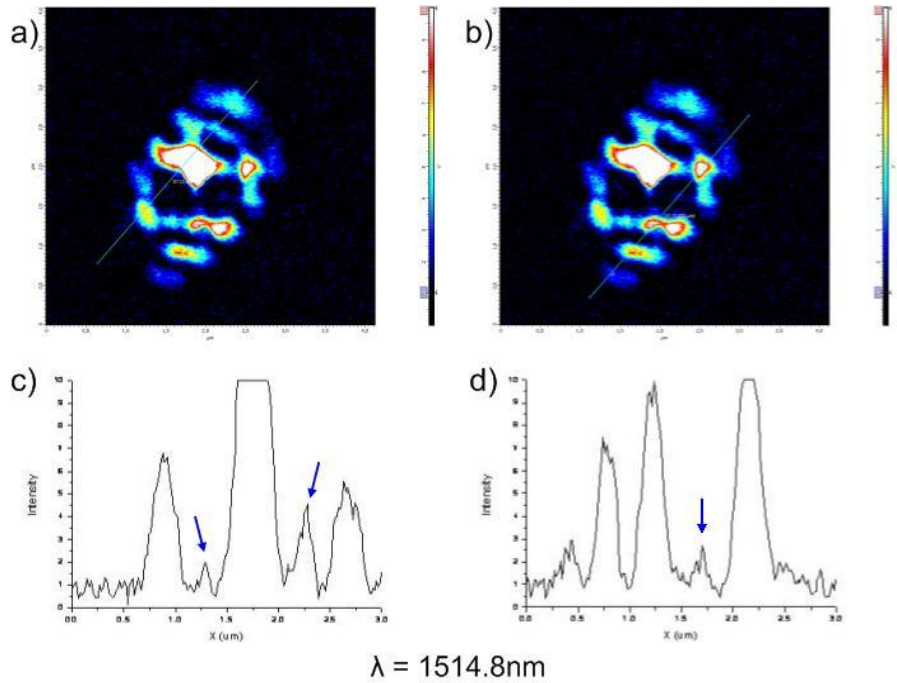

Figure 4.18 Cross sections of S2-CL5-1.8e-Hxs at λ = 1514.8nm. a), b), the schemas of the cross sections; c), d), the light intensities of the cross sections. The blue arrows show the optical field near the NA.

To avoid the signal saturation, the monochromator slits are reduced to a suit value. Figure 4.19 shows the new maps and the new cross sections of optical field $Ey^2$ at λ = 1514.8 nm of S2-CL5-1.8e-Hxs in this situation. On this figure, we clearly see that the NA signal dominates the light intensity in the cavity. Comparing with the intensity record of cross section of optical field $Ex^2$ measured at λ = 1514.39 nm of structure S2-CL5-1.9e-NO, one can see for the structure S2-CL5-1.8e-Hxs, the intensity of optical signal of NA is stronger than the signals of other optical field $Ex^2$. But for the structure S2-CL5-1.8e-NO, the intensities of the optical fields $Ex^2$ are same. It indicates that the NA enhances the optical signal in near-field level.

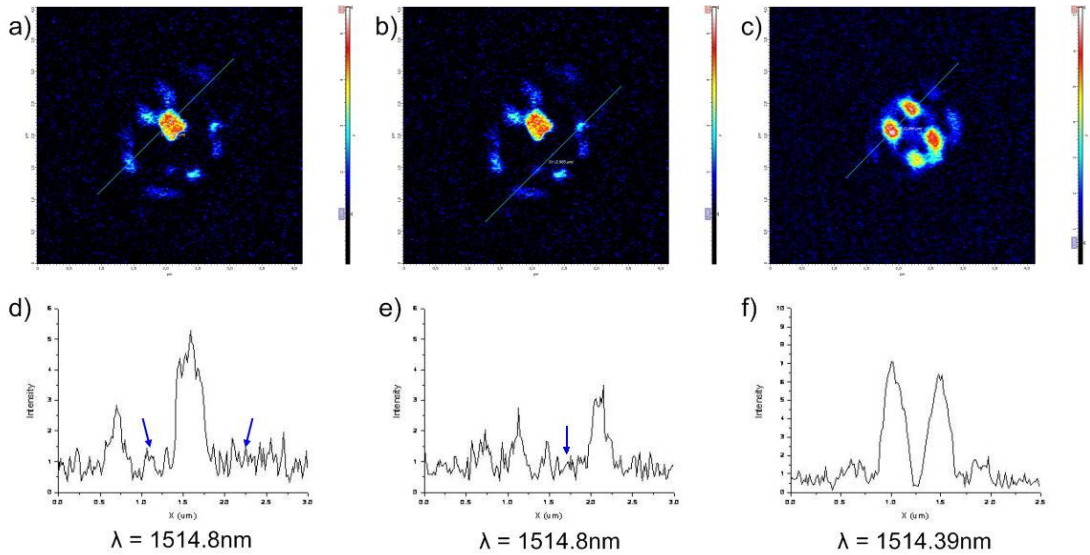

*Figure 4.19 Cross sections of S2-CL5-1.8e-Hxs at λ = 1514.8nm, compare with S2-CL5-1.9e-NO at λ = 1514.39nm. a), b), c), the schemas of the cross sections; d), e), f), the light intensities of the cross sections. The blue arrows show the optical field near the NA.*

## 4.3 Near-field characterizations of CL7 structures

### 4.3.1 Spectroscopic results of the CL7 structures

For CL7 structures the recorded wavelength were measured firstly. These structures were not observed for by MEB-Tescan before the near-field measurements. Figure 4.20 shows the photoluminescence spectra of structure S2-CL7-1.5c-NO recorded by the SNOM in far-field and near-field at room temperature. The slit is full open (slit = 200). The laser peak obtained at λ = 1564nm corresponding to the fundamental mode of PC cavity in both far-field (black curve) and near-field (red curve) measurement. Conversely to CL5 structures, no extraordinary red-shift was observed for these structures. The near-field spectrum shows several emission peaks at other wavelength corresponding to other modes [14].

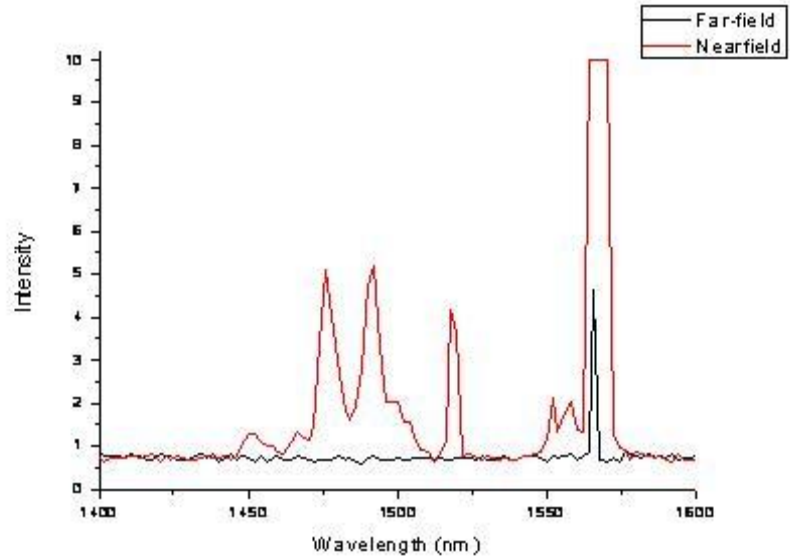

*Figure 4.20 Laser emission spctra of S2-CL7-1.5c-NO in far-field and near-field.*

The other CL7 structures do not show the wavelength tuning phenomenon neither, including different structures (PC cavities with and without NAs) with different design (location of NA) and fabrication parameter (dose of E-beam lithography). Figure 4.21 shows the measurement results of structureS2-CL7-2.0c-NO (a)), S2-CL7-2.0c-Hxs (b)) and S1-CL7-1.8c-Hxc (b)) for instance. From the spectra of structure S1-Hxc-CL7-18c, one can see the laser emission of fundamental cavity mode at 1562nm, and there is another cavity mode laser emission at $\lambda = 1542$nm [2].

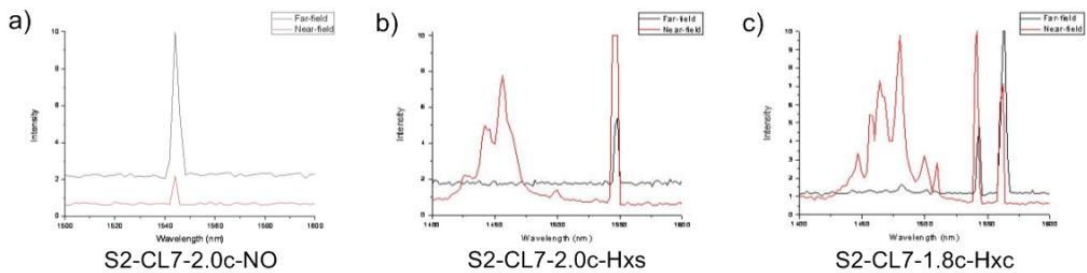

*Figure 4.21 Laser emission spctra of S2-CL7-2.0c-NO (a)), S2-CL7-2.0c-Hxs (b)) and S2-CL7-1.8c-Hxc (c)) in far-field and near-field.*

### 4.3.2 Far-field images of the laser mode

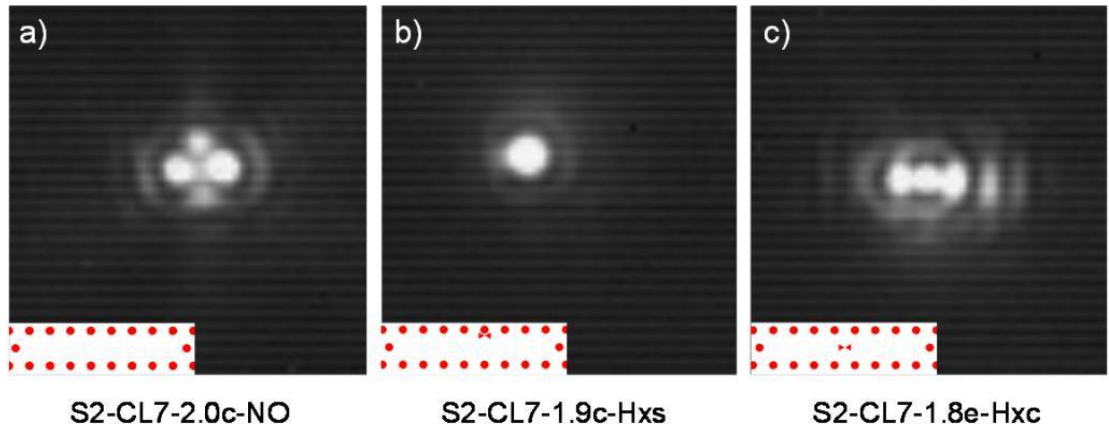

**S2-CL7-2.0c-NO**          **S2-CL7-1.9c-Hxs**          **S2-CL7-1.8e-Hxc**

*Figure 4.22 Far-field images of structure S2-CL7-2.0c-NO (a)), S2-CL7-1.9c-Hxs(b)) and S2-CL7-1.8e-Hxc(c)).*

Figure 4.22 shows the far-field optical images of structure S2-CL7-2.0c-NO, S2-CL7-1.9c-Hxs and S2-CL7-1.8e-Hxc. It shows that the presence of the NA modifies the far-field pattern of the PC cavity. The far-field pattern of S2-CL7-1.8e-NO presents two bright spots in the pattern. The far-field patterns of S2-CL7-1.9c-Hxs and is concentrated and smaller. It has only one bright spot. The pattern of S2-CL7-1.8e-Hxc presents three bright spots in. One additional bright spot is in the centre. This spot may due to the presence of NA. However, since there are two laser modes detected in the spectroscopic measurement, it is probably the combination of these two modes.

### 4.3.3 Near-field maps of the CL7 structures

### 4.3.3.1 Pure PC cavity

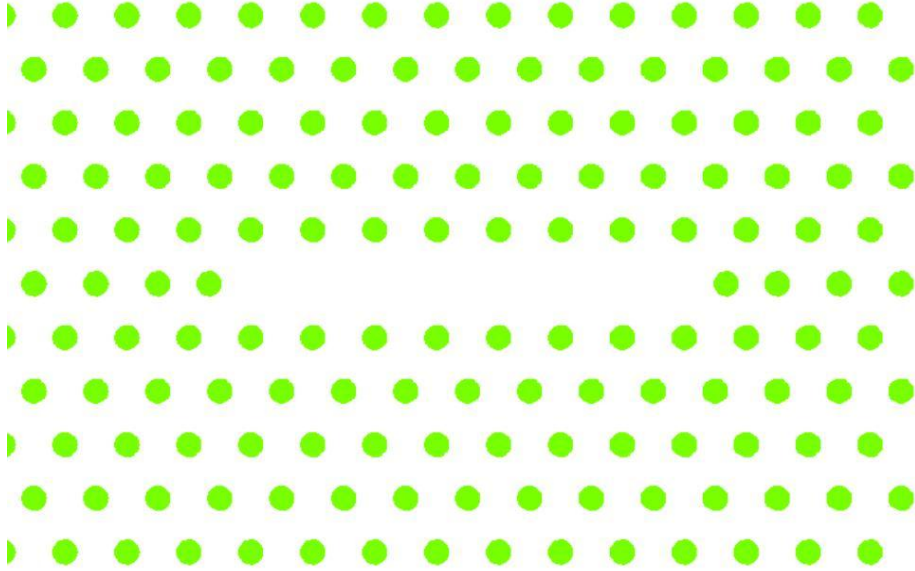

*Figure 4.23 Structure S2-CL7-2.0c-NO.*

Concerning the CL7 micro-cavities without NA, we choose to present the results of structure S2-CL7-2.0c-NO (Figure 4.23) because it is quite representative. The wavelengths of fundamental cavity mode measured in the centre of the cavity is $\lambda$ = 1543.00nm. For this cavity, the interaction with the SNOM-tip was sufficiently weak to neglect the tuning effect presented in the previous section. The reason is probably the effect of the cavity size (or mode volume) with respect to the probe size. Moreover, the structures have not been observed by SEM, and the wavelength keeps stable both in far-field and in near-field.

Figure 4.24 shows the SNOM measurement result with the wavelength set at $\lambda$ = 1543.00nm. From the topography image, one can see the cavity clearly. The optical field distribution is shown by figures c). From the figures one can see that this mode is indeed the fundamental mode of the cavity. The optical field distribution patterns are in very good agreement with the simulations.

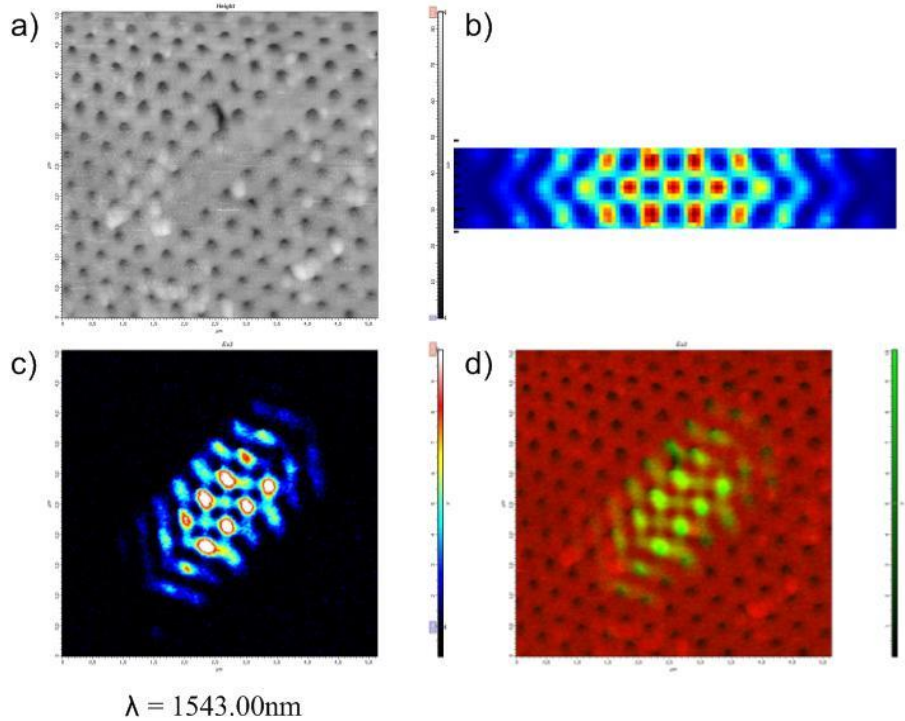

$\lambda = 1543.00nm$

*Figure 4.24 SNOM measurement result of S2-CL7-2.0c-NO. a), topography of the PC cavity; b), numerical simulation of the optical field map; c), optical field maps at fixed wavelength of $\lambda = 1543.00nm$; d), Superposition of the optical pattern with the topography correspond to c).*

As described in chapter 1, for the fundamental cavity mode, the field in the centre along the long axis of the cavity is mainly due to the contribution of Eypolarization, and the field at the sides of the cavity is mainly due to the contribution of Ex polarization. Figure 4.25 shows the cross sections of optical field $Ey^2$ in the centre and $Ex^2$ at the sides at $\lambda = 1543.00nm$. The cross sections are shown in tiles a), b) and c). And tiles d), e) and f) recorded the intensities correspond to them. The intensities of both the optical fields $Ey^2$ and $Ex^2$ decrease from centre to theside. Note that the intensity of optical field $Ey^2$ is much weaker than the optical field $Ex^2$, only about 1/2 of them. It is because the optical field $Ex^2$ is scattered by the holes around the cavity, so the signal of optical field $Ex^2$ are much stronger.

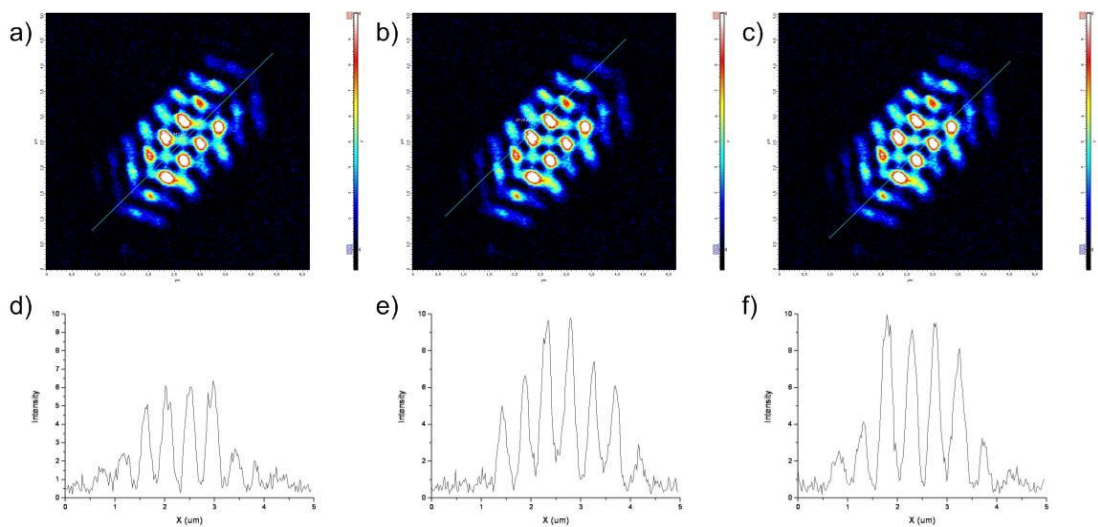

*Figure 4.25 Cross sections along the long axis and the sides of S2-CL7-2.0c-NO.*

The recorded optical field distributions of other CL7 PC cavities show the same results. Figure 4.26 shows the near-field investigation results of structure S3-CL7-1.7d-NO and S3-CL7-1.6c-NO.

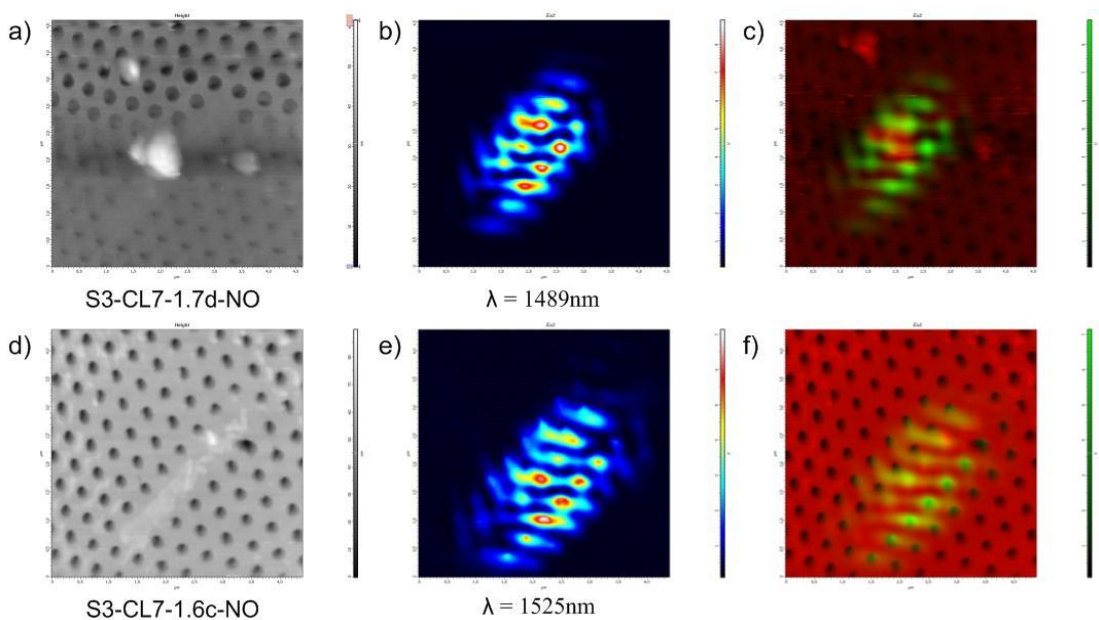

*Figre 4.26 SNOM measurement results of S3-CL7-1.7d-NO (a), b), c)) and S3-CL7-1.6c-NO (d), e), f)).*

### 4.3.3.2 Hybrid structures

**a. NA in the direction of Ex at the side of cavity**

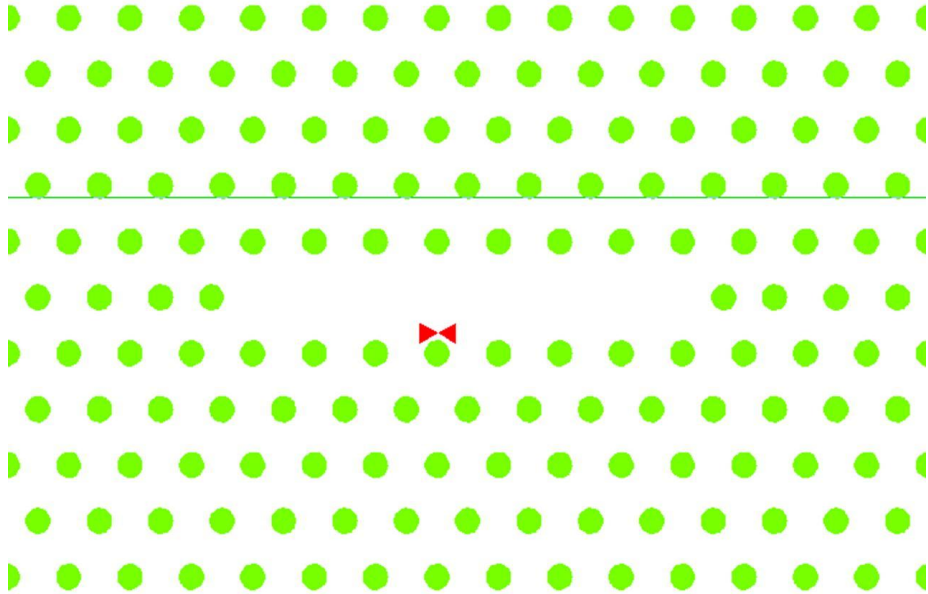

*Figure 4.27 Structure S2-CL7-2.0e-Hxs*

CL7 PC cavity with NA had been measured. Here the measurement results of structure S2-CL7-2.0c-Hxs (Figure 4.27) are presented. Figure 4.28 shows the SNOM measurement result of the structure. The recorded wavelength is λ = 1545.58 nm. The topography image shows the feature of the cavity and the NA. The optical field distributions are shown in figures c). As for CL5 structures, the patterns of the intensity distributions are different from the PC cavity without NA : itmeans that the NA modifies the mode. The position of the NA is bright and the regions close to the NA are very dark.

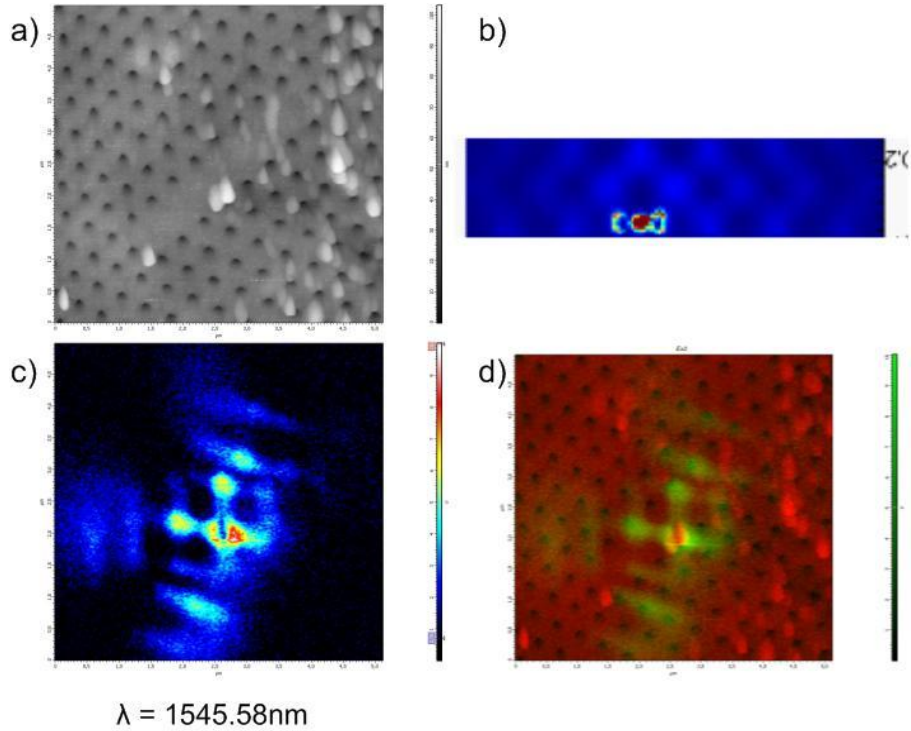

λ = 1545.58nm

*Figure 4.28 SNOM measurement result of S2-CL7-2.0c-Hxs. a), topography of the PC cavity; b), numerical simulation of the optical field map; c), optical field maps at fixed wavelength of λ = 1545.58nm; d), superpositions of the optical pattern with the topography correspond to c).*

Figure 4.29 shows the cross sections of optical field of structure S2-CL7-2.0c-Hxs at $\lambda_1$ = 1545.58nm (a) and b)). Comparing with the result measured at $\lambda_2$ = 1543.00nm of structure S2-CL7-2.0c-NO (e) and f)). We can see that the NA signal dominates the light intensity in the cavity. For the structure S2-CL7-2.0c-Hxs, the intensity of optical signal of NA is stronger than the signals of other optical fields.

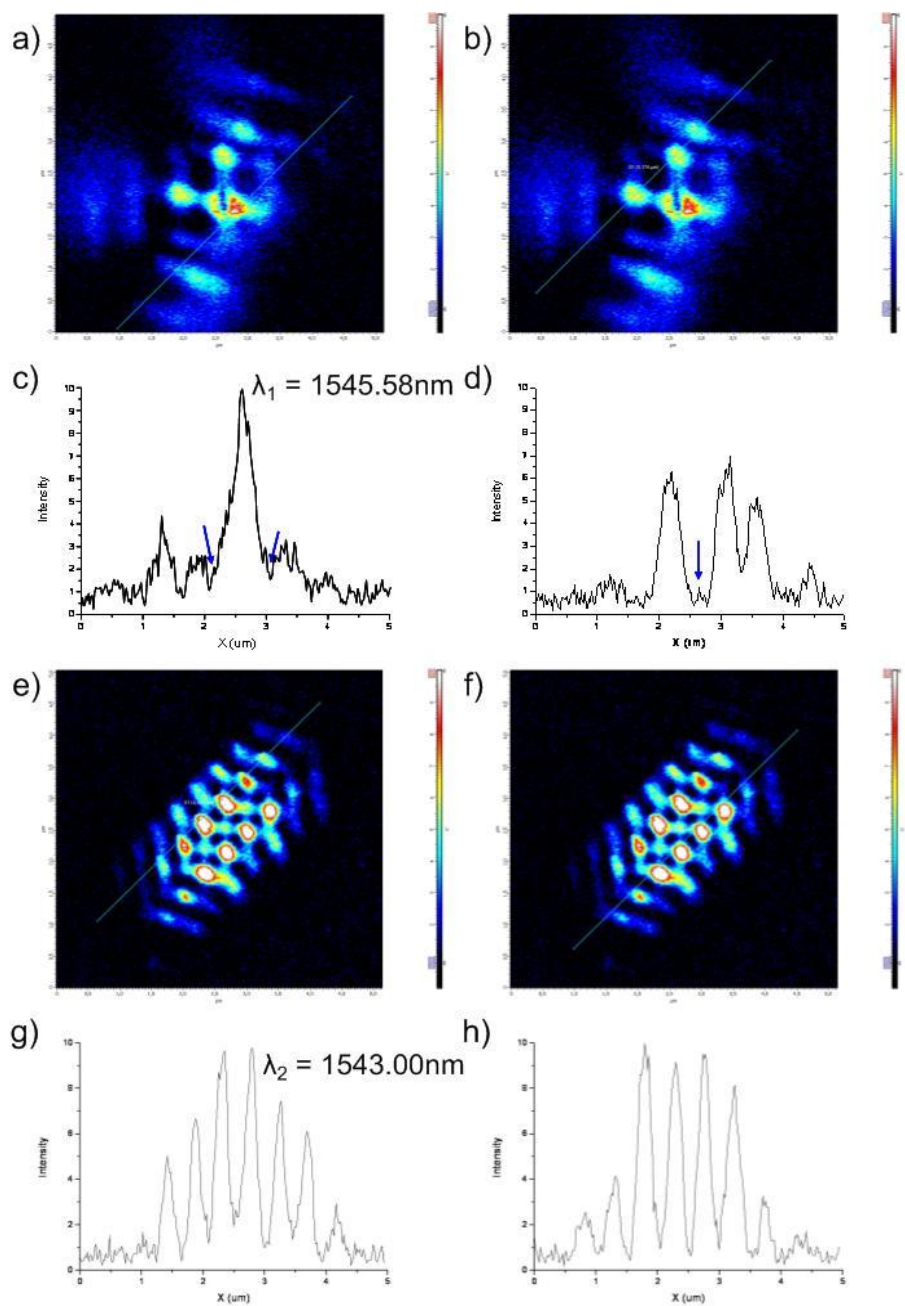

*Figure 4.29 Cross sections of Ex² S2-CL7-2.0c-Hxs (a), b), c), d)) and S3-CL7-2.0c-NO (e),f),g),h)).*

## b. NA in the direction of Ex in the centre of cavity

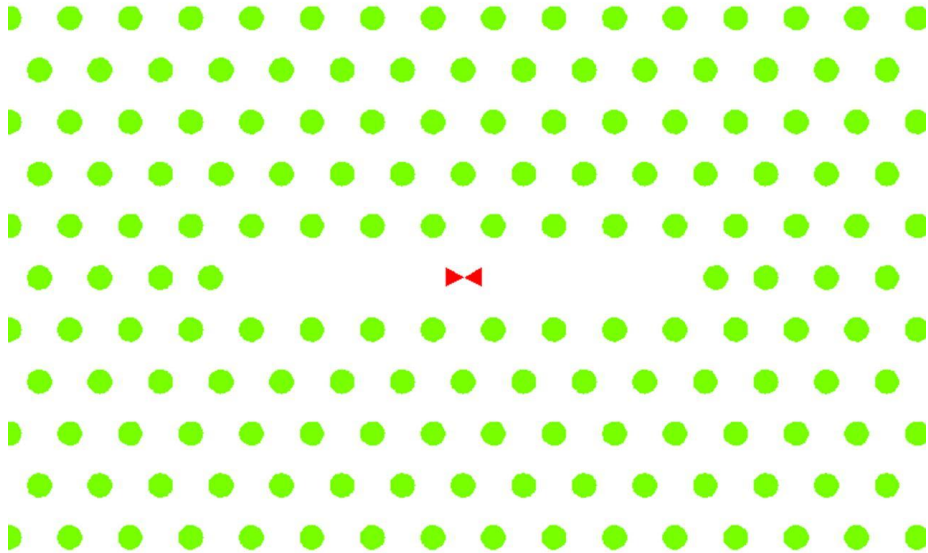

*Figure 4.30 Structure S3-CL7-1.9d-Hxc.*

The measurement results of structure S3-CL7-1.9d-Hxc (Figure 4.30) are given. Since the resonance of NA is sensitive to the polarization of the exciting source. The orientation of the NA is optimized for a coupling with the Ex component. As there is no $Ex^2$ component at this location, this is a test of the polarization sensitivity of the NA. Figure 4.31 shows the SNOM measurement result of the structure at $\lambda$ = 1485.02nm. The topography image shows the cavity the NA. Figure c) shows the optical field distributions. In this case, the four regions of optical fields $Ex^2$ near the corner of the NA are very bright. The intensity of signal from the gap of the NA is weak, indicating the polarization sensitivity of the hybrid structure. The result agrees with the theory expect [13] and the simulation result. This behaviour is reproducible and has been recorded on several identical structures.

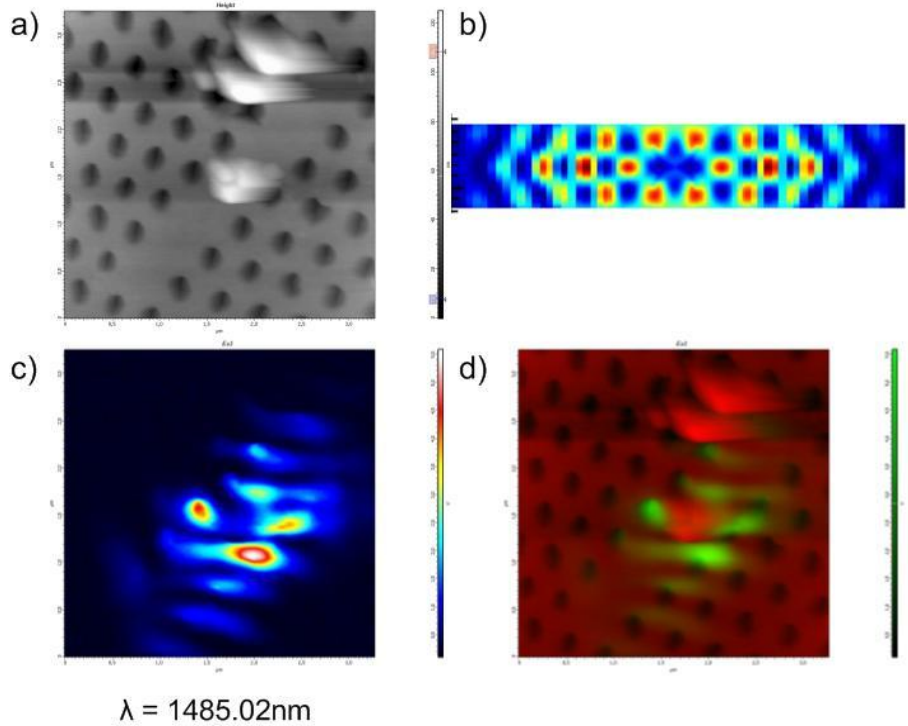

$\lambda = 1485.02nm$

*Figure 4.31 SNOM measurement result of S3-CL7-1.9d-Hxc. a), topography of the PC cavity; b), numerical simulation of the optical field map; c), optical field maps at fixed wavelength of $\lambda = 1485.02nm$; d), superpositions of the optical pattern with the topography correspond to c).*

## 4.4 First step of investigation of the optical properties of graphite hybrid nanodevices

### 4.4.1 Far-field invistigation

The optical properties of graphite PC-NA hybrid structures are investigated as well. In the far-field measurement, most of the equipment set-up is same as the CL5 and CL7 PC-NA hybrid devises. The differences are mainly about the excitation situation. The wavelength of the pulsed laser diode emitting is 780 nm. The pulse with was 15 ns, with a 7.5% duty cycle. The achromatic objective lens is 10X, since the graphite structure should be excited in a large area [3].

Figure 4.32 shows a far-field measurement result of a hybrid structure. Figure 4.32 a) presents the typical emission spectrum corresponding to the laser peak of the mode of the hybrid graphite structures. Figure 4.32 b) shows the variation of the laser peak intensity versus the effective incident optical pump power. Here, we consider the incident mean power and we assume, according to our simulations, that only 35% of the optical pump power is effectively absorbed by the InP barriers. A laser threshold of 2 mW can be inferred from the data.

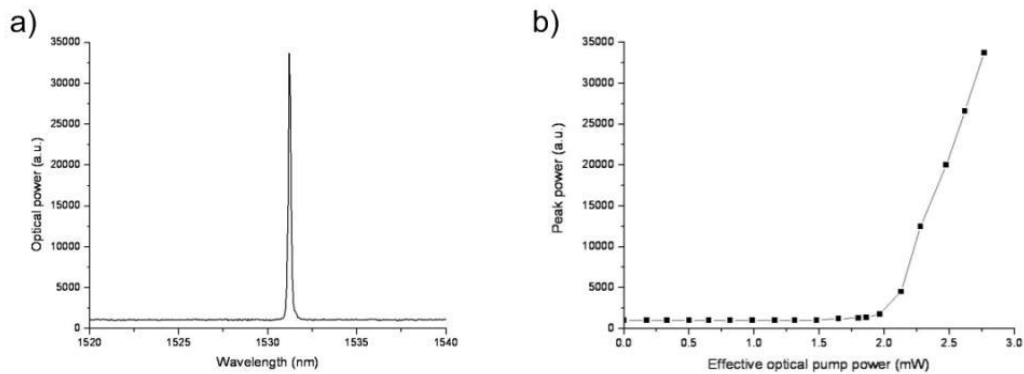

*Figure 4.32 Laser spectrum (a)) and threshold (b)) of a graphite hybrid structure.*

### 4.4.2 Near-field investigation

The set-up of SNOM for the graphite PC structure is same as the CL5 and CL7 PC structures.the only difference is the near-field probe. Since the optical field of graphite PC structure radiates a lot, the silica near-field probe is not adapted the near-field investigation for the optical field distribution of graphite hybrid structure (cf TP Vo PhD thesis) [1]. The aperture metalized-probe is employed.

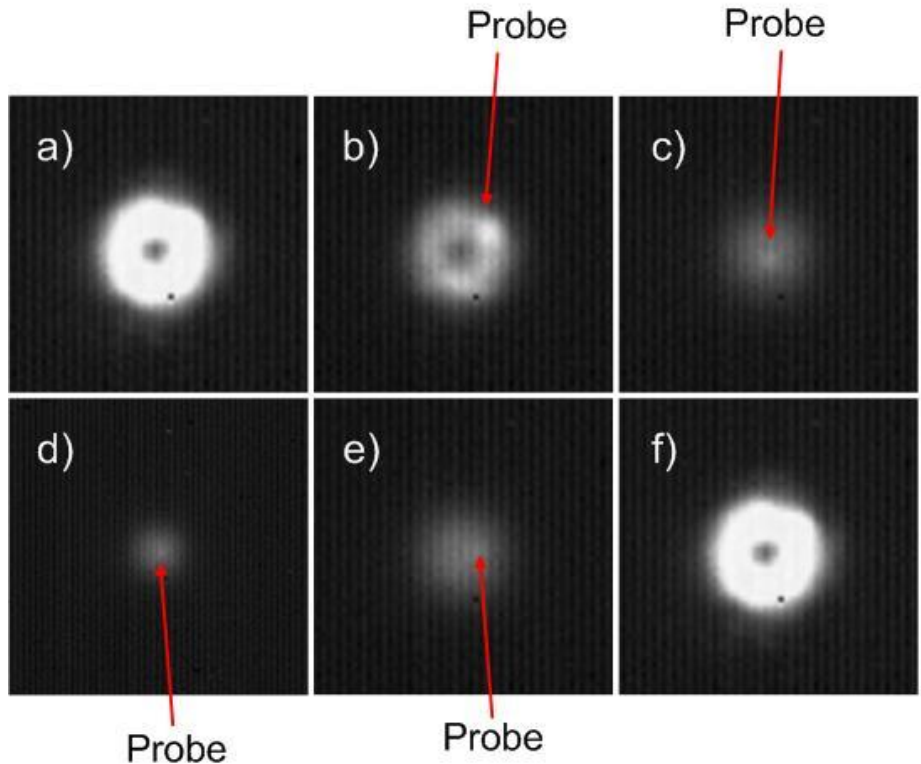

*Figure 4.33 A round-trip of the scanning process on a graphite hybrid structure.*

However, when the probe is approached in to near-field, there is a very strong interaction between the near-field probe and the PC-structure, and the laser mode is killed. Therefore, during the near-field optical field distribution investigation, while the probe scans across the emission area, the laser mode is killed, and while the probe passes the emission region, the laser mode restores. This phenomenon can be obtained via the far-field images recorded by the IR-camera. Figure 4.33 shows a round-trip of the scanning process. In figure 4.33 a), the probe does not enter the emission area. The far-field pattern of the laser mode is very bright and the shape is like a doughnut, same as the pattern which is got in previous studies [1]. Figure 4.33 b) to e) show the probe round-trip process. The red arrows point out the position of the probe. In the images, the probe is a little white spot. When the probe is at the edge of the emission area, the laser mode becomes weak. And when the probe crosses the centre of the emission area, the laser mode disappears. After the probe leaves the emission area, the laser mode restores totally. This phenomenon was not observed in the previous studies [1]. The reason of the phenomenon is not clear. It is maybe due to some chemical residues of the fabrication process change the surface property of the PC structure. And this change yields the interaction between the probe and the PC structure. Therefore, the near-field optical field distribution of this kind of structure can not be obtained. This investigation should be continued with some other suitable near-field probes which have not interaction with the PC-structures and can get high resolution.

## Conclusion

In summary, near-field properties of the PC-NA hybrid structures are investigated. The measurement is difficult to reach. First reason is because the structures are easy to be destroyed by the shear force of the near-field probe. The value of the shear-force is tested before the optical characterization. Second reason is some times there is a huge interaction between the near-field probe and the structures, which induce the measurement difficult.

For the CL5 and CL7 hybrid structures, the measurements present that the coupling between the NA and PC structure modifies the optical field distribution of the laser mode. The modification depends on the position and direction of the NA and sensitive to the polarization of the optical field. If the direction of polarization and the NA is same, the NA can localize and enhance the laser signal in the near-field zone. The investigation demonstrates a plasmonic-photonic crystal laser in nanoscale.

The investigation also presents that a long period observation by MEB Tescan will induces an extraordinary wavelength tuning to the laser mode. The reason is not clear. An interdisciplinary research should be taken to study the phenomenon.

The far-field and near-field properties of the graphite PC-NA hybrid structures are investigated in first step. However, because of the huge interaction between the near-field probe and PC structure, the near-field optical field distribution can not be got.

**Bibliography**

[1] T-P. Vo. *Optical Near-Field Characterization of Slow-Bloch Mode Based Photonic Crystal Devices, doctoral dissertation,* Ecole Centrale de Lyon (2009).

[2] T-P. Vo, A. Rahmani, A. Belarouci, C. Seassal, D. Nedeljkovic and S. Callard. *Near-field and far-field analysis of an azimuthally polarized slow Bloch mode microlaser,* Optics Express 18(26) : 26879-26886 (2010)

[3] D. Turner. *US Patent 4,469,554* (1984)

[4] P. Lambelet, A. Sarah, M. Pfeffer, C. Philipona and F. Marquis-Weible. *Chemically Etched Fiber Tips for Near-Field Optical Microscopy: A Process for Smoother Tips,* Applied Optics, 37(31) : 7289-7292 (1998).

[5] R. Bachelot, C. Ecoffet, D. Deloeil, P. Royer and D-J. Lougnot. *Integration of Micrometer-Sized Polymer Elements at the End of Optical Fibers by Free-Radical Photopolymerization,* Appl. Opt. 40 : 5860-5871 (2001)

[6] R. Bachelot, A. Fares, R. Fikri, D. Barchiesi, G. Lerondel and P. Royer. *Coupling semiconductor lasers into single-mode optical fibers by use of tips grown by photopolymerization,* Opt. Lett. 29 : 1971-1973 (2004)

[7] M. Hocine, R. Bachelot, C. Ecoffet, N. Fressengeas, P. Royer and G. Kugel. *End of the fiber polymer tip: manufacturing and modeling,* Synthetic Metals, 127(1-3) : 26 313-318 (2002).

[8] P.D. Bear. *Microlenses for coupling single-mode fibers to single-mode thin-film waveguides,* Appl. Opt. 19 : 2906-2909 (1980).

[9] B. Hecht, B. Sick, U.P. Wild, V. Deckert R. Zenobi, O.J.F. Martin and D.W. Pohl. *Scanning near-field optical microscopy with aperture probes: Fundamentals and applications,* The Journal of Chemical Physics, 112(18) : 7761-7774 (2000)

[10] B. Cluzel, L. Lalouat, P. Velha, E. Picard, E. Hadji, D. Peyrade, F. de Fornel. *Extraordinary tuning of a nanocavity by a near-field probe,* Photonics and Nanostructures - Fundamentals and Applications, 9 : 269-275 (2011)

[11] G. Le Gac, A. Rahmani, C. Seassal, E. Picard, E. Hadji, S. Callard. *Tuning of an active photonic crystal cavity by an hybrid silica/silicon near-field probe,* Optics Express, 17(24) : 21672-21679 (2009)

[12] S. Mujumdar, A.F. Koenderink, T. Sünner, B.C. Buchler, M. Kamp, A. Forchel and V. Sandoghdar. *Near-field imaging and frequency tuning of a high-Q photonic crystal membrane microcavity,* Optics Express, 15(25) : 17214-17220 (2007)

[13] M. Barth, S. Schietinger, S. Fischer, J. Becker, N. Nüsse, T. Aichele, B. Löchel, C. Sönnichsen and O. Benson. *Nanoassembled Plasmonic-Photonic Hybrid Cavity for Tailored Light-Matter Coupling,* Nano Lett., 10 : 891-895 (2010).

[14] G. Le Gac. *Etude de l'impact d'une pointe SNOM sur les propriétés des modes optiques d'une cavité à base de cristaux photoniques,* doctoral dissertation, Ecole Centrale de Lyon (2009).

# Conclusion and Perspectives

## Conclusion

In this thesis, we design and realize a novel nano-optical device based on the use of a photonic crystal (PC) cavity to generate an efficient coupling between the external source and a nanoantenna (NA). The research work includes nanodevice design, fabrication and characterization.

The PC structures are formed in an InP-based membrane. Four InAsP quantum wells are in the centre of the membrane to act as an optical gain material of laser mode. The PC structures include defect mode PC structures and Bloch mode PC structures. Defect mode PC structures are triangular array of cylindrical holes. 7 holes or 5 holes are omitted to form the defects: CL7 and CL5 cavities. The Bloch mode PC structures are large squares of regular hexagon array of cylindrical holes. The bowtie NAs are placed on the backbone of the PC structures. For the CL7 and CL5 PC cavities, the NAs are placed in three positions on the cavities to see whether the NA can couple to the laser mode of the PC cavity. And for Bloch mode PC structures the NAs are placed in four positions to couple to the monopolar mode of the PC structures.

The PC structures are formed in an active high refractive index InP-based membrane with four InAsP quantum wells in the centre. A $SiO_2$ layer is deposited on the top of the sample as a mask layer for the etching process. Processing of the PC is done by electron beam lithography. Reactive ion beam etching (RIBE) is used to transmit the patterns of PC structures into the InP layer. Individual metallic nano-antennas (NAs) are then deterministically positioned on the backbone of the PC structures by a second e-beam exposure followed by a lift-off process. Overlay measurements showed that the deviation in the alignment error could be as small as 50nm.

Optical properties of the hybrid structure are investigated in both far-field and near-field. The far-field measurement is performed by micro-photoluminescence spectroscopy at room temperature. The results show that the NA couple to the PC cavity and a laser signal can be detected. The coupling increases the lasing threshold. The wavelength of the laser is also impacted.

Near-field scanning optical microscopy (SNOM) has employed to investigate the near-field optical field distribution. However, since the NAs are easily be destroyed by the shear-force of the near-field probes, the investigation is complex. A suitable value of shear force should be set first. The measurement results show that the NA modifies the mode of the structure and localizes the optical field under it. The modification depends on the position of the NA and sensitive to the polarization of the optical field. This novel system may open the route to applications in integrated opto-plasmonic devices for quantum information processing, as efficient single photon sources or nanolasers, or as sensing elements for bio-chemical species.

The far-field and near-field investigation also presents that a long period observation by MEB Tescan will induces an extraordinary wavelength tuning to the laser mode of the CL5 PC cavities and hybrid structures. The reason is not clear. A possible explanation is given here. An interdisciplinary research should be taken to study the phenomenon.

The far-field and near-field properties of the graphite PC-NA hybrid structures are investigated in first step. However, there is a huge interaction between the near-field probe and PC structures, the near-field optical field distribution cannot be got. Further studies on this kind of structure will be taken.

**Perspectives**

For the future research, we will continue to study this kind of nanodevices. Since the research has only just begun, we just got the preliminary results of the properties of this kind of nanodevice. There are still a lot of properties and applications which can be investigated.

In near future, the interaction between the defect PC structures and the NA should be researched in detail. We should understand whether the NA really can be driven more efficiently via adding the PC cavity and whether the NA enhances the intensity of the optical signal from the PC cavity. Hence it is necessary to investigate the interactions between different NAs and different PC cavities. We should note that the position, direction of the NA, and the distance between the NA and the PC cavity will impact the interaction of them. After these researches we can get the plasmonic-photonic crystal hybrid system with an efficient coupling and low losses.

The extraordinary wavelength tuning phenomenon will be investigated in detail. The surface characterization of the sample and the impact of the SEM observation will be studied.

In far future, the interactions between the plasmonic-photonic hybrid structure and other nanostructures will be investigated. For instance, quantum dots (QDs). The NA can confine the optical signal from the PC cavity in a nanometer scale. This scale just matches the size of the QDs. And the NA can enhance the exciting light intensity to provide an efficient exciting of the QDs. Therefore, if the NA does not quench the QDs, we could provide a route to more efficiently exciting the QDs.

Several other applications of this kind of plasmonic-photonic hybrid nanodevices will be extended including efficient single photon sources and single molecule detection.

# Appendix

**Appendix A: Patterns for Autoalignment E-Beam lithography**

For each group of hybrid structures, four write field alignment marks are put at the corners of the write field (Figure 1). The alignment marks are crosses. Each arm of the crosses is 5µm long and the head of the arms are 1µm wide.

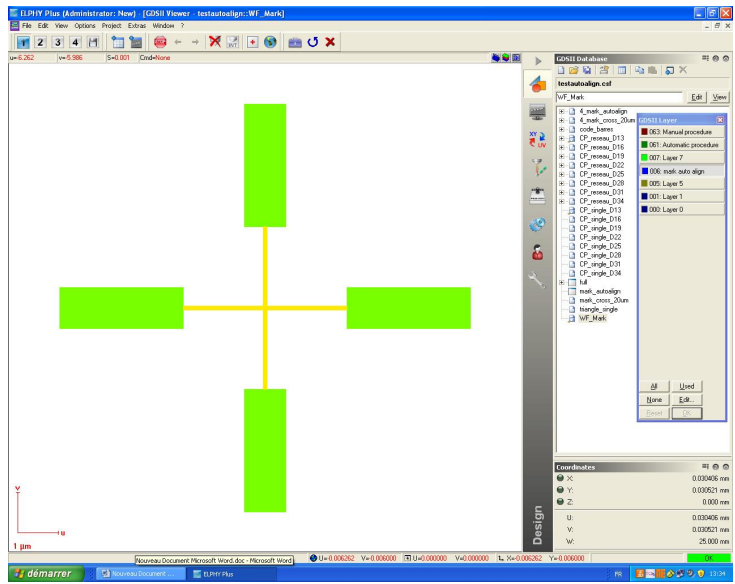

*Figure 1 Detail of an alignment mark*

Four large crosses are put at the four corners of the full pattern. During the second and third steps of lithography, they will be used for looking for the working area, and will be used in the step of stage to sample adjustment. The size of each cross is 60µm by 60 µm (Figure 2).

*Figure 2 Detail of an large mark*

In each full pattern there is also a group of vernier pattern (Figure 3). There are two groups of rods, which are used to measure the mismatch of the alignment between two steps of lithography. One group is in horizontal (X) direction and another is in vertical (Y) direction. Each group contains two rows of rods, side by side. Each row consists of 19 rods. The width of each rod is 200nm. The two rods in the centre are longer than the others and they are exactly head to head. The distance between each rod is 1µm for the lower and right row, and 1.01µm for upper and left row. So there will be an alignment mismatch from -90nm to 90nm between the rods lower and upper, and the same to the right and left (Figure 4)[1]. From the centre to the edge, for the first pair of rods, the missing is 10nm. For the second pair of rods, the missing is 20nm, until the last pair of rods, the missing is 90nm. The upper rods and left rods are outward. The upper row and left row are set in the same layer with photonic crystals. And the rows which are opposite to them are set in the same layer with the autolignment marks. So if there is an alignment mismatch between the write fields of the two steps of lithography for photonic crystal and autolignment marks, the rods in the centre will be not aligned, and another pair of rods will be aligned, then the mismatch can be measured. For instance, in the horizontal rows, if the first pair of rods in right side is aligned very well, it means the write field of photonic crystals has a 10nm shift to left, corresponds to the write field of autoalignment marks. As shown in Figure 5.

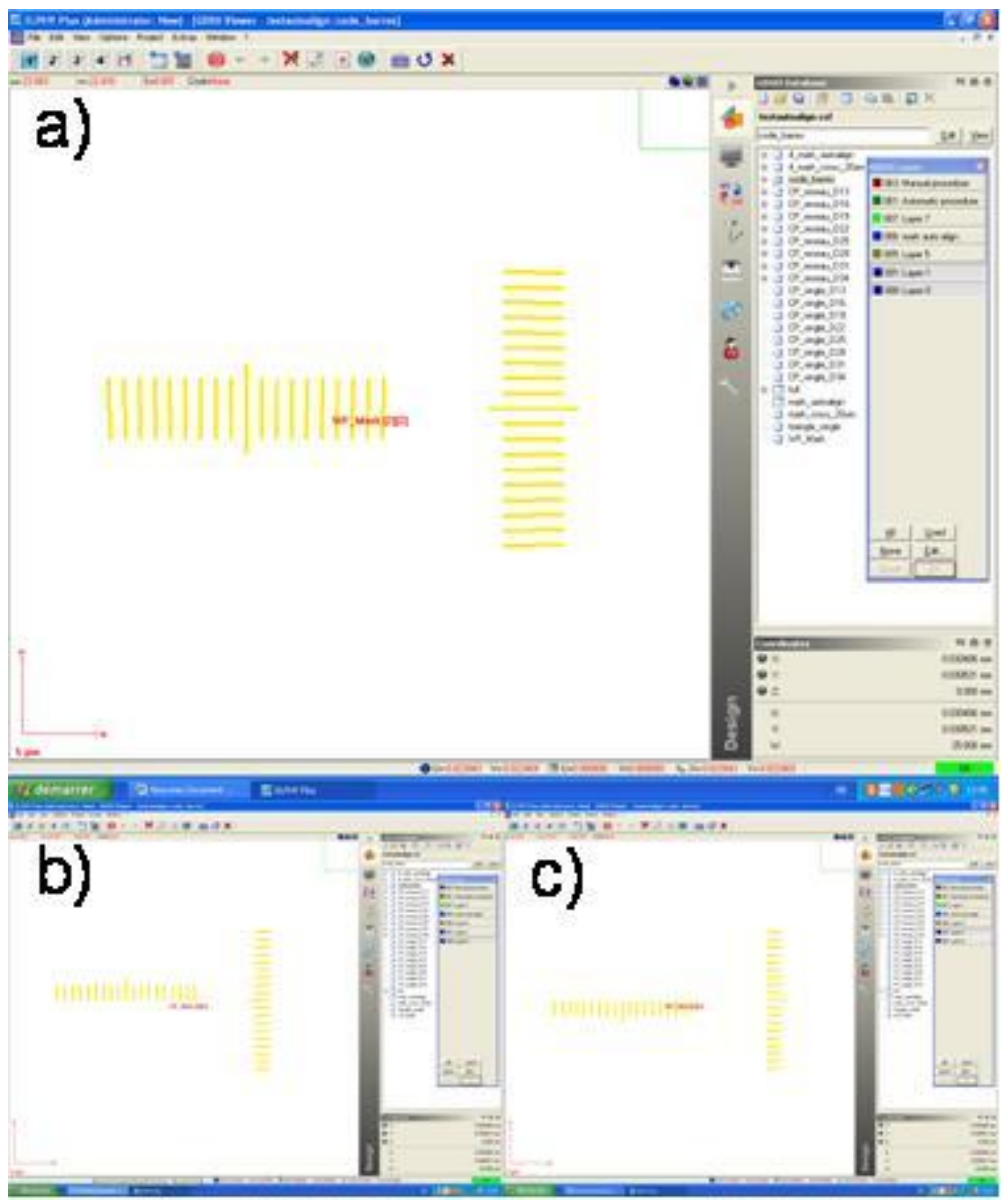

*Figure 3 Over view of vernier pattern. a) full pattern; b) upper and left rows; c) lower and right rows.*

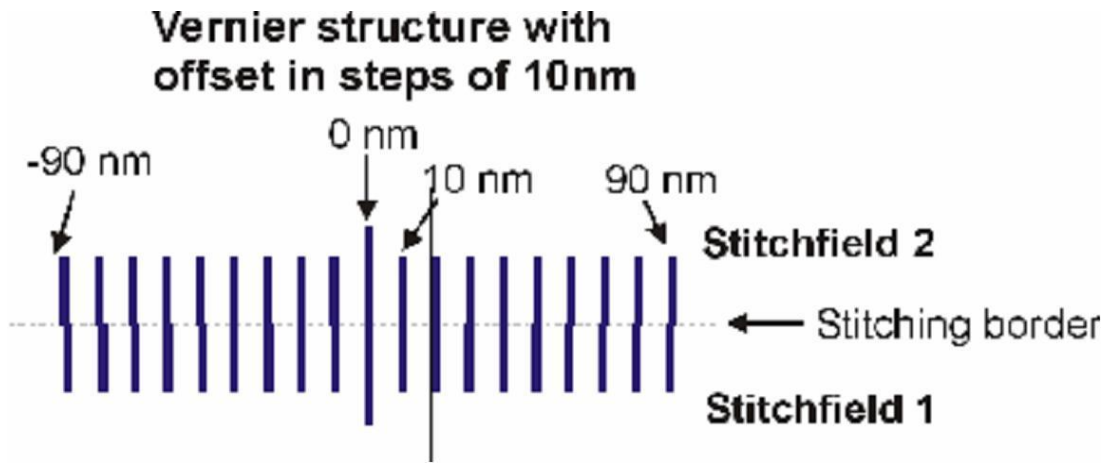

*Figure 4 Detail of the vernier pattern. Figure from reference [1]*

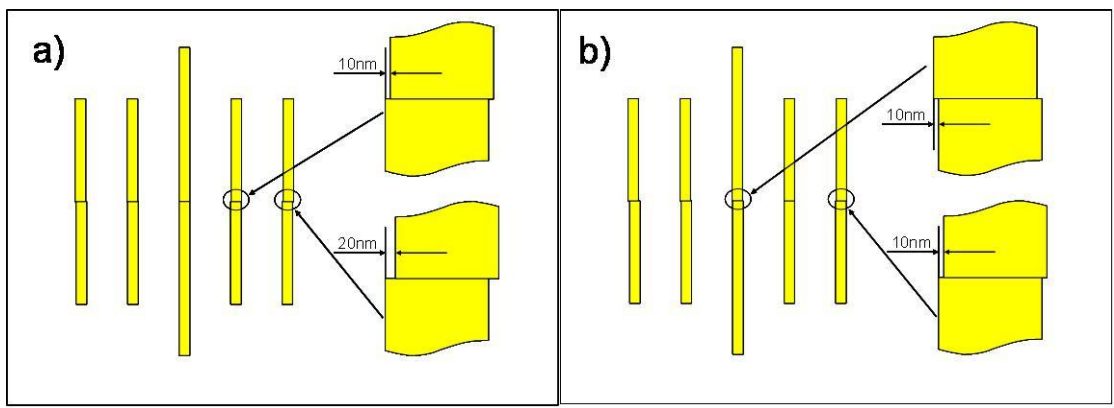

*Figure 5 Diagram of measuring the mismatch of write field by the vernier structure*

**Bibliography**

[1] *Demo Pattern and Performance Test For Raith Lithography Systems,* (Raith GmbH).

## Appendix B: Stage to sample adjustment and Write field alignment of E-Beam lithography

The steps of "stage to sample adjustment" and "write field alignment" must be operated very carefully. The accuracy of the autoalignment lithography depends on the accuracy of the positions of the alignment marks. The global transformation should be chosen to transform the coordinate systems from the stage (X, Y, Z) to the sample (U, V, W). And after origin correction and angle correction (repeat one time for accuracy), 3 points transformation was selected. It is used to calculate the shift, two rotation angles and two scaling factors between the two systems [1]. In this step, some dusts were found at three locations on the sample. The dusts were far from each other and not at the working areas. For the first dust, after got a good focus on it, a contamination point (Figure 1) was made by the focused e-beam near the dust. After got good focus on the contamination point, the location in the (X, Y, Z) system of the point was read, the location in the (U, V, W) system of the point was input. The coordinate system transformation was proceed (Figure 2). Then the same process to the other two locations. The process of 3 points transformation should be repeated several times with the contamination points, until got an accurate transformation. For the origin correction and write field alignment, a contamination point could be made and used also.

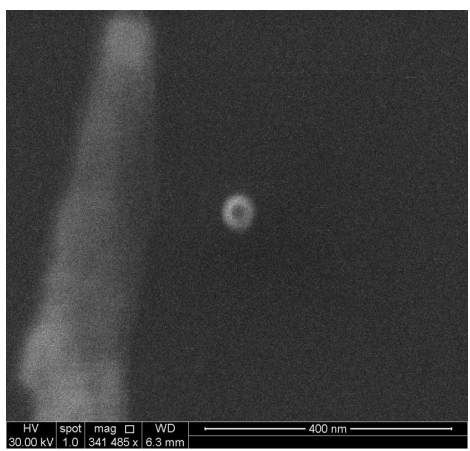

*Figure 1 SEM image of a contamination point*

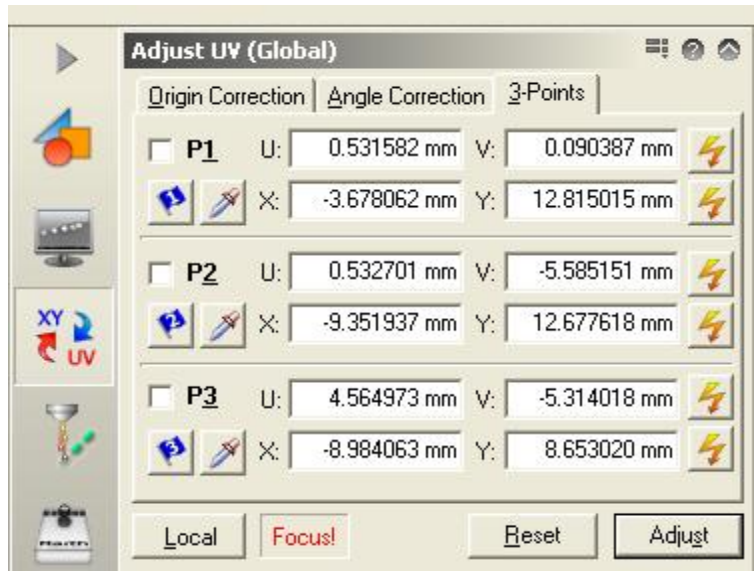

*Figure 2 Adjust UVW window of E-Beam lithography*

For the write field alignment, since the system is equipped with a laser interferometer stage, the laser stage could be used as measure standard. A contamination point was selected as a mark (for the equipment with laser interferometer stage, it only needs one mark) [1]. In this research the write field should be set as 100μm by 100μm and the magnification was set as x1100. The alignment procedure was selected, and then scanned over the mark. In the mark window, there was a green cross in the centre of the image (Figure 3). The green cross defined the expected location of the mark [1]. The cross was drag and drop on to the mark in each image and the correction values were applied. The system would calculate the write field transformation. This sequence should be repeated with scan fields from 25μm decreased to 500nm for accuracy. A very precise alignment could be got due to the precision of the laser stage [1].

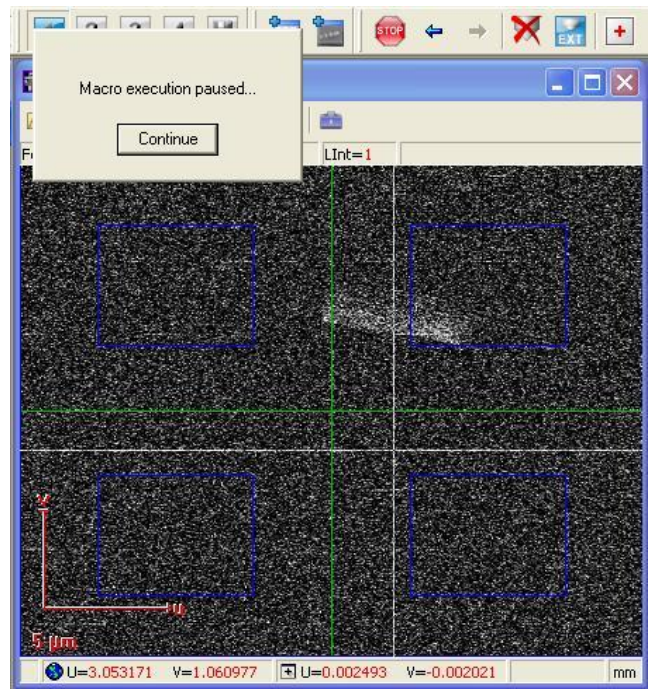

*Figure 3 Mark window of E-Beam lithography*

In the position list, we put the full patterns as a 7 lines and 7 columns matrix. The period of the matrix was 500µm, same as the length of each full pattern. For the scanning pattern, only layer 5 and 6 were scanned to put the autoalignment marks and the large crosses.

| ID | U/mm | V/mm | Attribute | Template | Comment | O |
|---|---|---|---|---|---|---|
| 0 | 0.000000 | 0.000000 | S5 | UV | full | |
| 1 | 0.500000 | 0.000000 | S5 | UV | full | |
| 2 | 1.000000 | 0.000000 | S5 | UV | full | |
| 3 | 1.500000 | 0.000000 | S5 | UV | full | |
| 4 | 2.000000 | 0.000000 | S5 | UV | full | |
| 5 | 2.500000 | 0.000000 | S5 | UV | full | |
| 6 | 3.000000 | 0.000000 | S5 | UV | full | |

*Figure 4 Position list of E-Beam lithography*

The contamination point can also be used to check the astigmatism correction of the e-beam, via the shape of the contamination point. If the shape of the point is very round, it means the astigmatism correction is very good. Conversely, if the point is elliptic, the astigmatism correction is not good.

**Bibliography**

[1] *ELPHY Plus Software Reference Manual Version 5.0,* (Raith GmbH).

## Appendix C: Stage to sample adjustment and Write field alignment of autoalignment E-Beam lithography

In this step of lithography, the step of "stage to sample adjustment" was different from Appendix B. After found the matrix of the marks, a large cross at a suitable location was chosen to do the origin correction. For instance the large cross which was at the left down corner, noted it as cross 1. Note that the left down corner of this cross should be set as the origin in both global transformation and local transformation. The location need not to be very accurate. The angle correction was also proceed in both global transformation and local transformation. A position at the edge of a horizontal arm of cross 1 was set as label 1. Then another large cross in the same line in horizontal direction was chosen, noted as cross 2. Cross 2 should be far from Cross 1. A position at the edge of the same horizontal arm of Cross 2 was set as label 2, and then adjusted the angle correction. This step should be repeated two times to get a very accurate angle correction. For the 3 point correction, the local transformation must be set. A contamination point was made in the centre of cross 1, and then got good focus on it and set it as point 1. Because the cross is at the left down corner of the write field, and the size is 60μm by 60μm, so the location of point 1 must be set as U = 0.03mm and V = 0.03mm. And then read the location in (X, Y, Z) system and did the coordinate system transformation. To set point 2, another large cross which is far from the origin could be directly arrived via stage control destination tab. For instance, the cross at the left down corner of the 7th full structure in the same horizontal line, it could directly arrived by input U = 3.03mm and V = 0.03mm in stage control command tab. The position actually arrived should be not far from the centre of the cross. Another contamination point was made in the centre of the cross and the location of the contamination point in (U, V, W) system was set as U = 3.03mm and V = 0.03mm. Then read the location in (X, Y, Z) system and did the coordinate system transformation. For point 3, the process as same as point 2 was taken. The locations of the 3 point should be far from each other. And the location reading and system transformation process with the points was also repeated until got an accurate transformation in this time of lithography (Figure 1).

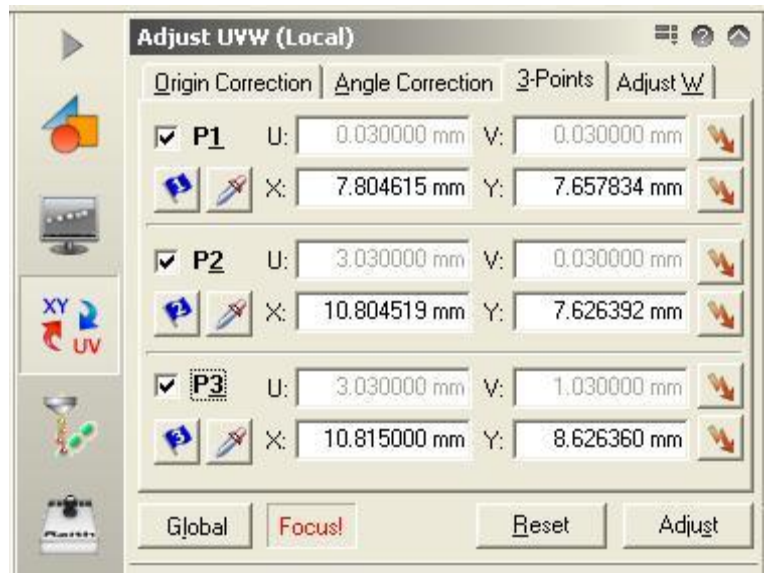

*Figure 1 Adjust UVW window of autoalignment E-Beam lithography*

For the write field alignment, a contamination point was made in the centre of a large cross. The process of the write field alignment was same as the first time of lithography. After the write field alignment, it was the process of checking the accuracy of the autolignment system. The first full pattern was dragged and dropped to the position list. The position of the pattern was set at the centre of the first write field of the full pattern, U = 0.05mm and V = 0.05mm. The layer 63 was set. it is the manual alignment process. After scanned, the autoalignment marks would be shown in the mark window one by one in order of the write fields. The green cross need to be dragged and dropped to the centre of the autoalignment mark (Figure 2). Just like the write field alignment process. If the autoalignment marks were absent in the mark window, it meant the autolignment system did not work.

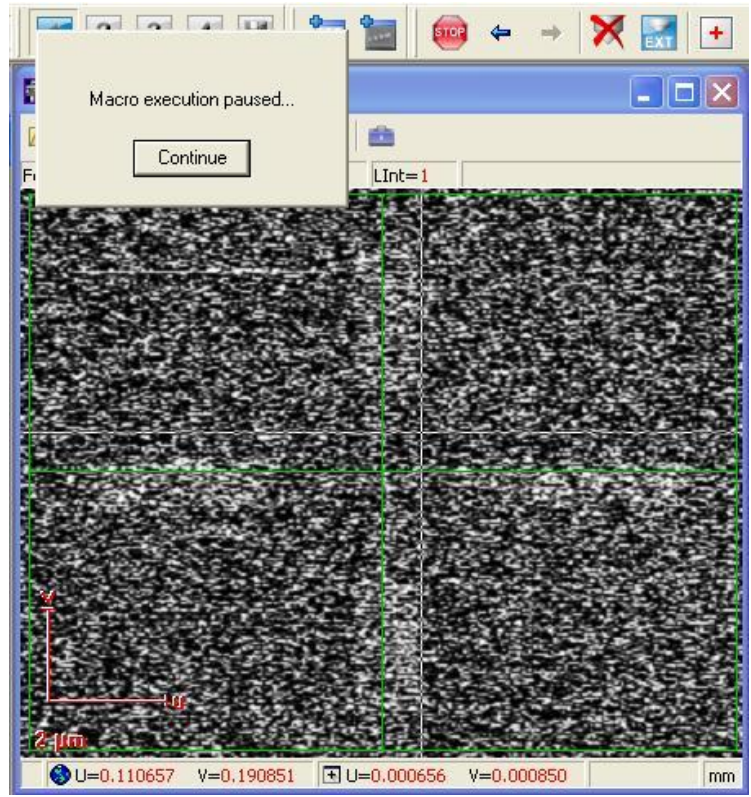

*Figure 2 Mark window of autoalignment E-Beam lithography*

After the manual alignment process, the layer 61 was set to test the autoalignment process. After the system started scan, the system would did several times of line scan over the arms of the alignment marks, in order of write fields for each full pattern. In each write field, the system did line scan on two arms of the alignment marks at the four corners. For each mark, one line scan was on a horizontal arm, another line scan was on a vertical arm (Figure 3). The software could record and display the intensity of the secondary electron as a function of the position [1]. The intensity of the marks is much higher than it of the substrate. So the system could find the location of the alignment marks and set the location of the write field same as the first time of lithography via the difference of intensity (Figure 4).

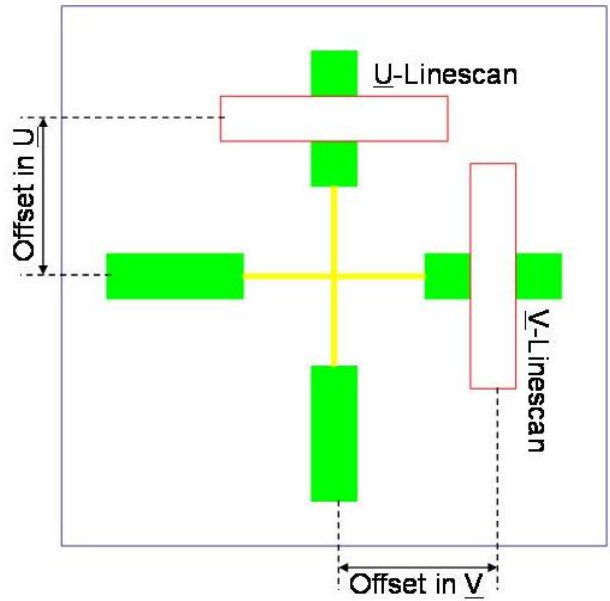

*Figure 3 Line scans on the arms of an alignment mark*

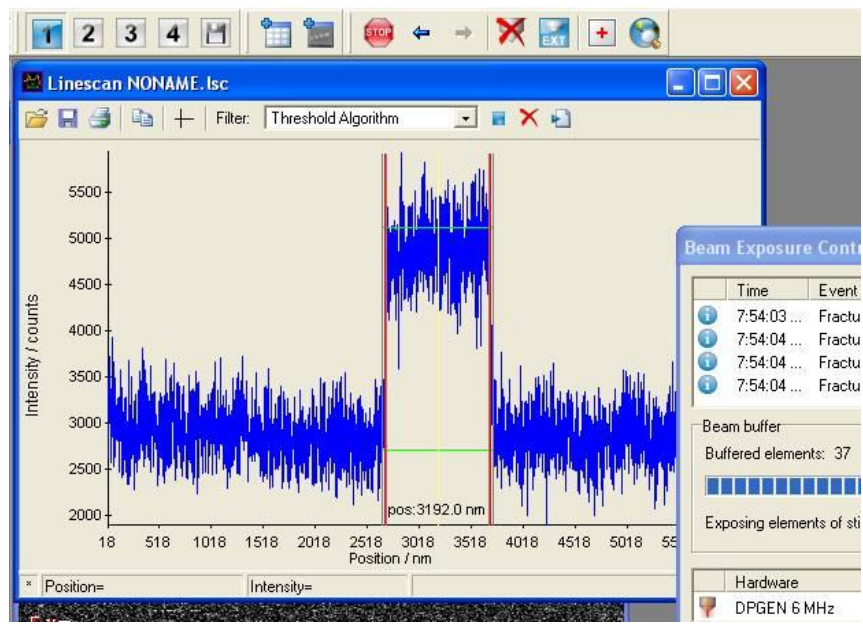

*Figure 4 The intensity of the secondary electron of a line scan*

After checking the autoalignment system, the full patterns which would be scanned were put into the position list and input the locations of each patterns were input. The locations which were input must be the centre locations of the first write field of the full patterns (Figure 5). And the positions of working areas must same as the marks which we put in the first lithography. Layer 61 was set for doing the autoalignment and layer 0 for scanning photonic crystals. During the lithography, the system would first do the autoalignment and then scan the photonic crystals, in order of write fields. A big square with size 50µm x 50µm should be wrote for the label of RIE.

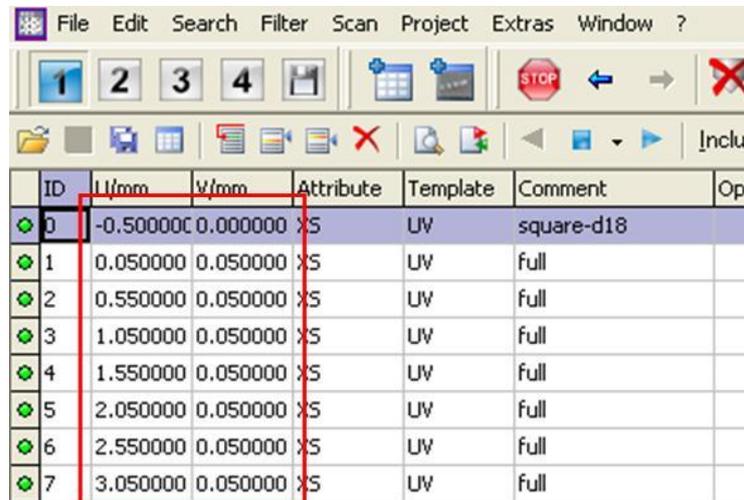

*Figure 5 Position list of autoalignment E-Beam lithography*

**Bibliography**

[1] *ELPHY Plus Software Reference Manual Version 5.0,* (Raith GmbH).

## Appendix D: The procedure of the calibration of pump power

In the far-field measurement for the CL7 and CL5 structures, the period of the pumping laser is 233 ns and the wide of the pulse is 22.5 ns, hence the duty cycle is about 10%.The objective is 20X. During the far-field measurement on the structures, the effective incident optical pump power on the structures should be given. To do this, before the far-field measurement on the structures, a range of the powers of the laser source, the position infrant of the objective and the plane of the sample were measured. The power of the laser source was measured by a keithley 197A autoranging microvolt DMM linked to the laser source. The unit is µA. The power in front the objective and the surface of the sample was also measured. They were measured by a Newport power meter model 1918-C. The mode of the power meter was CW continue mode. The unit is mW.

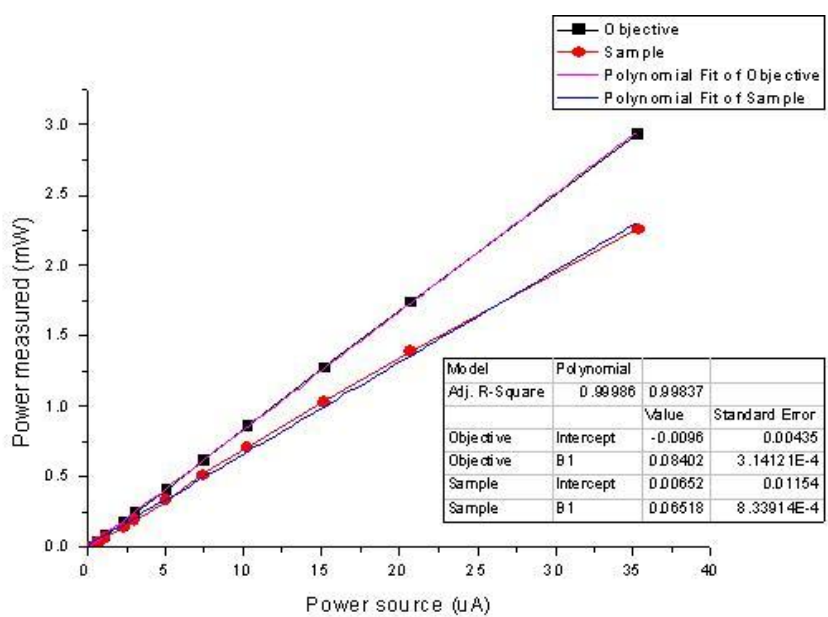

*Figure 1 The relation between the three powers*

The relation between these three powers is shown in Figure 1. From this line graph, after filter by the Orign software, it shows that the relationship between the pumping powers on the sample surface and the laser source is $P_{sample}$ (µw) = 65.18 x $P_{source}$ (µA).

Therefore, during the far-field measurement, the effective incident optical pump power on the structures can be calculate directly from the powers of laser source via the relationship.

# Plasmonic-photonic hybrid nanodevice

Metallic nano-particles or nano-antennas (NAs) provide a strong spatial confinement down to the subwavelength regime. However, a key challenge is to address and collect light from those nano-scale systems. The tiny active area of the NA is both an advantage for its miniaturization, and a real limit for the level of the collected signal. Therefore, one needs to reconsider how to drive efficiently such NA. Here, we propose to tackle this important issue by designing and realizing a novel nano-optical device based on the use of a photonic crystal cavity (PC cavity) to generate an efficient coupling between the external source and a NA. In this thesis, we design and realize a novel nano-optical device based on the coupling engineering of a photonic crystal (PC) cavity and a nanoantenna (NA). The research work includes nanodevice design, fabrication and characterization.

The PC structures are formed in an InP-based membrane with four InAsP quantum wells are in the centre of the membrane to act as an optical gain material of laser mode. The PC structures include defect mode PC structures and Bloch mode PC structures. The bowtie NAs are placed on the backbone of the PC structures. The fabrication of the PC is done by electron beam lithography. Reactive ion beam etching (RIBE) is used to transmit the patterns of PC structures into the InP layer. The NAs are then deterministically positioned on the PC structures by a second e-beam exposure followed by a lift-off process. Overlay measurements showed that the deviation in the alignment error could be as small as 20nm.

Optical properties of the hybrid structure are investigated in both far-field and near-field. The far-field measurement shows that the NA increases the lasing threshold of the PC cavity. The wavelength of the laser is also impacted. Near-field scanning optical microscopy (SNOM) has employed to investigate the near-field optical field distribution. The measurement results show that the NA modifies the mode of the structure and localizes the optical field under it. The modification depends on the position and orientation of the NA.